\documentclass[11pt, a4paper]{article}
\usepackage{jheparxiv,integrable}
\usepackage[latin1]{inputenc}
\usepackage[compat=1.1.0]{tikz-feynman}
\usetikzlibrary{arrows,positioning,arrows.meta}
\tikzfeynmanset{ with arrow/.style = {
   decoration={
     markings,
     mark=at position 0.5
          with {\arrow[xshift=1mm]{stealth'[width=2mm,length=3mm]}}
     },
   postaction=decorate}
}
\usepackage{amsmath}
\usepackage{amsfonts}
\usepackage{amssymb}
\usepackage{latexsym}
\usepackage{mathrsfs}
\usepackage{graphicx}
\usepackage{color}
\usepackage{slashed}
\usepackage[all]{xy}
\usepackage[compat=1.1.0]{tikz-feynman}
\usepackage{multirow}
\usepackage{mathtools}
\usepackage{tcolorbox}
\usepackage{subcaption}

\subheader{\hfill \texttt{DAMTP-2019-7}}

\title{Gauge Theory and Boundary Integrability}

\author{Roland Bittleston and David Skinner}

\affiliation{Department of Applied Mathematics \& Theoretical Physics \\
        University of Cambridge \\
        Wilberforce Road \\
        Cambridge CB3 0WA, United Kingdom}

\emailAdd{[r.bittleston, d.b.skinner]@damtp.cam.ac.uk}

\abstract{We study the mixed topological / holomorphic Chern-Simons theory of Costello, Witten and Yamazaki on an orbifold $(\Sigma\times\C)/\Z_2$, obtaining a description of lattice integrable systems in the presence of a boundary. By performing an order $\hbar$ calculation we derive a formula for the the asymptotic behaviour of $K$-matrices associated to rational, quasi-classical $R$-matrices. The $\Z_2$-action on $\Sigma\times \C$ fixes a line $L$, and line operators on $L$ are shown to be labelled by representations of the twisted Yangian. The OPE of such a line operator with a Wilson line in the bulk is shown to give the coproduct of the twisted Yangian. We give the gauge theory realisation of the Sklyanin determinant and related conditions in the $RTT$ presentation of the boundary Yang-Baxter equation.}

\begin{document}
 
\maketitle

\newpage

\section{Introduction}

\subsection{The Yang-Baxter Equation}
Interactions of an integrable spin chain are determined by an object known as an $R$-matrix. This is a linear map
\[
R(z,z') : V\otimes V'\to V\otimes V'
\]
for $V$, $V'$ a pair of complex vector spaces and $z$, $z'$ complex spectral parameters on which the $R$-matrix depends meromorphically.  The spin chain is integrable if the $R$-matrix obeys the Yang-Baxter equation (YBE)
\[
R_{12}(z_1,z_2)R_{13}(z_1,z_3)R_{23}(z_2,z_3) 
= R_{23}(z_2,z_3)R_{13}(z_1,z_3)R_{12}(z_1,z_2)\,,
\]
where $R_{12}(z_1,z_2)$ acts with $R(z_1,z_2)$ on the first two factors in the tensor product $V_1\otimes V_2\otimes V_3$ and with the identity on the third, and the action of a general $R_{ij}(z_i,z_j)$ is defined similarly. This equation is  more easily understood geometrically as
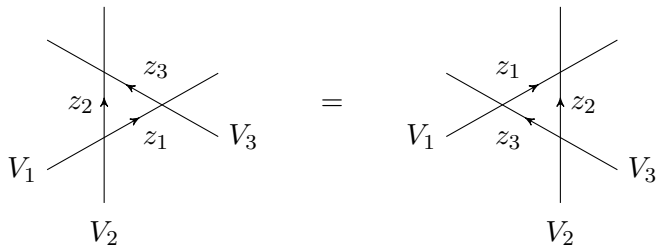
\begin{figure}[h!]
	\centering
	\begin{tikzpicture}[baseline]
	\begin{feynman}
	\vertex (o){$=$};
	\vertex[left=3cm of o] (l);
	\vertex[above=1.3cm of l] (lbf);
	\vertex[below=1.3cm of l] (lbi);
	\vertex[left=0.75cm of l](ll);
	\vertex[above=0.86cm of ll](lcf);
	\vertex[below=0.86cm of ll](lai);
	\vertex[right=1.5cm of l](lr);
	\vertex[above=0.42cm of lr](laf);
	\vertex[below=0.42cm of lr](lci);
	\vertex[right=3cm of o] (r);
	\vertex[above=1.3cm of r] (rbf);
	\vertex[below=1.3cm of r] (rbi);
	\vertex[right=0.75cm of r](rr);
	\vertex[above=0.86cm of rr](raf);
	\vertex[below=0.86cm of rr](rci);
	\vertex[left=1.5cm of r](rl);
	\vertex[above=0.42cm of rl](rcf);
	\vertex[below=0.42cm of rl](rai);
	\diagram*{{[edges=fermion] (lai) --[edge label'=$z_1$] (laf), (lbi) --  [edge label=$z_2$](lbf), (lci) --  [edge label'=$z_3$] (lcf), (rai) -- [edge label=$z_1$] (raf), (rbi) --  [edge label'=$z_2$](rbf), (rci) --  [edge label=$z_3$] (rcf)},};
	\vertex[left=0cm of lai]{$V_1$};
	\vertex[below=0.1cm of lbi]{$V_2$};
	\vertex[right=0cm of lci]{$V_3$};
	\vertex[left=0cm of rai]{$V_1$};
	\vertex[below=0.1cm of rbi]{$V_2$};
	\vertex[right=0cm of rci]{$V_3$};
	\end{feynman}
	\end{tikzpicture}
\end{figure}

\vspace{-0.5cm}\noindent where whenever two lines cross we act on the vector spaces associated to these lines with the $R$-matrix, with the arrows indicating the order in which the $R$-matrices act. 

Naively, the YBE spectacularly over-determines the $R$-matrix. For example, in the simplest case that all three vector spaces are the copies of the same $V$,  the YBE consists of $\mathcal{O}\big((\dim V)^3\big)$ equations whereas the $R$-matrix itself has only $\mathcal{O}\big((\dim V)^2\big)$ components. It is therefore remarkable that the YBE admits many non-trivial solutions. Integrable spin chain interactions are thus very special, and it is natural to ask if they have a common underlying origin.

\medskip

In~\cite{Costello:2013zra} Costello has shown that a large class of solutions of the YBE can be understood using a variant of Chern-Simons theory, defined on the product $\Sigma\times C$ of a 2-surface $\Sigma$ and a Riemann surface $C$, so as to be topological on $\Sigma$ but only holomorphic in $C$. Costello's theory, and its further development by Costello-Witten-Yamazaki~\cite{Costello:2017dso,Costello:2018gyb}, will be reviewed in section~\ref{sec:Theory}, but in brief, lines in the YBE are associated with Wilson lines in $\Sigma$ that sit at points $z\in C$. Topological invariance in $\Sigma$ means that the only Feynman diagrams which contribute to the correlator of an array of such line operators are those which join points that coincide in $\Sigma$. Thus the correlation function leads to a local $R$-matrix at each crossing. Furthermore, since the line operators are generically at distinct locations in $C$, one encounters no singularity in moving between the configurations representing the left- and right-hand sides of the YBE, so the $R$-matrices generated by the gauge theory are  integrable by construction.

The mixed topological-holomorphic Chern-Simons theory of~\cite{Costello:2013zra,Costello:2017dso,Costello:2018gyb} is defined only perturbatively, so the $R$-matrices obtained this way have an expansion
\begin{equation}
\label{eq:qcrm} 
R_\hbar(z,z') = {\bf 1}_{V\otimes V'} + \hbar\,r_{V\otimes V'}(z,z') 
+ \mathcal{O}(\hbar^2)\,.
\end{equation}
around the identity ${\bf 1}_{V\otimes V'}$. $R$-matrices admitting such an expansion are called {\it quasi-classical}, and $r(z,z')$ is known as the classical $r$-matrix. This classical $r$-matrix takes values in $\fg\otimes\fg$ for $\fg$ a finite dimensional, complex, simple Lie algebra, acting in representations associated to $V\otimes V'$.  Expanding the YBE to second order in $\hbar$ shows that the classical $r$-matrix obeys the classical Yang-Baxter equation,
\[
[r_{23}(z_2,z_3),r_{13}(z_1,z_3)] + [r_{13}(z_1,z_3),r_{12}(z_1,z_2)] 
+ [r_{23}(z_2,z_3),r_{12}(z_1,z_2)] = 0\,.
\]
Solutions of the classical Yang-Baxter equations were classified by Belavin \& Drinfeld~\cite{belavin1982solutions}, under the (mild) assumption that the $r$-matrix is non-degenerate. The solutions can be separated into three families, distinguished by whether the classical $r$-matrix can be written in terms of rational, trigonometric, or elliptic functions. In this work we will concentrate on the rational case. 

Each family of solutions to is associated with an algebra, which in the rational case is the Yangian $\cY(\fg)$. Indeed, in~\cite{drinfeld1990hopf} it was demonstrated that all rational solutions of the YBE of the form~\eqref{eq:qcrm} determine a representation of the Yangian on $V$, and are themselves determined by such a representation.  Accordingly, the line operators in Costello's theory are actually associated with representations of $\cY(\fg)$. These include ordinary Wilson lines in certain representations of $\fg$  itself, but also more general line operators. As shown in~\cite{Costello:2013zra,Costello:2017dso,Costello:2018gyb}, the more general line operators arise even in the OPE of two ordinary Wilson lines.

Since the YBE is homogeneous, its solutions are defined only up to multiplication by a function $f(z_1,z_2)$ of the spectral parameters. This degeneracy can be removed by required that the $R$-matrix obeys a constraint known as the quantum determinant condition. In the rational case,  for $V$ the defining vector representation of $\fg=\mathfrak{sl}_n(\C)$,  the quantum determinant~\cite{Kulish:1981bi} condition is enough to guarantee existence of a unique quasi-classical $R$-matrix in this representation for a given classical $r$-matrix. Similar results exist for different choices of representation and $\fg$. The quantum determinant naturally arises in the $RTT$ presentation of the Yangian~\cite{drinfeld1986quantum}, and in~\cite{Costello:2018gyb} was interpreted in terms of networks of Wilson lines in gauge theory.

\subsection{The Boundary Yang-Baxter Equation} 
\label{subsec:Kmxintro}

In this paper we will be concerned with extending the gauge theory approach of~\cite{Costello:2013zra,Costello:2017dso,Costello:2018gyb} to the case where $\Sigma$ has a boundary. Investigations of integrability--preserving boundary conditions date back to work of Skylanin~\cite{sklyanin1988boundary}, Olshanski~\cite{olshanskii1992twisted,molev1996yangians}, and they have since been extensively studied in the context of open spin chains ({\it e.g.}~\cite{kulish1996yang}), arising for example in the Hubbard model~\cite{zhou1996quantum,guan1997lax,shiroishi1997integrable} and giant magnon interactions D-branes in AdS/CFT~\cite{berenstein2005integrable,hofman2007reflecting,regelskis2012quantum}. 

In the simplest case, boundary conditions on a spin chain are encoded in a $K$-matrix $K(z):V\to V$, which again depends meromorphically on the spectral parameter associated to the vector space $V$. The boundary conditions preserve integrability if the $K$-matrix obeys the boundary Yang-Baxter equation (bYBE), given by
\[
R_{12}(z_1,z_2)K_1(z_1)R_{21}(z_2,-z_1)K_2(z_2) = K_2(z_2)R_{12}(z_1,-z_2)K_1(z_1)R_{21}(-z_2,-z_1)
\]
in the rational case. As before, the indices just tell us on which factors of the tensor product $V\otimes V$ the various operators act. This equation also has a much clearer geometric interpretation:
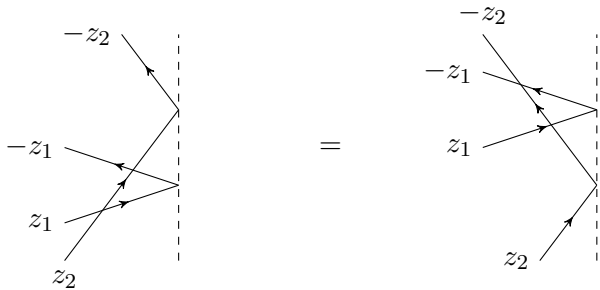
\begin{figure}[th]
	\centering
	\begin{tikzpicture}[baseline]
	\begin{feynman}
	\vertex (o){$=$};
	\vertex[left=2cm of o](l);
	\vertex[above=0.5cm of l](lb);
	\vertex[below=0.5cm of l](la);
	\vertex[above=1.5cm of l](ln);
	\vertex[below=1.5cm of l](ls);
	\vertex[left=1.5cm of ls](lsw);
	\vertex[left=1.5cm of l](lw);
	\vertex[below=1cm of lw](lai);
	\vertex[left=0.75cm of ln](lbf);
	\vertex[right=3.5cm of o](r);
	\vertex[above=0.5cm of r](ra);
	\vertex[below=0.5cm of r](rb);
	\vertex[above=1.5cm of r](rn);
	\vertex[below=1.5cm of r](rs);
	\vertex[left=1.5cm of rn](rnw);
	\vertex[left=1.5cm of r](rw);
	\vertex[above=1cm of rw](raf);
	\vertex[left=0.75cm of rs](rbi);
	\diagram*{{[edges=fermion] (lai) --  (la), (la) --  (lw), (lsw) --  (lb), (lb) --  (lbf), (rw) -- (ra), (ra) -- (raf), (rbi) -- (rb), (rb) -- (rnw)},{[edges=scalar] (ls) -- (ln), (rs)--(rn)},};
	\vertex[below=0cm of lsw]{$z_2$};
	\vertex[left=0cm of lai]{$z_1$};
	\vertex[left=0cm of lw]{$-z_1$};
	\vertex[left=0cm of lbf]{$-z_2$};
	\vertex[left=0cm of rbi]{$z_2$};
	\vertex[left=0cm of rw]{$z_1$};
	\vertex[left=0cm of raf]{$-z_1$};
	\vertex[above=0cm of rnw]{$-z_2$};
	\end{feynman}
	\end{tikzpicture}
\end{figure}\\

\vspace{-0.5cm}\noindent where we note that (in the rational case) when a line reflects of the boundary its spectral parameter changes sign. This diagram makes it clear that the bYBE should also be amenable to Costello's approach. Exactly as in the YBE, one encounters no singularities when moving line operators at different locations $z_i\in C$ from the configuration on the left- to the right-hand side of the bYBE. 

\medskip

In this paper we explore the mixed topological-holomorphic Chern-Simons theory on an orbifold $\widetilde{M}$. The goal is to understand how the gauge theory generates integrable $K$-matrices and their associated algebraic structure. In section~\ref{sec:Theory} we study the gauge field $A$ and its action, imposing conditions on $A$ that ensure the theory remains well-defined on $\widetilde{M}$. We study line operators on the orbifold, highlighting the possibility a line operator living along the orbifold singularity.

In section~\ref{sec:Kmx} we show interactions among such line operators generate integrable $K$-matrices. Since we work perturbatively, we must introduce a notion of a quasi-classical $K$-matrix.  This needs to be defined with some care: unlike the quasi-classical $R$-matrices, the leading ($\hbar=0$) term in a $K$-matrix is generically not the identity. For example, the first known~\cite{cherednik1984factorizing} solution to the bYBE in the rational case has $\fg=\mathfrak{sl}_n(\C)$ and $V$ the defining vector representation, and is given by
\[
K(z) = \tau_{V} + \frac{\lambda}{z}{\bf 1}_{V} 
\]
where $\tau^2\in {\mathbb Z}_n$, the centre of $SL_n(\C)$, and $\lambda\in\C$ is a free parameter. All correlators in Costello's theory depend on $z$ only through the ratio $\hbar/z$, so even allowing for an overall $z$-dependent rescaling, if $\tau_V\neq {\bf 1}_V$, this family of $K$-matrices cannot be obtained via perturbative corrections to the identity.  Motivated by this, we will take a {\it quasi-classical} $K$-{\it matrix} to be a $K$-matrix admitting an expansion
\[
K_\hbar(z) = \tau_V + \hbar\, k(z) + \cO(\hbar^2)
\]
where $k(z)$ is the `classical $k$-matrix' and $\tau_V^2\in Z(G)$. We shall see that such $\tau_V$ naturally arise as part of the specification of gauge invariant operators in the classical theory on $\widetilde M$, while $k(z)$ is generated by quantum corrections. If both the $R$- and $K$-matrices are quasi-classical, we can expand the bYBE order-by-order in $\hbar$. A non-trivial condition we call the {\it classical boundary Yang-Baxter equation} (cbYBE) arises at order $\cO(\hbar^2)$.

\medskip

In generating solutions to the bYBE from gauge theory, a key role will be played by line operators supported a certain distinguished line $L =\del\Sigma\times\{z=0\}$. The simplest example of such a `boundary line operator' is a Wilson line in a representation $W$ of a certain subalgebra $\fh\subset\fg$ determined by the choice of boundary conditions. In section~\ref{sec:Kmx}, we obtain more general $K$-matrices 
\[
K_\hbar(z) : V\otimes W \to V\otimes W
\]
with the explicit form
\[
K_\hbar(z)  = \tau_V\otimes{\bf 1}_W 
+ \frac{\hbar}{4z}(t_a \tau t^a)_V \otimes {\bf 1}_W 
+ \frac{2\hbar}{z} (\tau t_\alpha)_V\otimes t^\alpha_W + \cO(\hbar^2)
\]
to leading order. (Here, the $t_a$ form a basis of $\fg$ while the $t_\alpha$ form a basis of $\fh$).  The most frequently studied $K$-matrices have $\dim W=1$, but by including a boundary line operator, the gauge theory is equally capable of providing $K$-matrices with $\dim W$ arbitrary. Such $K$-matrices are important {\it e.g.} in studying integrable scattering off impurities that possess internal degrees of freedom, arising for example in the Kondo problem (see {\it e.g.}~\cite{Fendley:1993my,Saleur:1998hq}). 

Just as bulk $R$-matrices are associated to representations of the Yangian, so too $K$-matrices satisfying the bYBE give rise to a rich algebraic structure of their own, known as the {\it twisted Yangian}~\cite{olshanskii1992twisted,molev1996yangians,delius2001boundary,mackay2002rational,mackay2003boundary}.  This is a left co-ideal subalgebra $\cB(\fg,\fh)$ of $\cY(\fg)$, meaning that there is an algebra homomorphism $\Delta_\hbar : \cB(\fg,\fh) \to \cY(\fg)\otimes \cB(\fg,\fh)$ analogous to the coproduct on $\cY(\fg)$. (See {\it e.g.}~\cite{MacKay:2004tc,ChariPressleyBook} for a readable introduction to Yangians and twisted Yangians.) In section~\ref{sec:Coprod}, generalizing similar considerations for the bulk coproduct in~\cite{Costello:2017dso}, we show that this homomorphism emerges in the gauge theory as the OPE between the line operator on $L$ and a parallel line operator in the bulk. Similar to the Yangian line operators of~\cite{Costello:2013zra,Costello:2017dso}, this calculation reveals that, at the quantum level, boundary line operators are actually labelled by representations $W$ of $\cB(\fg,\fh)$.

In section~\ref{sec:RTT}, following~\cite{Costello:2018gyb}, we examine the $RTT$ presentation of the twisted Yangian, in particular explaining why the gauge theory leads to boundary transfer matrices $B_W(z)$ that obey a unitarity condition, and understanding the role of the Sklyanin determinant condition~\cite{sklyanin1988boundary,olshanskii1992twisted,molev2002representations}. Importantly, we prove that, given a classical $k$-matrix obeying the cbYBE and the Sklyanin determinant condition, there is a unique quasi-classical $K$-matrix in the defining representation of a classical $\fg$. Our proof follows the same argument as~\cite{Costello:2018gyb} for quasi-classical $R$-matrices in the bulk.

\subsection{Notation}

Before starting we review some of the notation used in the text. We take $\fg$ to be a finite dimensional, complex, simple Lie algebra, and $G$ to be the unique connected and simply connected complex Lie group with Lie algebra $\fg$. We let $\{t_a\}_{a=1}^{\dim\fg}$ be a basis of $\fg$, and define structure constants with respect to this basis by $[t_a,t_b] = f_{ab}^{\ \ c} t_c$, where the summation convention is in force. We choose a non-degenerate, invariant bilinear symmetric form $\kappa$ (or $\Tr$) on $\mathfrak{g}$, but, following~\cite{Costello:2017dso},  we do not normalize this to be the Killing form.  Instead, the normalization of $\kappa$ is fixed by requiring that in the adjoint representation $f_{ab}^{\ \ c}f_{c}^{\ bd} = (\kappa^{-1})^{be}f_{ab}^{\ \ c} f_{ec}^{\ \ d}=-2{\bf h}^\vee\delta_a^{\ d}$, where ${\bf h}^\vee$ is the dual Coxeter number of $\fg$.  Indices are raised and lowered with $\kappa$ and $\kappa^{-1}$ in the standard way. We will frequently abuse notation by writing $V$ for the representation $\rho:\fg\to \text{End}(V)$. We indicate that we're evaluating a Lie algebra element, $X\in\fg$, in the representation $V$ using a suffix, $X_V$. We write $U(\fg)$ for the universal enveloping algebra of any Lie algebra $\fg$.

\section{The Classical Theory}
\label{sec:Theory}

In this section we study Costello-Witten-Yamazaki's mixed holomorphic-topological Chern-Simons theory~\cite{Costello:2013zra,Costello:2017dso,Costello:2018gyb} (henceforth, CWY theory) on an orbifold.

\subsection{Costello-Witten-Yamazaki Theory} 
\label{subsec:action}

First we review CWY theory. Let $M = \Sigma \times C$, where $\Sigma$ is some 2-dimensional real manifold and $C$ is a Riemann surface admitting a closed, holomorphic 1-form $\omega$. We will often use real coordinates $(x,y)$ on $\Sigma$, and complex coordinates $(z,\bar z)$ on $C$. We sometimes denote these coordinates collectively by $w\in M$.  

The dynamical field for the theory is a partial connection
\[
	A(w) = A_x(w) \d x + A_y(w) \d y + A_{\bar z}(w)\d\bar z
\]
on a $G$-bundle over $M$. Note that $A$ is only a partial connection since there is no $A_z$ component. We will sometimes write $A_i$ for the spatial components of $A$, where the index $i$ takes values in the set $\{x,y,\bar z\}$. Under an infinitesimal gauge transformation with parameter $\varepsilon$, the transformation of $A$ is
\begin{equation}
\label{GaugeTransform} 
\delta\!A = (\d_{\Sigma} + \bar\partial_C)\varepsilon + [A,\varepsilon]\,,
\end{equation}
where $\d_{\Sigma}$ is the exterior derivative on $\Sigma$, and $\bar\partial_C$ is the  Dolbeault operator on $C$. CWY choose the action to be
\begin{equation}
\label{CSAction}
S_M[A] = \frac{1}{2\pi}\int_{M}\omega \wedge 
\Tr\bigg(A\wedge \diff A + \frac{2}{3}A\wedge A\wedge A\bigg) 
= \frac{1}{2\pi}\int_{M}\omega\wedge\text{CS}(A)\,.
\end{equation}
The fact that the Chern-Simons form is wedged against $\omega$ has two important consequences. Firstly, while the theory remains topological in $\Sigma$, it is only holomorphic in $C$. In particular, the field equations $\omega\wedge F=0$ imply that the partial connection $A$ is flat on $\Sigma$, but only holomorphic along $C$. Fortunately, the equations $\omega\wedge F=0$ are strong enough to ensure that any possible gauge-invariant counterterms must vanish on-shell, and so can be removed by field redefinitions. Thus, despite being non-renormalizable by naive power counting, Costello proved in~\cite{Costello:2013zra} that the theory is in fact both renormalizable and IR free. The fact that it is IR free plays an important role in ensuring that the quantum theory generates local $R$-matrices.

To understand the second consequence of wedging with $\omega$, consider the more standard case of Chern-Simons theory on a compact three-manifold $N$. To handle non-trivial $G$-bundles on $N$, where a connection $A$ does not globally exist, one picks a four-manifold $N'$ with boundary $N$ and writes $\int_N {\text CS}(A) = \int_{N'} \Tr (F\wedge F)$. Provided the coefficient of this action is appropriately quantized, the path integral is independent of the choice of $N'$ and extension of $F$ over $N'$. In the case of CWY theory, as in holomorphic Chern-Simons theory~\cite{RThomasThesis}, the periods of $\omega$ on $C$ are not naturally quantized, and there is no canonical way to define $S[A]$ if the bundle is not trivial. Thus the action~\eqref{CSAction} for CWY theory as it stands makes sense only perturbatively.  (The factor of $1/2\pi$ in front of the integral is chosen for convenience and leads more readily to agreement with conventions in the integrable systems literature.) A non-perturbative definition of CWY theory has recently been given in~\cite{Ashwinkumar:2018tmm,Costello:2018txb} in terms of a brane construction of twisted six-dimensional Yang-Mills theory.

Since the path integral weight $\e^{\im S[A]/\hbar}$ depends on $\omega$ only through the combination $\omega/\hbar$, poles of $\omega$ correspond to regions of weak coupling $\hbar\to0$ while zeros of $\omega$ are at strong coupling. Such zeros must therefore be excluded in a perturbative treatment, and this constrains $(C,\omega)$ to be either $(\C,\d z)$, $(\C^*,\d z/z)$, or $(\C/(\Z+\tau\Z),\d z)$ up to biholomorphy. In \cite{Costello:2013zra} it was demonstrated that the these three cases correspond to the rational, trigonometric, and hyperbolic quasi-classical solutions of the YBE respectively. In this paper we'll consider just the simplest case of $(C,\omega) = (\C,\d z)$, leading to rational $R$- and $K$-matrices that are invariant under a simultaneous scaling of $\hbar$ and the spectral parameter $z$. We also choose $\Sigma=\R^2$, though since the theory is both topological in $\Sigma$ and IR free, our arguments should apply more generally.

\subsection{CWY Theory on an Orbifold} 
\label{subsec:bcs}

We now consider how to extend  CWY theory to allow for boundary $K$-matrices. Our approach is motivated by the fact that, in the rational case, the bYBE changes the sign of the spectral parameter of lines reflecting off the boundary. 

Let $\widetilde{M}=M/{\mathbb Z}_2$ be the orbifold defined by the equivalence relation
\begin{equation}
\label{Orbifold}
(x,y,z,\bar{z})\sim (-x,+y,-z,-\bar{z})\,,
\end{equation}
of points on $M=\R^2\times\C$. We will often think of this in terms of a map 
\[
\cP:(x,y,z,\bar z)\mapsto (-x,+y,-z,-\bar z)
\]
acting as reflection in the $y$-axis, together with rotation through $\pi$ around the origin in $\C$. $\cP$ reverses the orientation of $M$; that is, $\cP(M)=-M$. Note that $\widetilde{M}$ has no boundary, being topologically $\R\times\mathscr{C}$, where $\mathscr{C}$ is a cone over $\RP^2$. In particular, there is an orbifold singularity $L=\{x=z=0\}$ which is fixed pointwise by the $\Z_2$ action on $M$. This distinguished line will play an important role in what follows.

\medskip

To place a gauge theory on $\widetilde{M}$ we must extend this $\Z_2$ action to the space of gauge fields. To do so, we choose an involutive automorphism $\sigma:\mathfrak{g}\to\mathfrak{g}$, {\it i.e.} a homomorphism from $\Z_2$ to ${\rm Aut}(\fg)$. Since $\sigma$ is an involution, it splits $\fg$ into its postive and negative eigenspaces as $\fg=\fh\oplus\fm$. The positive eigenspace $\fh$ is a Lie subalgebra of $\fg$ (see appendix~\ref{app:symspa} for details). The choice of $\sigma$ is part of what is meant by a bundle on $\widetilde{M}$. Non-perturbatively, we would expect to sum over all such choices of $\sigma$ as part of the quantum gauge theory. However, for reasons explained above, in this paper we work only perturbatively. In perturbation theory, it makes sense to fix a choice of $\sigma$.

Given any gauge field on $\widetilde{M}$, denoted $\widetilde A$, its pullback to $M$, which we shall denote $A$, obeys
\begin{equation}
\label{OrbifoldConditions}
A =  \sigma  \cP^* \!A\,.
\end{equation}
Explicitly, this is 
\begin{equation}
\label{ExplicitOrbifoldConditions}
\begin{aligned}
A_x(x,y,z,\bar z) &= - \sigma(A_x(-x,y,-z,-\bar z)\big)\,,\\ 
A_y(x,y,z,\bar z) &= + \sigma\big(A_y(-x,y,-z,-\bar z)\big)\,,\\ 
A_{\bar z}(x,y,z,\bar z) &= -\sigma\big(A_{\bar z}(-x,y,-z,-\bar z)\big)
\end{aligned}
\end{equation}
in terms of the components of $A$. Similarly, the pullback to $M$ of a gauge transformation on $\widetilde{M}$ respects
\begin{equation}
\label{OrbifoldGauge}
\varepsilon=\sigma \cP^*\varepsilon\,,
\end{equation}
which ensures that
\begin{equation}
\label{desperation}
\d_\Sigma \varepsilon + \delbar_C\varepsilon +[A,\varepsilon] 
=\sigma \cP^*\big(\d_\Sigma \varepsilon + \delbar_C\varepsilon+ [A,\varepsilon]\big)
\end{equation}
guaranteeing that gauge transformations are consistent with~\eqref{OrbifoldConditions}.

Along the orbifold singularity $L$, the condition~\eqref{OrbifoldConditions} implies that ${A_y}\big|_L=\sigma {A_y}\big|_L$, whereas ${A_x}\big|_L=-\sigma {A_x}\big|_L$ and ${A_{\bar z}}\big|_L = -\sigma{A_{\bar z}}\big|_L$. Thus, along $L$, the component of the gauge field tangent to $L$ is restricted to lie in $\fh$, whilst the normal components live in $\fm$. We also impose the condition $A\to0$ at infinity. Gauge parameters $\varepsilon$ which tend to a constant at infinity correspond to global transformations, not gauge transformations. At infinity along $L$, such $\varepsilon$ must lie in $\fh$, and since they are constant, placing the theory on $\widetilde{M}$ with a non-trivial choice of $\sigma$ breaks the global symmetry algebra from $\fg$ to $\fh$. This reduction of the global symmetry algebra will be responsible for the fact that solutions to the bYBE are not in general $\mathfrak{g}$-invariant. 

\medskip

It will often be convenient to think of this slightly differently. We can pick a representative  of $\widetilde{M}$ inside $M$ by removing a codimension 1 surface $\pi$ chosen so that $M\backslash\pi = M_{\rm L}\sqcup M_{\rm R}$ such that $\cP(M_{\rm L}) = -M_{\rm R}$. For example, we can take $\pi$ to be the plane $x=0$, and then $x<0$ in $M_{\rm L}$ while $x>0$ in $M_{\rm R}$. (Explicitly, we have $M_{\rm L} \cong \Sigma_{\rm L}\times\C$, where $\Sigma_{\rm L} = \{(x,y)\in\R^2 \,:\, x>0\}$.) Since $\cP(x)=-x$, every point in $\widetilde M$ has at least one representative in  $\overline{M}_{\rm L} = M_{\rm L}\sqcup \pi\cong\overline{\Sigma}_{\rm L}\times\C$. However, $\cP$ acts non-trivially on $\pi$ itself, identifying $(y,z)\sim (y,-z)$, so $\pi$ is itself an orbifold. 

The conditions~\eqref{OrbifoldConditions} mean the gauge field on $M$ is determined by its restriction to $\overline{M}_{\rm L}$. Whilst $A$ is unconstrained in $M_{\rm L}$, on 
$\pi$ it must obey the non-local `boundary' conditions 
\begin{equation}
\label{ExplicitBoundaryConditions}
A_i(w)\bigg|_\pi = \mathbb{P}_i^{\ j}\sigma\,A_j\big(\cP(w)\big)\bigg|_\pi\\
\end{equation}
inherited from the constraints~\eqref{OrbifoldConditions}. (Here we have introduced the symbol $\mathbb{P}_i^{\ j} = {\rm diag}(-,+,-)$ for later convenience, where the indices $i,j,\ldots=(x,y,\bar{z})$.) Similarly, along $\pi$ the gauge transformations inherit the conditions
\begin{equation}
\label{ExplicitBoundaryGauge}
\del_i\varepsilon(w) \bigg|_\pi 
= -\sigma\,\mathbb{P}_i^{\ j}\big(\del_j\varepsilon\big(\cP(w))\big)\bigg|_\pi
\end{equation}
from~\eqref{desperation}.

\medskip

We take the action on the orbifold to be
\[
S_{\widetilde{M}}[\widetilde{A}]
=\frac{1}{|\Z_2|}S_M[A]
=\frac{1}{4\pi}\int_{M} \omega\wedge \Tr\bigg( A \wedge\d A+\frac{2}{3}A\wedge A\wedge A\bigg)
\]
defined in terms of the pullback of $\widetilde A$ to $M$.  Note that this action is invariant under diffeomorphisms of $M$ commuting with its $\mathbb{Z}_2$ action and leaving $\omega$ invariant. In the quantum theory we should integrate over all field configurations on $\widetilde{M}$, modulo gauge transformations on $\widetilde{M}$. To do this, we use the fact that each gauge field on $\widetilde{M}$ has a unique representative on $M_{\rm L}$, obeying the boundary conditions~\eqref{ExplicitBoundaryConditions}. In particular we have
\[
S_M[A] = S_{\overline{M}_{\rm L}}[A] + S_{M_{\rm R}}[A]\,.
\]
For any gauge field pulled back from $\widetilde{M}$, the condition~\eqref{OrbifoldConditions} allows us to write 
\[
S_{M_{\rm R}}[A] = S_{M_{\rm R}}[\sigma\cP^*  \!A] 
=\frac{1}{2\pi}\int_{M_{\rm R}} \omega \wedge{\rm CS}(\sigma\cP^*\!A)
=\frac{1}{2\pi}\int_{\cP(M_{\rm R})}\cP^*\omega\wedge{\rm CS}(\sigma A)\,.
\]
Using the facts that $\cP^*\omega = -\omega$, that $\cP(M_{\rm R}) = -M_{\rm L}$ and that $\sigma$ preserves the invariant form ${\rm Tr}$, this is $S_{M_{\rm R}}[A] = S_{M_{\rm L}}[A]$ and therefore
\begin{equation}
S_{\widetilde{M}}[\widetilde{A}] = \frac{1}{2} S_{M}[A] = S_{\overline{M}_{\rm L}}[A]\,.
\end{equation}
To summarise, we have shown that the action on the orbifold can be expressed in terms of an action for a gauge field $A$ that is unconstrained on $M_{\rm L}$, and obeys the conditions~\eqref{ExplicitBoundaryConditions} on the boundary $\pi$ of $\overline{M}_{\rm L}$. The advantage of doing this is that we know how to do quantum field theory on $\overline{M}_{\rm L}$. In particular,  the path integral is taken over all gauge fields on $M_{\rm L}$ obeying the boundary conditions~\eqref{ExplicitBoundaryConditions}, modulo gauge transformations on $M_{\rm L}$ obeying~\eqref{ExplicitBoundaryGauge} on the boundary.

The conditions~\eqref{ExplicitBoundaryConditions}  ensure that the boundary terms obtained when varying the action vanish. To see this, consider an arbitrary variation $\delta\!A$ on $\overline{M}_{\rm L}$ that also obeys~\eqref{ExplicitBoundaryConditions}. One has 
\[
2\pi\delta S_{\overline{M}_{\rm L}} 
= 2\int_{\overline{M}_{\rm L}}\omega\wedge\Tr\big(\delta\!A\wedge F\big) 
- \int_{\pi}\omega\wedge\Tr\big(\delta\!A\wedge A\big)\,.
\]
Since $\cP$ {\it preserves} the orientation of $\pi$, the same argument as above shows that
the boundary term 
\[
\int_{\pi} \omega\wedge{\rm Tr}\big(\delta\!A\wedge A\big)
=\int_\pi \cP^*\omega\wedge{\rm Tr}\big(\sigma\cP^*(\delta\!A)\wedge \sigma\cP^*A\big) 
= -\int_{\pi} \omega\wedge{\rm Tr}\big(\delta\!A\wedge A\big)\,,
\]
and so vanishes.

The action $S_{\overline{M}_{\rm L}}[A]$ is invariant under diffeomorphisms of $\overline{M}_{\rm L}$ preserving $\omega$, but the boundary conditions~\eqref{ExplicitBoundaryConditions} are not. They are invariant under any such diffeomorphism which descends from a diffeomorphism of $M$ commuting with its $\mathbb{Z}_2$ action and fixing $\pi$. This restriction that diffeomorphisms should preserve $\pi$ is an artefact of representing the orbifold theory as a theory on $\overline{M}_{\rm L}$, and is not fundamental. See also~\cite{horava1996chern} for a treatment of real Chern-Simons theory on a 3-dimensional orbifold.

\subsection{Line Operators on the Orbifold} 
\label{sec:LineOperators}

As explained in~\cite{Costello:2013zra,Costello:2017dso,Costello:2018gyb}, at the classical level the simplest class of operators on $M$ are Wilson lines\footnote{Following the convention used in \cite{Costello:2017dso} Wilson lines are defined by $(\d_{\Sigma} + \delbar_\C + A)\cW_V[\gamma(w);A]^{-1} = 0$. This means that the plus sign in the definition of the Wilson line is correct, but the convention for path ordering is from left to right. This was done to make the connection to the integrable systems literature as clear as possible.} 
\[ 
	\cW_V[\gamma;A] = P\exp\bigg(\int_{\gamma}A_V\bigg)\,
\]
along a curve $\gamma\subset M$, for $V$ a representation of $\fg$. Since we only have a partial connection, Wilson lines along curves varying in $C$ are not gauge invariant, so  we allow only Wilson lines along curves that are supported at a point $z\in C$. Since $A$, $\varepsilon\to0$ at infinity, Wilson lines extending to infinity in $\Sigma$ are allowed and are gauge invariant without taking a trace. 

A crucial feature of CWY theory is that its quantization forces the line operators to correspond to representations of the Yangian $\cY(\fg)$, rather than $\fg$ itself. (A representation of $\fg$ is permitted iff it lifts to a representation of $\cY(\fg)$.) We shall consider these Yangian line operators in section~\ref{sec:Coprod}. In addition, if $\gamma$ curves in $\Sigma$ (with respect to some reference framing), then in the quantum theory $\cW_V[\gamma;A]$ is anomalous and requires special treatment. We will consider these framing anomalies in section~\ref{sec:RTT}.

\medskip

On the orbifold $\widetilde{M}$, the simplest class of line operators are again Wilson lines on straight lines in $\widetilde{M}$. To treat these in the quantum theory, we'll interpret them as line operators on $\overline{M}_{\rm L}$. To do this, suppose we have a line operator on $M$, where the gauge field obeys $A=\sigma\cP^*\!A$ having been pulled back from the orbifold. Then
\begin{equation}
\label{WilsonLineReflection}
	\cW_V[\gamma;A] = \cW_V[\gamma;\sigma\cP^*\!A] 
	= \cW_V[\cP(\gamma);\sigma A] = \cW_{V^\sigma}[\cP(\gamma);A]\,,
\end{equation}
where $V^\sigma$ is the representation of $\fg$ defined by composing the representation $V$ with the map $\sigma$, {\it i.e.} $X_{V^\sigma} = (\sigma X)_V$. We can see that the operators $\cW_V[\gamma;A]$ and $\cW_{V^\sigma}[\cP(\gamma);A]$, are equivalent. This is consistent with the fact that on $\widetilde M$, $\gamma$ and $\cP(\gamma)$ are identified.

At this point, for simplicity, we fix $\sigma$ to be an {\it inner} automorphism of $\fg$. Such inner automorphisms preserve representations, so the original and reflected line operators will both be in isomorphic representations. An inner automorphism $\sigma$ acts as conjugation by some element $\tau\in G$. The condition that $\sigma$ be involutive means that we must also have $\tau^2\in Z(G)$. In this case we can simplify the right hand side of \eqref{WilsonLineReflection} slightly, with $\cW_{V^\sigma}[\cP(\gamma),A] = \cW_V[\cP(\gamma),\tau A\tau^{-1}] = \tau_V\cW_V[\cP(\gamma),A]\tau^{-1}_V$, so that
\begin{equation}
\label{InnerWilsonLineReflection}
	\cW_V[\gamma;A] = \tau_V\cW_V[\cP(\gamma);A]\tau_V^{-1}
\end{equation}
when $\sigma\in {\rm Inn}(\fg)$. 

We can exploit~\eqref{InnerWilsonLineReflection} to write any Wilson line on $\widetilde{M}$ in terms of Wilson lines on $\overline{M}_{\rm L}$. Firstly, if $\gamma$ already happens to lie purely in $ M_{\rm L}$, the the line operator $\cW_V[\gamma,A]$ is clearly of the required form immediately. Conversely, if $\gamma\subset M_{\rm R}$ then we use~\eqref{InnerWilsonLineReflection} to express it as Wilson line supported on $\cP(\gamma)\subset M_{\rm L}$, as required. The interesting case is where $\gamma$ crosses $\pi$. Suppose this happens at most once (which will be true for straight lines) and let $\gamma = \gamma_{\rm L}\cup\gamma_{\rm R}$ where $\gamma_{\rm L}\in \overline{M}_{\rm L}$ and $\gamma_{\rm  R}\subset M_{\rm R}$. We can decompose any such Wilson line as
\begin{equation}
\label{DescendingWilsonLine}
	\cW_V[\gamma;A] = \cW_V[\gamma_{\rm L};A]\cW_V[\gamma_{\rm R};A] 
	= \cW_V[\gamma_{\rm L};A]\tau_V\cW_V[\cP(\gamma_{\rm R});A]\tau_V^{-1}\,,
\end{equation}
and since $\cP(\gamma_{\rm R})\subset M_{\rm L}$, we have written our orbifold line operator in terms of a representative on $\overline{M}_{\rm L}$. We can view this configuration as being made up of an incident Wilson line $\cW_i = \cW_V[\gamma_{\rm L};A]$ terminating at some point $(y,z)\in\pi$, and a reflected Wilson line $\cW_r = \cW_V[\cP(\gamma_{\rm R});A]$, emerging from $(y,-z)\in\pi$. Though separated in $\C$, from the point of view of $\overline{\Sigma}_{\rm L}$ these operators appear to emerge from the same point on the boundary. 

In fact, we will actually consider the closely related operator
\begin{equation}
\label{KMatrixWilsonLine}
	\cW_V[\gamma;A]\tau_V = \cW_i\tau_V\cW_r\,,
\end{equation}
where we've removed the $\tau_V^{-1}$ acting at infinity on $\cW_r$. From the perspective of $\overline{\Sigma}_{\rm L}$, we can represent this configuration by the diagram
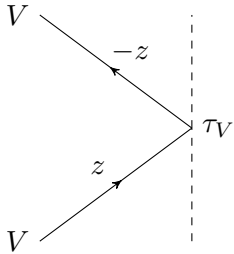
\begin{figure}[th]
	\centering
	\begin{tikzpicture}[baseline]
	\begin{feynman}
	\vertex (nw);
	\vertex[below=3cm of nw](sw);
	\vertex[below=1.5cm of nw](w);
	\vertex[right=2cm of w](e);
	\vertex[above=1.5cm of e](ne);
	\vertex[below=1.5cm of e](se);
	\vertex[left=2cm of e](l);
	\vertex[left=0cm of sw]{$V$};
	\vertex[left=0cm of nw]{$V$};
	\vertex[right=0cm of e]{$\tau_V$};
	\vertex[above=0.75cm of se](bW);
	\diagram*{{[edges=fermion]  (sw) -- [edge label=$z$](e), (e) --[edge label'=$\!\!\!-z$] (nw)}, {[edges = scalar] (se) -- (e), (e) -- (ne)},};
	\end{feynman}
	\end{tikzpicture}
\end{figure}

\vspace{-0.2cm}\noindent Because it just represents a line on the orbifold, extending to infinity at both ends, it is clear that $\cW_i\tau\cW_r$ is invariant (at least classically) under any gauge transformation on $\widetilde{M}$ that vanishes at infinity. From the perspective of $\overline{M}_{\rm L}$, under such a gauge transformation we have
\[
\begin{aligned}
\delta\big(\cW_i\,\tau_V\,\cW_r\big) 
&= \cW_i\,\varepsilon_V(0,y,z)\,\tau_V\,\cW_r - \cW_i\,\tau_V\,\varepsilon_V(0,y,-z)\,\cW_r \\
&= \cW_i\,\varepsilon_V(0,y,z)\,\tau_V\,\cW_r - \cW_i\,\varepsilon_V(0,y,z)\,\tau_V\,\cW_r\\
& = 0\,,
\end{aligned}
\]
as expected, where in the second line we've used the condition~\eqref{ExplicitBoundaryGauge} that $\varepsilon(0,y,z) = \tau \varepsilon(0,y,-z)\tau^{-1}$ on $\pi$. 

Note that whilst diffeomorphisms can move such configurations around on $\pi$, they can never remove them. From the point of view of $\overline{\Sigma}_{\rm L}$, a (single) line that reflects once off $\{x=0\}$ can never be detached from the boundary. Nonetheless, from the point of view of the orbifold,  there was nothing special about the choice of $\pi$ and it is possible to move Wilson lines through $\pi$ as long as this is done in a way consistent with the $\mathbb{Z}_2$ action. In particular, a line operator that curves so as to pass through $\pi$ an even number of times can be removed from the boundary altogether, simply by choosing a different representative plane $\pi'$ with $\gamma\cap \pi'=\emptyset$. It should also be clear how to write general arrangements of Wilson lines on $\overline{M}$ in terms of representatives on $\overline{M}_{\rm L}$: whenever any lift to $M$ of a curve $\gamma\subset\widetilde{M}$ intersects $\pi$ transversely, the associated  line operator on $\overline{M}_{\rm L}$ will reflect off the boundary. 

\medskip

There is one exceptional class of line operators we can admit: those lying along the orbifold singularity $L= \{x=z=0\}$. Recalling that components of $A$ along $L$ are required to take values $\fh$, Wilson lines along $L$ can effectively be built from a gauge field for $\fh$:
\[ 
	\cW_W[L;A] = P\exp\bigg(\int_L A_{W}\bigg)\,,
\]
where, classically,  $W$ is a representation of $\fh$. Just as quantum corrections on $M$ mean that bulk line operators are really labelled by representations of the Yangian $\cY(\fg)$~\cite{Costello:2013zra,Costello:2017dso,Costello:2018gyb} rather than just to $\fg$, so too quantum corrections on $\widetilde{M}$ will mean that boundary line operators are really labelled by representations $W$ of the twisted Yangian $\cB(\fg,\fh)$, rather than $\fh$ itself. This will be explored in section~\ref{sec:TwistedYangians}. From the perspective of $\overline{M}_{\rm L}$, this line operator lies entirely within the boundary orbifold plane $\pi$. To illustrate the presence of a line operator on $L$, we can decorate the previous picture as
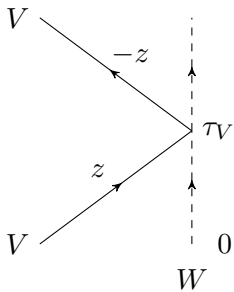
\begin{figure}[th]
	\centering
	\begin{tikzpicture}[baseline]
	\begin{feynman}
	\vertex (nw);
	\vertex[below=3cm of nw](sw);
	\vertex[below=1.5cm of nw](w);
	\vertex[right=2cm of w](e);
	\vertex[above=1.5cm of e](ne);
	\vertex[below=1.5cm of e](se);
	\vertex[left=2cm of e](l);
	\vertex[left=0cm of sw]{$V$};
	\vertex[left=0cm of nw]{$V$};
	\vertex[right=0cm of e]{$\tau_V$};
	\vertex[above=0.75cm of se](bW);
	\vertex[right=0.2cm of se]{$0$};
	\vertex[below=0.2cm of se]{$W$};
	\diagram*{{[edges=fermion]  (sw) -- [edge label = $z$] (e), (e) -- [edge label'=$\!\!\!-z$](nw)}, {[edges = charged scalar] (se) -- (e), (e) -- (ne)},};
	\end{feynman}
	\end{tikzpicture}
\end{figure}

\vspace{-0.2cm}\noindent where arrows along boundary indicate the presence of $\cW_W[L,A]$. We shall see that the boundary Wilson line plays an important role in generating $K$-matrices from the gauge theory. The boundary Wilson line will generically be needed even for the simplest case where $K\in {\rm End}(V)$ rather than ${\rm End}(V\otimes W)$. This case occurs when $\fh$ has non-trivial one-dimensional representations $W=\C$. The boundary line is then labelled by its `charge' $q$.

\section{Rational $K$-matrices from Gauge Theory}
\label{sec:Kmx}

In this section we explain how to use the quantum gauge theory on $\widetilde{M}$ to construct rational $K$-matrices, extending the calculation of $R$-matrices in~\cite{Costello:2017dso}. The $K$-matrices we obtain are quasi-classical, in that they admit an expansion $K_\hbar(z) = \tau + \hbar k(z) + \cO(\hbar^2)$. In this section we compute just the leading-order correction $k(z)$. We consider the extension to all orders in $\hbar$ in section~\ref{sec:RTT}.

\subsection{The Propagator} 
\label{subsec:prop}

To study the quantum theory, the first ingredient we need is the propagator $\widetilde{\Delta}$ for the theory on the orbifold.

In~\cite{Costello:2017dso}, Costello {\it et al.} wrote the propagator on $M$ around the vacuum $A=0$ in the gauge $D^iA_i=0$, where $D^i = (\partial_x,\partial_y,4\partial_z)$. It is clear that the vacuum $A=0$  obeys~\eqref{OrbifoldConditions}, so since this was an isolated solution (modulo gauge) on $M$, it will also be an isolated solution (modulo gauge) on the orbifold. We thus need to check that the condition $D^iA_i=0$ completely fixes the gauge for the orbifold theory. To see that this is so, note that varying the condition $D^iA_i=0$ gives
\[ 
D^i\partial_i\varepsilon=\Delta\varepsilon=0
\]
on $M_{\rm L}$. Standard results about Laplace's equation tell us that $\varepsilon$ is determined throughout $\overline{M}_{\rm L}$ by its values on $\pi$. However, any gauge parameter obeying the first condition in~\eqref{ExplicitBoundaryGauge} on $\pi$ will lead to a solution $\varepsilon_{\rm L}$ with the symmetry $\epsilon_{\rm L}(x,y,z,\bar z) = \sigma\big(\varepsilon_{\rm L}(x,y,-z,-\bar z)\big)$ throughout $\overline{M}_{\rm L}$. On the other hand, if $\varepsilon_{\rm L}\neq0$, differentiating it in the normal direction to $\pi$ leads to a contradiction with~\eqref{ExplicitBoundaryGauge}, forcing $\varepsilon=0$. Hence the conditions~\eqref{ExplicitBoundaryConditions} \&~\eqref{ExplicitBoundaryGauge} ensure that the gauge-fixing condition $D^iA_i=0$ completely fixes the gauge on $\overline{M}_{\rm L}$.

Using the gauge-fixing condition on $D^iA_i$, the standard BRST procedure leads to the gauge-fixed action 
\begin{equation}
S + S_{\text{gf}} 
= \frac{1}{2\pi}\int_{M}\d^4w\,
\Bigg(\varepsilon^{ijk}\Tr\bigg(A_i\partial_jA_k + \frac{2}{3}A_iA_jA_k\bigg)
-\xi^{-1}\Tr\Big((D^iA_i)^2\Big)\Bigg)\,,
\end{equation}
where $\xi$ is a gauge-fixing parameter. The propagator $\Delta_{ij}(w,w')$ on $M$ satisfies 
\begin{equation}
\frac{1}{\pi}(\varepsilon^{ijk}\partial_j + \xi^{-1}D^iD^k)\Delta_{k\ell}(w,w') 
= c\,\delta^i_{\,\ell}\,\delta^{(4)}(w-w')\,,
\end{equation}
where $c=t_a\otimes t_b\,(\kappa^{-1})^{ab} \, = t_a\otimes t^a \in \mathfrak{g}^{\otimes 2}$, and we recall that $\kappa_{ab}=\Tr(t_at_b)$.  In Landau gauge ($\xi=0$), the propagator is
\begin{equation}
\label{PropagatoronM}
\Delta_{ij}(w,w') = -\frac{c}{4\pi}\varepsilon_{ijk}D^k\bigg(\frac{1}{\|w-w'\|^2}\bigg)\,,
\end{equation}
which is the propagator used in~\cite{Costello:2017dso}. Note that $\Delta$ is symmetric under simultaneous exchange of its arguments, indices, and the factors in the tensor product, as expected since the gauge field is bosonic.

It is now straightforward to construct the propagator $\widetilde{\Delta}$ on $\overline{M}_{\rm L}$ from $\Delta$ using the method of images. The conditions~\eqref{ExplicitBoundaryGauge} imply that $\widetilde{\Delta}$ should obey
\begin{equation}
\label{PropagatorBoundaryConditions}
\begin{aligned}
\widetilde\Delta_{ij}(w,w') &= \mathbb{P}_j^{\ \ell}({\rm id}\otimes\sigma)
\widetilde\Delta_{i\ell}(w,\cP w')\,,\\
\widetilde\Delta_{ij}(w',w) &= \mathbb{P}_i^{\ k}(\sigma\otimes{\rm id})
\widetilde\Delta_{kj}(\cP w',w)\,,
\end{aligned}
\end{equation}
for $w\in M_{\rm L}$ and $w'\in\pi$. Here $\sigma\otimes\rm id$ acts on the first factor of $\fg^{\otimes 2}$ with $\sigma$ and trivially on the second, and ${\rm id}\otimes\sigma$ is defined similarly. The combination
\[
\Delta_{ij}(w,w') 
+ \mathbb{P}_j^{\ \ell}(\text{id}\otimes\sigma)\Delta_{i\ell}\big(w,\cP(w')\big)\,,
\]
manifestly obeys the required boundary condition in its second argument, {\it i.e.} the first condition in~\eqref{PropagatorBoundaryConditions}. To see that it in fact behaves correctly in both arguments, note that
\[
\mathbb{P}_j^{\ \ell}(\text{id}\otimes\sigma)\Delta_{i\ell}\big(w,\cP(w')\big) 
= \mathbb{P}_i^{\ k}(\sigma\otimes\text{id})\Delta_{kj}\big(\cP(w),w'\big)\,,
\]
where we've used the fact that since any automorphism of $\mathfrak{g}$ preserves the Killing form, $(\text{id}\otimes\sigma)\,c= (\sigma^{-1}\otimes\text{id})\,c$, and $\sigma^{-1}=\sigma$.  Therefore, the propagator on $\overline{M}_{\rm L}$ is
\begin{equation}
\label{Propagator}
\widetilde{\Delta}_{ij}(w,w') = \Delta_{ij}(w,w') 
+ \mathbb{P}_j^{\ \ell}(\text{id}\otimes\sigma)\Delta_{i\ell}\big(w,\cP(w')\big)\,.
\end{equation}
Written out explicitly, this is  
\[
\widetilde{\Delta}^{ab}_{ij}(w,w') 
= - \frac{c^{ab}}{4\pi}\varepsilon_{ijk}D^k\bigg(\frac{1}{\|w-w'\|^2}\bigg) 
- \frac{(c^\sigma)^{ab}}{4\pi}\mathbb{P}_j^{\ \ell}\varepsilon_{i\ell k}
D^k\bigg(\frac{1}{\|w-\cP(w')\|^2}\bigg)\]
in terms of its components with respect to our basis of $\mathfrak{g}$, where $c^\sigma = (\text{id}\otimes\sigma)\,c$.

\subsection{Computing the $k$-matrix}
\label{subsec:compkmx}

Working on $M$, one of the key points of~\cite{Costello:2013zra,Costello:2017dso} is that infra-red freedom means that the interactions between two line operators are weak when they are far apart. Topological invariance in $\Sigma$ then allows us to treat operators $\cW_{V_1}[\gamma_1;A]$ and $\cW_{V_2}[\gamma_2;A]$ supported at distinct points $z_1,z_2\in\C$ as arbitrarily far apart, except at points in $\Sigma$ where they cross. Thus the Feynman diagrams contributing to the correlator of two such operators build up a quasi-classical $R$-matrix
\begin{equation}
\label{RationalR}
\begin{aligned}
	R_\hbar(z_1-z_2) = \la\cW_{V_1}[\gamma_1;A]\, \cW_{V_2}[\gamma_2;A]\rangle
	= {\bf 1}_{V_1\otimes V_2} + \frac{\hbar}{z_1-z_2} c_{V_1\otimes V_2} + \cO(\hbar^2)
\end{aligned}
\end{equation}
\noindent associated to the crossing. The order $\hbar$ term involves the colour factor
$c_{V_1\otimes V_2} = t^a_{\ V_1}\otimes t_{aV_2}$ and comes from the Feynman diagram~\cite{Costello:2017dso}\\

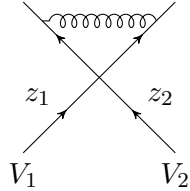
\begin{figure}[h]
	\centering
	\begin{tikzpicture}[baseline]
	\begin{feynman}
	\vertex(o);
	\vertex[right=1cm of o](e);
	\vertex[left=1cm of o](w);
	\vertex[right=1cm of o](oe);
	\vertex[left=1cm of o](ow);
	\vertex[above=1cm of e](ne);
	\vertex[below=1cm of e](se);
	\vertex[above=1cm of w](nw);
	\vertex[below=1cm of w](sw);
	\vertex[below=1cm of oe](ose);
	\vertex[below=1cm of ow](osw);
	\vertex[below=0cm of se]{$V_2$};
	\vertex[below=0cm of sw]{$V_1$};
	\vertex[right=0.75cm of o](ooe);
	\vertex[left=0.75cm of o](oow);
	\vertex[above=0.75cm of ooe](gi);
	\vertex[above=0.75cm of oow](gf);
	\diagram*{(o) -- [fermion] (ne), (se) -- [fermion	, edge label'=$z_2$
	] (o), (sw) -- [fermion, edge label = $z_1$
	] (o), (o) -- [fermion] (nw), (gi) -- [gluon] (gf)};
	\end{feynman}
	\end{tikzpicture}
\end{figure}

\noindent  built using the propagator $\Delta$ on $M$. The rational $r$-matrix here is indeed the one found by Belavin and Drinfeld in~\cite{belavin1982solutions}. 

We obtain the same $R$-matrix between line operators on $\overline{M}_{\rm L}$ whose crossing takes place away from $\pi$. Topological invariance again means the difference
\[
	\widetilde\Delta_{ij}(w,w') - \Delta_{ij}(w,w') =  \mathbb{P}_j^{\ \ell}({\rm id}\otimes \sigma)\Delta_{i\ell} (w,\cP(w'))
\]
between the propagators on $M$ and $M_{\rm L}$ has no effect on the $R$-matrix associated to the crossing, as $\cP(w')\in M_{\rm R}$ and may be treated as arbitrarily far away from $w$. We thus recover all the results of~\cite{Costello:2013zra,Costello:2017dso,Costello:2018gyb} for line operators in the bulk.

\medskip

In general, quasi-classical $K$-matrices will be generated in correlators whenever a Wilson line reflects off the boundary in $\overline{M}_{\rm L}$. As explained in section~\ref{sec:LineOperators}, the basic such configuration
\[
 K_\hbar(z) = \la \cW_i\tau_V\cW_r\otimes \cW_b\ra \,,
\]
involves a line that reflects off the boundary after being acted on by $\tau$ and replacing $z\to-z$. We also include the possibility of a line operator $\cW_b$ supported on $L$. Just like in the bulk, interactions between line operators supported at different points in $\C$ can be made arbitrarily weak, except at points in $\overline{\Sigma}_{\rm L}$ where they cross. In the configuration described, line operators only cross in $\overline{\Sigma}_{\rm L}$ at the point of reflection, and its interactions are localised near this point.

We now proceed to calculate the $\cO(\hbar)$ term, $k(z)$, in the $K$-matrix. We'll start with the simple case where there is no line operator on $L$, including this later. In the absence of a boundary line operator, the order $\hbar$ contribution to $\la \cW_i\,\tau_V\,\cW_r\ra$ comes from the three Feynman diagrams
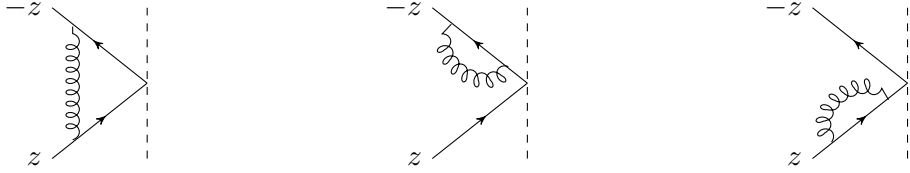
\begin{figure}[th]
	\centering
	\begin{tikzpicture}[baseline]
	\begin{feynman}
	\vertex (nw1);
	\vertex[below=2cm of nw1](sw1);
	\vertex[below=1cm of nw1](w1);
	\vertex[right=1.25cm of w1](e1);
	\vertex[above=1cm of e1](ne1);
	\vertex[below=1cm of e1](se1);
	\vertex[left=0.98cm of e1](l1);
	\vertex[above=0.75cm of l1](gf1);
	\vertex[below=0.75cm of l1] (gi1);
	\vertex[left=0cm of sw1]{$z$};
	\vertex[left=0cm of nw1]{$-z$};
	\vertex[right=5cm of nw1](nw2);
	\vertex[below=2cm of nw2](sw2);
	\vertex[below=1cm of nw2](w2);
	\vertex[right=1.25cm of w2](e2);
	\vertex[above=1cm of e2](ne2);
	\vertex[below=1cm of e2](se2);
	\vertex[left=0.25cm of e2](l2);
	\vertex[above=0.22cm of l2](gi2);
	\vertex[left=1cm of e2](r2);
	\vertex[above=0.782cm of r2](gf2);
	\vertex[left=0cm of sw2]{$z$};
	\vertex[left=0cm of nw2]{$-z$};
	\vertex[right=5cm of nw2](nw3);
	\vertex[below=2cm of nw3](sw3);
	\vertex[below=1cm of nw3](w3);
	\vertex[right=1.25cm of w3](e3);
	\vertex[above=1cm of e3](ne3);
	\vertex[below=1cm of e3](se3);
	\vertex[left=0.25cm of e3](l3);
	\vertex[below=0.22cm of l3](gf3);
	\vertex[left=1cm of e3](r3);
	\vertex[below=0.782cm of r3](gi3);
	\vertex[left=0cm of sw3]{$z$};
	\vertex[left=0cm of nw3]{$-z$};
	\diagram*{{[edges=fermion]  (sw1) -- (e1), (e1) -- (nw1), (sw2) -- (e2), (e2) -- (nw2), (sw3) -- (e3), (e3) -- (nw3)}, {[edges = scalar] (se1) -- (ne1), (se2) -- (ne2), (se3) -- (ne3)}, {[edges = gluon] (gi1) -- (gf1), (gi2) -- [half left] (gf2), (gi3) -- [half left] (gf3)},};
	\end{feynman}
	\end{tikzpicture}
\end{figure}\\
The second two diagrams arise from the interaction of a Wilson line with itself. Such self-interactions of a line operator are usually interpreted as renormalizing it, but here the the line operator can interact with its image when it gets close to the boundary. We'll see that we need to take this into account in order to get a $K$-matrix that is invariant under diffeomorphisms of $\widetilde M$ fixing $\d z$.

The first diagram contributes
\[
  t_a\tau t_b\int_{-\infty}^{0}\d s\int_{0}^\infty\d t\ \frac{\d\gamma^i_1}{\d s} \,
  \frac{\d\gamma^j_2}{\d t} \ \widetilde{\Delta}^{ab}_{ij}(\gamma_1(s),\gamma_2(t))
\]
for $\gamma_1(s)=(s\cos\theta,s\sin\theta,z,\bar{z})$ and $\gamma_2(s)=(-t\cos\theta,t\sin\theta,-z,-\bar{z})$, where $\theta$ is the angle of incidence to the normal\footnote{Note that a line on the orbifold that is straight (with respect to some framing of $\widetilde{M}$) will appear in $\overline{\Sigma}_{\rm L}$ to have equal angles of incidence and reflection with the normal to the boundary, again defined wrt a framing of $\overline{M}_{\rm L}$.}.  Using the explicit form~\eqref{Propagator} of the propagator, this diagram is
\[
 \begin{aligned}
 	& t_a\tau t^a\frac{4\sin\theta\cos\theta}{\pi}
 	\int_{-\infty}^{0}\int_{0}^\infty\d s\,\d t\,
 	\frac{\bar z}{\big(\cos^2{\theta}\,(t+s)^2+\sin^2{\theta}\,(t-s)^2+4|z^2|\big)^2}\\
 	&\qquad\qquad=\frac{t_a\tau t^a}{\pi}\int_{A}\d u\,\d v\,
 	\frac{2\bar z}{\big(u^2+v^2+4|z^2|\big)^2}\,,
 \end{aligned}
\]
where in the second line we have introduced $u=-s-t\cos{2\theta}$ and $v=t\sin{2\theta}$, and $A\subset\R^2$ is a sector subtended by an angle $\pi-2\theta$. This integral can be performed directly, giving a contribution
\[
	  \frac{t_a\tau t^a}{4z}\bigg(1-\frac{2\theta}{\pi}\bigg)\,.
\]
to the $k$-matrix from the first Feynman diagram above. 

This contribution alone is not diffeomorphism invariant on $\overline{\Sigma}_{\rm L}$ as it depends on the angle of incidence $\theta$. Computations similar to the one above, included in appendix~\ref{app:selfinteractions}, show that the two self-interaction diagrams contribute
\[
	\frac{t_a\tau t^a}{4z}\frac{\theta}{\pi}
\]
each, so combining the results gives
\begin{equation} 
	k(z) = \frac{1}{4z}t_a\tau t^a\,,
\label{eq:classkmx} 
\end{equation}
which is independent of $\theta$. We refer to this as the classical $k$-matrix, by analogy with the classical $r$-matrix. To order $\hbar$,  it gives us a $K$-matrix
\begin{equation}
	K_\hbar(z) = \tau_{V} + \frac{\hbar}{4z}\,(t_a\tau t^a)_V + \mathcal{O}(\hbar^2)
\end{equation}
when the Wilson line is in the $V$ representation.

\medskip

We now consider including a boundary line operator. Recall that (classically) this lives in a representation of the subalgebra $\mathfrak{h}\subset\mathfrak{g}$, defined to be the positive eigenspace of the involutive automorphism $\sigma = \text{conj}_\tau$. The splitting $\mathfrak{g}=\mathfrak{h}\oplus\mathfrak{m}$ defined by $\sigma$ is orthogonal, and we refine our basis of $\mathfrak{g}$ as $\{t_a\}=\{t_\alpha,t_\mu\}$, where for $\alpha=1,\ldots, {\rm dim}\,\mathfrak{h}$,  $\{t_\alpha\}$ is a basis of $\mathfrak{h}$ and similarly for $\mu= 1,\ldots, {\rm dim} \,\mathfrak{m}$. 
 From now on we use indices from the beginning of the Greek alphabet $\alpha, \beta, \gamma, \dots$ to index the basis vectors in $\mathfrak{h}$, and indices from the middle of the Greek alphabet $\mu, \nu, \xi, \dots$ to index the basis vectors in $\mathfrak{m}$. Note that since $\mathfrak{h}$ and $\mathfrak{m}$ are orthogonal, raising and lowering indices with $\kappa$ respects which part of the Greek alphabet they're from. We will assume summation convention for repeated $\alpha$ and $\mu$ indices.

In the presence of a boundary Wilson line labelled by a representation $W$ of $\fh$, in addition to the Feynman diagrams of the previous section, there are two new tree-level contributions to the $k$-matrix arising from interactions between the bulk and boundary Wilson lines:
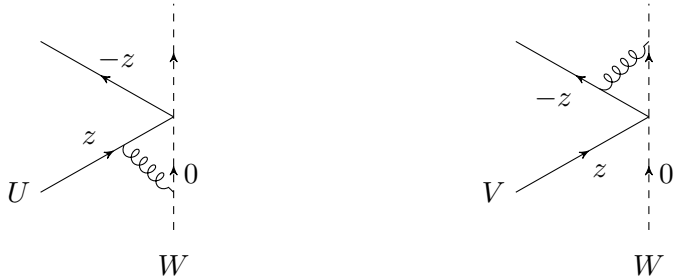
\begin{figure}[th]
	\centering
	\begin{tikzpicture}[baseline]
	\begin{feynman}
	\vertex (nw1);
	\vertex[below=2cm of nw1](sw1);
	\vertex[below=1cm of nw1](w1);
	\vertex[right=1.75cm of w1](e1);
	\vertex[above=1.5cm of e1](ne1);
	\vertex[above=1cm of e1](u1);
	\vertex[below=1.5cm of e1](se1);
	\vertex[below=1cm of e1](d1);
	\vertex[left=0.65cm of e1](l1);
	\vertex[above=0.375cm of l1](gf1);
	\vertex[below=0.375cm of l1] (gi1);
	\vertex[left=0cm of sw1]{$U$};
	\vertex[below=0.2cm of se1]{$W$};
	\vertex[right=4.5cm of ne1] (t);
	\vertex[below=0.5cm of t](nw2);
	\vertex[below=2cm of nw2](sw2);
	\vertex[below=1cm of nw2](w2);
	\vertex[right=1.75cm of w2](e2);
	\vertex[above=1.5cm of e2](ne2);
	\vertex[above=1cm of e2](u2);
	\vertex[below=1.5cm of e2](se2);
	\vertex[below=1cm of e2](d2);
	\vertex[left=0.65cm of e2](l2);
	\vertex[above=0.375cm of l2](gf2);
	\vertex[below=0.375cm of l2] (gi2);
	\vertex[left=0cm of sw2]{$V$};
	\vertex[below=0.2cm of se2]{$W$};
	\diagram*{{[edges=fermion] (sw1) -- [edge label=$z$](e1), (e1) -- [edge label'=$\!\!\!\!-z$] (nw1), (sw2) -- [edge label'=$z$](e2), (e2) -- [edge label=$-z$](nw2)}, {[edges = charged scalar] (se1) -- 
	[edge label'=$0$](e1), (e1) -- (ne1), (se2) -- 	[edge label'=$0$]
	(e2), (e2) -- (ne2)}, {[edges = gluon] (gf2) -- (u2), (gi1) -- (d1)},};
	\end{feynman}
	\end{tikzpicture}
\end{figure}\\
Since the gauge field along $L$ is restricted to lie in $\mathfrak{h}$, the boundary line couples as
\[
	 t_\alpha\int_{-\infty}^\infty\d y\,A_y^\alpha(0,y,0,0)\,.
\]
Hence the first diagram contributes
\[
\begin{aligned}
	 \delta k(z) &= t_a\tau\otimes t_{\beta}
	 \int_{-\infty}^0\diff s\int_{-\infty}^{\infty}\d y\,
	 \frac{\d\gamma^i}{\d s}\widetilde{\Delta}_{iy}^{a\beta}(\gamma_1(s),(0,y,0,0))\\
	&= t_a\tau\otimes t_{\beta}\,\cos{\theta}
	\int_{-\infty}^0\d s\int_{-\infty}^{\infty}\d y\,
	\widetilde{\Delta}_{xy}^{a\beta}\big((s\cos{\theta},s\sin{\theta},z,\bar{z}),(0,y,0,0)\big)\,,
\end{aligned}
\]
where the generator in the first factor of the tensor product corresponds to the bulk Wilson line, while the second factor corresponds to the boundary line operator. Here again, $\gamma_1(s) = (s\cos{\theta},s\sin{\theta},z,\bar{z})$ and we've used the fact that $\widetilde{\Delta}_{yy}=0$. To simplify the propagator further, note that $t_a\otimes\tau t^a\tau^{-1} = t_\alpha\otimes t^\alpha - t_\mu\otimes t^\mu$ implies $(c^\sigma)^{a\beta} = c^{a\beta}$, so for $w'\in L$ we have 
\[
	\widetilde{\Delta}_{xy}^{a\beta}(w,w') 
	= c^{a\beta}\Delta_{xy}(w-w') + (c^{\sigma})^{a\beta}\mathbb{P}_y^{\ y}
	\Delta_{xy}(w-\cP(w')) 
	= 2c^{a\beta}\Delta_{xy}(w- w')\,.
\]
This vanishes for $a=\mu$ since $\fh$ and $\fm$ are orthogonal.
Substituting the explicit form~\eqref{PropagatoronM} of $\Delta_{xy}(w- w')$ into the integral gives
\[
\begin{aligned}
	 \delta k(z) &= \tau t_\alpha\otimes t^\alpha\, \frac{2\bar z\cos{\theta}}{\pi}
	 \int_{-\infty}^0\d s\int_{-\infty}^{\infty}\d y\,
	 \frac{1}{(s^2\cos^2{\theta} + (s\sin{\theta} - y)^2 + |z|^2)^2}\\
	 &= \tau t_\alpha\otimes t^\alpha\,\bar z\cos{\theta}\int_{-\infty}^0\d s
	 \frac{1}{(s^2\cos^2{\theta} + |z|^2)^{3/2}} \\
	 &= \frac{\tau t_\alpha\otimes t^\alpha}{z}
\end{aligned}
\]
as the contribution from the first diagram above.

To find the contribution from the second diagram we need to swap the sign of $\theta$, which does nothing, swap the sign of $z$, giving an overall sign, and replace $t_a$ by $-t_a$, cancelling this sign\footnote{Changing the direction on a Wilson line amounts to replacing a representation by its dual. This is achieved by mapping $t_a\mapsto -t_a^\text{t}$. The reason we don't get the transpose is that we've reversed what we mean by incoming and outgoing.}. The $\tau$ will also act before the $t_a$, not after it, but in the above calculation we found $a=\alpha$, and $\tau$ commutes with $t_\alpha$ by definition. Hence the overall contribution from the two diagrams is 
\[ \delta k(z) = \frac{2}{z}\tau t_\alpha\otimes t^\alpha\,.\]
Combining this with equation~\eqref{eq:classkmx} we obtain the more general expression 
\begin{equation}
\label{eq:ourKmx}
	K_{\hbar}(z) = \tau_V\otimes{\bf 1}_W 
	+ \frac{\hbar}{4z}(t_a\tau t^a)_{V}\otimes{\bf 1}_W 
	+ \frac{2\hbar}{z}(\tau t_{\alpha})_{V}\otimes t^\alpha_{\ W} 
	+ \mathcal{O}(\hbar^2)\,,
\end{equation}
when the reflecting Wilson line is in the representation $V$ of $\fg$, and the boundary line operator is in the representation $W$ of $\fh$. This is the formula we highlighted in the introduction: it gives the asymptotic behaviour of any semi-classical, rational $K$-matrix at first order in $\hbar$. In appendix~\ref{app:cbYBEproof} we verify that this is indeed a solution of the bYBE up to second order in $\hbar$. The fact that perturbation theory allows one systematically to construct such explicit expressions (even for arbitrary $\dim V$ and $\dim W$) is a key virtue of the gauge theory approach. We do not believe this expression for the quasi-classical $K$-matrix has appeared in the literature before, though it can be derived from the intertwiner relation of Delius, MacKay \& Short~\cite{delius2003quantum,delius2001boundary,mackay2002rational}.
 
In section~\ref{sec:Coprod}, we'll see that at the quantum level, just as bulk line operators are labelled by representations $V$ of the Yangian $\cY(\fg)$ rather than  of $\fg$ itself, line operators on $L$ are actually labelled by representations $W$ of the twisted Yangian $\cB(\fg,\fh)$ rather than $\fh$.  The asymptotic form~\eqref{eq:ourKmx} of the $K$-matrix remains valid in such cases, at least for finite dimensional representations.

\subsection{Explicit $K$-matrices with $\dim W=1$} 
\label{sec:Kmatrix1}

A frequently studied case of~\eqref{eq:ourKmx} is for $W=\C$  a 1-dimensional representation of $\fh$. In this case, $K(z) : V\to V$ may be represented by a matrix with $\C$-valued entries. 

To restrict to this case, suppose $\fh$ contains a copy\footnote{Note that for all symmetric spaces the centre of $\fh$ is at most 1-dimensional.} of $\C$ and let $Q$ be its generator, normalized so that $\text{Tr}(Q^2)=1$. Then the representations of $\fh$ on $\C$ are given by $Q\mapsto q \in\C$ with all other generators vanishing. Our $K$-matrices are then parametrized by $q$, which can be thought of physically as the 'charge' associated to the boundary Wilson line. In this case, equation~\eqref{eq:ourKmx} becomes
\begin{equation}
\label{eq:1dWKmx}
	K_{\hbar}(z) = \tau_V + \frac{\hbar}{4z}(t^a\tau t_a)_{V} 
	+ \frac{2q\hbar}{z}(\tau Q)_V + \cO(\hbar^2)\,,
\end{equation}
where we used the fact that $V\otimes\C\cong V$.

To be completely explicit, let is now give a list the evaluations of equation~\eqref{eq:1dWKmx} for all classical, simple Lie algebras $\fg$ and all (non-trivial) inner automorphisms $\sigma$, specialising for simplicity for the case where $V$ is the defining vector representation\footnote{The defining vector representation of a classical Lie algebra always lifts to the associated Yangian. For further details see \cite{ChariPressleyBook}. The 1 dimensional representations of $\fh$ also always lift to the twisted Yangian $\cB(\fg,\fh)$.}.  We obtain

\begin{enumerate}
\item[$\bullet$]
$(\fg,\fh) = (\mathfrak{sl}_n(\C),\mathfrak{sl}_{n-k}(\C)\oplus \mathfrak{sl}_k(\C)\oplus \C)$ and 
	$\tau_V= \e^{\im\pi k/n} \,{\rm diag}({\bf 1}_{n-k},-{\bf 1}_k)$, for $k=1,\dots,\left \lfloor{n/2}\right \rfloor\,.$

In this case our solution gives
\[ 
	\bigg(1 + \frac{\hbar}{z}\Big( q'(2k-n) - \frac{1}{4n}\Big)\bigg)
	\bigg(\tau_{V} + \frac{\hbar}{z}\Big(\frac{n}{4} - \frac{k}{2} + q'n\Big)\e^{\im\pi k/n}
	{\bf 1}_{V}\bigg) + \mathcal{O}(\hbar^2)\,,
\]
where $q' = q/\sqrt{nk(n-k)}$ labels the charge of the boundary line under the $\C$ factor in $\fh$. Defining
\[ 
	\lambda = \bigg(\frac{n}{4} - \frac{k}{2} + q'n\bigg)\e^{\im\pi k/n}\,,
\]
we can see that at first order in $\hbar$ it is proportional to 
\[ 
	\tau_V + \frac{\hbar\lambda}{z}{\bf 1}_V~.
\]
This was the first solution of the boundary Yang-Baxter equation to be written down, identified by Cherednik in~\cite{cherednik1984factorizing}.

\item[$\bullet$] 
$(\fg,\fh) = (\mathfrak{so}_{2n+1}(\C),\mathfrak{so}_{2n+1-k}(\C)\oplus \mathfrak{so}_k(\C))$ and $\tau_V = (-)^k {\rm diag} ({\bf 1}_{2n+1-k},-{\bf 1}_{k})$, for $k=1,\dots,n$.
	
Here our solution gives
\[ 
	\Big(1-\frac{\hbar}{4z}\Big)\bigg(\tau_{V} + (-)^k\frac{\hbar(2n+1-2k)}{4z}{\bf 1}_{V}\bigg) 
	+ \cO(\hbar^2)
\]
whenever $k>2$. In the particular case $k=2$, $\fh$ contains a copy of $\mathfrak{so}_{2}(\C)\cong\C$, so the $K$-matrix is enhanced to 
\[ 
	\Big(1-\frac{\hbar}{4z}\Big)\bigg(\tau_{V} + (-)^k\frac{\hbar(2n-3)}{4z}{\bf 1}_{V}-(-)^k	
	\frac{2\im q \hbar}{z}M_{2n+1}\bigg) + \cO(\hbar^2)\,,
\]
where again the parameter $q$ labels the charge of the boundary line under the $\mathfrak{so}_2(\C)$ factor, and 
\[ 
	M_m  =\begin{pmatrix}
				{\bf 1}_{m-2} & 0 & 0 \\
				0 & 0 & 1 \\
				0 & -1 & 0
			\end{pmatrix}\,.
\]

\item[$\bullet$]
$(\fg,\fh) = (\mathfrak{sp}_{2n}(\C),\mathfrak{sp}_{2(n-k)}(\C)\oplus\mathfrak{sp}_{2k}(\C))$ and $\tau_V = {\rm diag}({\bf 1}_{n-k},  -{\bf 1}_{k}, {\bf 1}_{n-k}, -{\bf 1}_{k})$, for $k=1,\dots,\left \lfloor{n/2}\right \rfloor$.

We obtain the $K$-matrix
\[ 
	\Big(1+\frac{\hbar}{8z}\Big)\bigg(\tau_{U} + \frac{\hbar(n-2k)}{4z}{\bf 1}_{V}\bigg) 
	+ \mathcal{O}(\hbar^2)~.
\]

\item[$\bullet$]
$(\fg,\fh) = (\mathfrak{sp}_{2n}(\C),\mathfrak{sl}_{n}(\C)\oplus\C)$ and 
$\displaystyle{ 
	\tau_V = 
		\begin{pmatrix} 
			0 & {\bf 1}_{n} \\ -{\bf 1}_{n} & 0 
		\end{pmatrix} }$.

Here we obtain 
\[ 
	\Big(1-\frac{\hbar}{8z}\Big)\bigg(\tau_V - \frac{\im q'\hbar}{z}{\bf 1}_V\bigg) 
	+\mathcal{O}(\hbar^2)\,,
\]
where $q'=q\sqrt{2/n}$ again labels the charge of the boundary Wilson line under the $\C$ factor of $\fh$.

\item[$\bullet$]
$(\fg,\fh) = (\mathfrak{so}_{2n}(\C),\mathfrak{so}_{2n-k}(\C)\oplus\mathfrak{so}_{k}(\C))$
and $\tau_V={\rm diag}({\bf 1}_{2n-k},-{\bf 1}_{k})$, for $k=1,\dots,\left \lfloor{n/2}\right \rfloor$.

Our formula gives
\[ 
	\left(1-\frac{\hbar}{4z}\right)\bigg(\tau_{V} + \frac{\hbar(n-k)}{2z}{\bf 1}_{V}\bigg) 
	+ \mathcal{O}(\hbar^2)\,.
\]
Strictly speaking, for odd $k$ our choice of $\tau_V$ has determinant $-1$ and so is contained in $O_{2n}(\C)$ rather than $SO_{2n}(\C)$. Conjugation by this $\tau_V$ is actually an {\it outer} automorphism of $\fg = \mathfrak{so}_{2n}$. However, since this particular outer automorphism fixes the defining vector representation, equation~\eqref{eq:1dWKmx} still applied straightforwardly. Note also that $\mathfrak{so}_{2n}$ is not simple for $n=1,2$, and so the above result doesn't apply in these cases.

This solution also gets enhanced when $k=2$, and we find
\[ 
	\Big(1-\frac{\hbar}{4z}\Big)\bigg(\tau_{V} 
	+ \frac{\hbar(n-2)}{2z}{\bf 1}_{V}-\frac{2\im q\hbar}{z}M_{2n}\bigg) 
+ \mathcal{O}(\hbar^2)
\]
with $M_{2n}$ as above.

\item[$\bullet$]
$(\fg,\fh) = (\mathfrak{so}_{2n}(\C),\mathfrak{sl}_{n}(\C)\oplus\C)$ and 
$\displaystyle{
	 \tau_V = 
	 	\begin{pmatrix} 
	 		0 & {\bf 1}_{n} \\ -{\bf 1}_{n} & 0 
	 	\end{pmatrix}}$.

In this case we have
\[ 	
	\Big(1+\frac{\hbar}{4z}\Big)
		\bigg(\tau_V - \frac{2\im\hbar q'}{z}{\bf 1}_V\bigg) + \mathcal{O}(\hbar^2)
\]
where $q' = q/\sqrt{n}$ is a continuous, free parameter. Note that $\mathfrak{so}_{2n}$ is not simple for $n=1,2$, and so the above result doesn't apply in these cases.
\end{enumerate}

We have checked that these $K$-matrices agree, to first order in $\hbar$, with all the rational $K$-matrices appearing in~\cite{cherednik1984factorizing,Arnaudon:2004sd} for $\fg=\mathfrak{sl}_n(\C)$, and in~\cite{mackay2003boundary,delius2001boundary,arnaudon2003classification} for the remaining classical, semi-simple Lie algebras $\fg$, with $V$ the defining representation of $\fg$. (We are not aware of any rational $K$-matrices admitting a semi-classical expansion in the sense we have explained that are not contained in the list above, again for $V$ the defining representation of a semi-simple, classical $\fg$.) In making this comparison,  we allow ourselves to multiply the $K$-matrix by an arbitrary dressing function which, like the $R$- and $K$-matrices themselves, is a formal power series in $\hbar/z$. The possibility of such a dressing arises because the bYBE is homogeneous and does not fix the overall normalization of $K(z)$. The precise normalization of the $K$-matrices generated using gauge theory is not arbitrary, and will be explained in detail in section~\ref{sec:RTT}. Note that whilst $\fh$-invariance imposes strong constraints on our $K$-matrices, it is not enough to determine them uniquely.

\section{Twisted Yangians and the OPE}
\label{sec:Coprod}

It is well known~\cite{drinfeld1990hopf} that rational solutions of the Yang-Baxter equation are labelled by representations of the {\it Yangian} $\cY(\fg)$, rather than of $\fg$ itself. In the presence of a boundary of $\Sigma$, the Yangian is broken to a subalgebra $\cB(\fg,\fh)$ known as the {\it twisted Yangian}\footnote{When $\fg=\mathfrak {sl}_n$, some authors use the name {\it reflection algebra} to refer to $\cB(\fg,\fh)$ in the case $\fh =\mathfrak {sl}_{n-m}\times \mathfrak{sl}_m$, reserving {\it twisted Yangian} for the cases $\fh=\mathfrak{o}_{n/2}$ and $\fh=\mathfrak{sp}_{n/2}$ associated to outer automorphisms~\cite{olshanskii1992twisted}. We will refer to all of these as twisted Yangians.}~\cite{molev2002representations,delius2001boundary}, which can be viewed physically as the Yangian charges that are preserved by the boundary conditions. Rational solutions of the boundary Yang-Baxter equation are labelled by representations of $\cB(\fg,\fh)$. 
 
In this section, we begin by reviewing~\cite{Costello:2013zra,Costello:2017dso} how the Yangian emerges in CWY theory on $M$.  In particular, there is a pleasingly direct relation between the OPE of (bulk) line operators in the gauge theory and the coproduct on $\cY(\fg)$. We then perform an analogous computation for line operators on $\widetilde{M}$, showing how the structure of $\cB(\fg,\fh)$ as a left coideal of $\cY(\fg)$ arises from quantum contributions to the OPE between a bulk line operator and a `boundary' line operator supported on $L$. 

\subsection{The Yangian from the Bulk OPE}
\label{sec:Yangian}

The Yangian is a deformation of the universal enveloping algebra of the (positive) loop algebra $\fg[[u]]$ (see {\it e.g.}~\cite{MacKay:2004tc,molev2002representations,ChariPressleyBook} for a review). The loop algebra $\fg[[u]]$ is an infinite dimensional Lie algebra generated as a vector space by $t^{(m)}_a=t_au^m$, with $a=1,\ldots,{\rm dim}\,\fg$ and $m\in\Z_0^+$, satisfying
\[
	\big[t^{(m)}_a,t^{(n)}_b\big] = f_{ab}^{\ \ c}t^{(m+n)}_c\,.
\]
It's universal enveloping algebra $U(\fg[[u]])$ has a coproduct 
\[
	\Delta : U(\fg[[u]]) \to U(\fg[[u]])\otimes U(\fg[[u]])
\]
defined trivially by  \smash{$\Delta: t^{(m)}_a \mapsto t^{(m)}_a\otimes{\bf 1} + {\bf 1}\otimes t^{(m)}_a$} for all generators. The Yangian coproduct 
\[
	\Delta_\hbar : \cY(\fg) \to \cY(\fg)\otimes\cY(\fg)
\]
deforms this to become
\begin{equation}
\label{Yangiancoprod}
\begin{aligned}
	\Delta_\hbar : t^{(0)}_a &\mapsto t^{(0)}_a \otimes {\bf 1} + {\bf 1}\otimes t^{(0)}_a\\
	\Delta_\hbar:  t^{(1)}_a &\mapsto t^{(1)}_a \otimes {\bf 1} + {\bf 1}\otimes t^{(1)}_a 
	- \frac{\hbar}{2}f_a^{\ bc}t_b^{(0)}\otimes t_c^{(0)}
\end{aligned}
\end{equation}
so that (in particular) the coaction on the level-1 generators is non-trivial. In order for this deformed coproduct to be a coassociative algebra homomorphism ({\it i.e.} to ensure that $\Delta_\hbar([x,y]) =[\Delta_\hbar(x),\Delta_\hbar(y)]$ for all $x,y\in\cY(\fg)$) it is necessary that the algebra of the generators themselves is itself deformed\footnote{If $\hbar\neq0$,  rescaling $t^{(1)}_a\mapsto \hbar t^{(1)}_a$ in~\eqref{DrinfeldJ} causes $\hbar$ to drop out. This fails for $\hbar=0$, for which this algebra is isomorphic to $U\big(\fg[[u]]\big)$. In this sense the Yangian is a deformation of \smash{$U\big(\mathfrak{g}[[u]]\big)$}. It cannot be viewed as a deformation of \smash{$\mathfrak{g}[[u]]$} alone since the polynomial $Q_{abc}(t^{(0)})$ is only defined in the universal enveloping algebra. We will keep the factors of $\hbar$ present in the algebra so as to make manifest the connection to gauge theory.}, becoming
\begin{equation}
\label{DrinfeldJ}
\begin{aligned}
	&[t_a^{(0)},t_b^{(0)}] = f_{ab}^{\ \ c}t_c^{(0)}\,,\\
	&[t_a^{(0)},t_b^{(1)}] = f_{ab}^{\ \ c}t_c^{(1)}\,,\\
	&f_{ab}^{\ \ d}\,[t_c^{(1)},t_d^{(1)}] + f_{ca}^{\ \ d}\,[t_b^{(1)},t_d^{(1)}] + f_{bc}^{\ \ d}		
	[t_a^{(1)},t_d^{(1)}] = \hbar^2 Q_{abc}(t^{(0)})\,,
\end{aligned}
\end{equation}
for $a=1,\dots,\dim\fg$. The first relation defines a $U\big(\mathfrak{g}\big)$ subalgebra of $\cY(\fg)$, and the second tells us that  $t^{(1)}_a$ transforms in the adjoint representation of this subalgebra. The third relation, known as {\it Drinfeld's terrific relation}, involves a homogeneous third-order polynomial\footnote{If $\fg=\mathfrak{sl}_2(\C)$ this relation becomes trivial and needs to replaced with an alternative expression. This occurs because as representations of $\mathfrak{sl}_2$, $\wedge^2\mathfrak{sl}_2\cong\mathfrak{sl}_2$~\cite{ChariPressleyBook}.} 
\[ 
	Q_{abc}(t^{(0)}) = \frac{1}{24}f_{ad}^{\ \ g}f_{be}^{\ \ h}f_{cf}^{\ \ i}f^{def}\{t^{(0)}_g,t^{(0)}_h,t^{(0)}_i\}\,, 
\]
where
\[
	\{t^{(0)}_{i_1},t^{(0)}_{i_2},t^{(0)}_{i_3}\} 
	= \sum_{\sigma\in S_3}t^{(0)}_{i_{\sigma(1)}}t^{(0)}_{i_{\sigma(2)}}t^{(0)}_{i_{\sigma(3)}}\,.
\]
This realization of $\cY(\fg)$ is known as the Drinfeld $J$ presentation. We will discuss an alternative presentation in section~\ref{sec:RTT}. Notice that while any representation $V$ of $\fg$ determines a representation of the loop algebra $\fg[[u]]$ by $t^{(0)}_a \mapsto t_{a,V}$ and $t^{(m)}_a\mapsto 0$ for all $m>0$, not every such representation can be promoted to a representation of the Yangian. This is because 
if we set $t^{(1)}_a=0$, Drinfeld's terrific relation requires that the $t^{(0)}_a$ obey $Q_{abc}(t^{(0)})=0$.  Thankfully, this is true\footnote{It is not true for all fundamental representations of $B_n$ and $D_n$, but is true for their spin representations~\cite{ChariPressleyBook}. All representations of $\fg$ used in this paper do lift to the Yangian.} for the defining vector representations of classical, simple $\fg$.

\medskip

In Costello's approach, the Yangian coproduct arises via the OPE of line operators in the gauge theory~\cite{Costello:2013zra,Costello:2017dso}. Topological invariance in $\Sigma$ means that  two parallel Wilson lines may be treated as arbitrarily far apart in $\Sigma$, in which case infra-red freedom implies the quantum correlator transforms in the tensor product of two representations $V_1$ and $V_2$ of $\cY(\fg)$. On the other hand, topological invariance equally allows us to treat them as arbitrarily {\it close} in $\Sigma$. In this case, if the lines are each supported at the same $z\in\C$, the OPE allows us to replace them by a single line operator supported on the common limiting line. The OPE thus tells us how the tensor product $V_1\otimes V_2$ decomposes into representations of the Yangian. (As usual, taking representations reverses the direction of the coproduct.)

Following~\cite{Costello:2017dso}, let's see this at work in the case that the two original Wilson lines are in representations $U_1$ and $U_2$ of $\fg$ itself. At lowest order, an external gluon can couple independently to either of the two original line operators, giving a coupling $(t_{a,U_1}\otimes{\bf 1}_{U_2} + {\bf 1}_{U_1}\otimes t_{a,U_2}) \int_{\gamma}\diff\gamma^i\,A^a_i(\gamma)$ representing the tensor product of the two original operators. However, in the quantum theory there are further corrections. The first such correction arises from the Feynman diagram
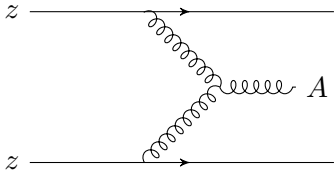
\begin{figure}[h!]
	\centering
	\begin{tikzpicture}[baseline]
	\begin{feynman}
	\vertex (nw);
	\vertex[right=4cm of nw](ne);
	\vertex[below=2cm of nw](sw);
	\vertex[right=4cm of sw](se);
	\vertex[right=1.5cm of nw](v1);
	\vertex[right=1.5cm of sw](v2);
	\vertex[right=2.5cm of nw](n);
	\vertex[below=1cm of n](v);
	\vertex[right=1cm of v](vA);
	\diagram*{{[edges=fermion] (nw) -- (ne), (sw) -- (se)},{[edges=gluon] (v) -- (v1), (v) -- (v2), (v) -- (vA)},};
	\vertex[left=0cm of nw]{$z$};
	\vertex[left=0cm of sw]{$z$};
	\vertex[right=0cm of vA]{$A$};
	\end{feynman}
	\end{tikzpicture}
\end{figure}\\
giving a correction~\cite{Costello:2017dso}
\begin{equation}
	 -\frac{\hbar}{2}\,t_{a,V_1}\otimes t_{b,V_2}\,f^{ab}_{\ \ c}\,
	 \int_\gamma\d\gamma^i\,\partial_zA^c_i
\label{eq:bcoprod1}
\end{equation}
at order $\hbar$. 

The $z$ derivative appearing in~\eqref{eq:bcoprod1} means we cannot interpret it as the coupling of a single external gluon to a line operator in any standard representation of $\fg$. Bringing the two Wilson lines together has resulted in an operator outside the class of operators we considered in~\ref{sec:LineOperators}. Thus, if we want the OPE to close, we must enlarge our class of line operators to those labelled by representations of $U(\fg[[u]])$. That is,  we take
\[
	A^{\fg[[u]]} = \sum_{m=0}^\infty \frac{t_a^{(m)}}{m!} \del^m_z A^a(w) 
		= \sum_{m=0}^\infty \frac{t_a\,u^m}{m!} \del^m_z A^a(w) 
\]
to formally be a $\fg[[u]]$-valued gauge field\footnote{We similarly expand  infinitesimal gauge transformations as
\[
	 \varepsilon^{\fg[[u]]}
	 = \sum_{m=0}^{\infty}\frac{t^{(m)}_a}{m!}\partial_z^m\varepsilon^a(w)
	  = \sum_{m=0}^{\infty}\frac{(u^mt_a)}{m!}\partial_z^m\varepsilon^a(w).
\]  
in terms of a formal power series in $u$.} and construct line operators 
\begin{equation}
\label{U(g)lineoperator}
	\cW_V[\gamma;A^{\fg[[u]]}] = P\exp\bigg(\int_\gamma\d\gamma^i\,A^{\fg[[u]]}_{i,V}\bigg)
\end{equation}
where, classically, $V$ is a representation\footnote{We restrict ourselves to finite dimensional representations with the property that there exists an $m_0$ such that $t^{(m)}_{a,V} = 0$ for all $m>m_0$.} of $\fg[[u]]$.

With this understanding, we recognise the quantum correction~\eqref{eq:bcoprod1} to the OPE as exactly the deformation term in $\Delta_\hbar$ when acting on representations of $\cY(\fg)$ obtained from representations of $\fg$ itself. Using this result, together with the associativity and certain braiding properties of the OPE, Costello proved~\cite{Costello:2013zra} that, at the quantum level, line operators actually correspond to representations of the Yangian. More generally, given any pair of line operators of the form~\eqref{U(g)lineoperator} in representations $\rho(V_1)$ and $\rho(V_2)$ of the Yangian, the OPE constructs a line operator in the tensor product representation $\rho(V_1\otimes V_2) = \rho(V_1)\otimes \rho(V_2) \circ \Delta_\hbar$ obtained via the coproduct.

\medskip

In fact, one can see that line operators in the full quantum theory really correspond to representations of $\cY(\fg)$ directly from their coupling to an external gauge field, without having to consider the OPE between pairs of line operators. Costello-Witten-Yamazaki
proved in~\cite{Costello:2017dso} that line operators suffer from a gauge anomaly, which appears in a two-loop diagram. This anomaly is cancelled if and only if the generators $t^{(0)}_a$ and $t^{(1)}_a$ obey the deformed algebra~\eqref{DrinfeldJ}\footnote{Strictly, they showed that the generators satisfy an algebra isomorphic to~\eqref{DrinfeldJ}.}, rather than the naive algebra of $\fg$. A further proof that line operators correspond to representations of $\cY(\fg)$, based on the RTT presentation of the Yangian, was given in~\cite{Costello:2018gyb} and will be reviewed in section~\ref{sec:RTT}.

\subsection{Twisted Yangians and the Bulk/Boundary OPE}
\label{sec:TwistedYangians}

The presence of a boundary breaks the Yangian symmetry of a quantum integrable system to a subalgebra $\cB(\fg,\fh)$ known as the 
{\it twisted Yangian}~\cite{belliard2017drinfeld,MacKay:2004tc,molev2002representations,guay2016twisted}. Let us first give a brief review of the twisted Yangians.

\medskip

We first use the splitting $\fg=\fh\oplus \fm$ to build an infinite-dimensional loop algebra $\fh[[u^2]]\oplus u\,\fm[[u^2]]$. As a vector space, $\fh[[u^2]]\oplus u\,\fm[[u^2]]$ is generated by $b^{(2m)}_\alpha= b_\alpha u^{2m}$ and $b^{(2m+1)}_\mu= b_\mu u^{2m+1}$ for $m\in\Z_0^+$, where $b_\alpha$ and $b_\mu$ are generators of $\fh$ and $\mathfrak{m}$, respectively. Its algebraic structure is given by the relations
\begin{equation}
\label{BoundaryLoopAlg}
\begin{aligned}
	&[b^{(2m)}_\alpha ,b^{(2n)}_{\beta}]= f_{\alpha\beta}^{\ \ \gamma} b^{(2m+2n)}_\gamma\,,\\
	&[b^{(2m)}_\alpha, b^{(2n+1)}_\mu] = f_{\alpha\mu}^{\ \ \nu} b^{(2m+2n+1)}_\nu\,,\\
	&[b^{(2m+1)}_\mu, b^{(2n+1)}_\nu] = f_{\mu\nu}^{\ \ \alpha}b^{(2m+2n+2)}_{\alpha}
\end{aligned}
\end{equation}
inherited from the underlying relations $[\fh,\fh]\subseteq\fh$, $[\fh,\fm]\subseteq\fm$ and $[\fm,\fm]\subseteq\fh$. Note also that any representation $W$ of $\fh$ can be lifted to a representation of \smash{$\fh[[u^2]]\oplus u\,\fm[[u^2]]$} by allowing the \smash{$b^{(0)}_\alpha$} to act on $W$ in the obvious way and setting all other generators to zero. Clearly $\fh[[u^2]]\oplus u\,\fm[[u^2]]$ can be viewed as a Lie subalgebra of $\fg[[u]]$.

The twisted Yangian $\cB(\fg,\fh)$ is a deformation of the universal enveloping algebra of \smash{$\fh[[u^2]]\oplus u\,\fm[[u^2]]$.} The natural coproduct on $U(\fh[[u^2]]\oplus u\,\fm[[u^2]])$, which using the fact that $\fh[[u^2]]\oplus u\,\fm[[u^2]]\subset\fg[[u]]$, can be viewed as a map $U(\fh[[u^2]]\oplus u\,\fm[[u^2]])\to\fg[[u]]\otimes\fh[[u^2]]\oplus u\,\fm[[u^2]]$, gets deformed to
\[
	\Delta_\hbar : \cB(\fg,\fh) \to \cY(\fg)\otimes \cB(\fg,\fh)\,.
\]
Thus the twisted Yangian appears as a left coideal of the Yangian. Explicitly, this deformation is~\cite{delius2001boundary,belliard2017drinfeld}
\begin{equation}
\label{twistedcoprod}
\begin{aligned}
	\Delta_\hbar(b^{(0)}_\alpha) &= t^{(0)}_\alpha\otimes {\bf 1} + {\bf 1}\otimes b^{(0)}_\alpha \\
	\Delta_\hbar(b^{(1)}_\mu) &= \bigg(t^{(1)}_\mu 
	+ \frac{\hbar}{4}f_{\mu}^{\ \beta\nu}\{t^{(0)}_\beta,t^{(0)}_\nu\}\bigg)\otimes {\bf 1} 
	+ {\bf 1}\otimes b^{(1)}_\mu- \hbar f_{\mu}^{\ \nu\gamma}t^{(0)}_\nu\otimes b^{(0)}_\gamma
\end{aligned}
\end{equation}
on the generators, where $\{t^{(0)}_\beta,t^{(0)}_\nu\}$ is the anticommutator. (We will slightly abuse terminology by continuing to refer to $\Delta_\hbar$ as a coproduct even when acting on the twisted Yangian.) At $\hbar=0$, this is just the (trivial) coproduct on \smash{$U(\fh[[u^2]]\oplus u\fm[[u^2]])$.}  Similar to the Yangian itself, in order for this coproduct to be a homomorphism one must deform some of the relations among the level-1 generators in~\eqref{BoundaryLoopAlg}, with concomitant deformations of the algebra for higher level generators\footnote{Specifically, 
the level-1 generators of the twisted Yangian must obey the `horrific relations'~\cite{belliard2017drinfeld}
\[
\begin{aligned}
	&\Phi_{\mu\nu}^{\ \ \rho\sigma}[b^{(1)}_{\rho},b^{(1)}_{\sigma}] 
	= \hbar^2\,\Phi_{\mu\nu}^{\ \ \rho\sigma} S_{\rho\sigma}\big(b^{(0)}\big)\,,\\
	&\Psi_{\mu\nu\lambda}^{\ \ \ \pi \rho\sigma}
	[[b^{(1)}_{\pi},b^{(1)}_{\rho}],b^{(1)}_{\sigma}]
	=\hbar^2 \,\Psi_{\mu\nu\lambda}^{\ \ \ \pi \rho\sigma}T_{\pi\rho\sigma}
	\big(b^{(0)},b^{(1)}\big)~,
\end{aligned}
\]
where
\[
	\Phi_{\mu\nu}^{\ \ \rho\sigma} = \delta_\mu^{\ \rho}\delta_\nu^{\ \sigma} 
	+ \sum_{\alpha}\bar{c}_{(\alpha)}^{-1}f_{\mu\nu}^{\ \ \alpha}f_{\alpha}^{\ \rho\sigma}
	\qquad\hbox{and}\qquad
	\Psi_{\mu\nu\lambda}^{\ \ \ \pi\rho\sigma} 
	= \delta_\mu^{\ \pi}\delta_\nu^{\ \rho}\delta_\nu^{\ \sigma} 
	+ 2c_{\fg}^{-1}f_{\mu\nu}^{\ \ \alpha}f_{\lambda\alpha}^{\ \ \pi}
	(\kappa^{-1})^{\rho\sigma}
\]
are projection operators and $S$ and $T$ are given by
\[\begin{aligned}
	S_{\rho\sigma}\big(b^{(0)}\big) 
	&= \frac{1}{18}f^{\alpha\mu}_{\ \ \rho}f^{\beta\nu}_{\ \ \sigma}
	f^{\gamma}_{ \ \mu\nu}\{b^{(0)}_\alpha,b^{(0)}_\beta,b^{(0)}_\gamma\}~,\\
	T_{\pi\rho\sigma}\big(b^{(0)},b^{(1)}\big) 
	&= \frac{1}{24}\Big(f_{\mu\nu}^{\ \ \gamma}f_{\pi}^{\ \alpha\mu}f_{\rho}^{\ \beta\nu}
	f_{\gamma\sigma}^{\ \ \xi} + f_{\pi\rho}^{\ \ \gamma}f_\gamma^{\ \alpha\delta}
	f_{\sigma}^{\ \beta\mu}f_{\delta\mu}^{\ \ \lambda}\Big)
	\{b^{(0)}_\alpha,b^{(0)}_\beta,b^{(1)}_\lambda\}\,.
\end{aligned}
\]
Here we employ notation introduced in appendix~\ref{app:symspa}. (The case $\fg=\mathfrak{sl}_2(\C)$ is again exceptional; the required deformations in this case can be found in~\cite{belliard2012coideal}.) Following the results of~\cite{Costello:2017dso} for $\cY(\fg)$, we expect these $\cO(\hbar^2)$ terms arise from a 2-loop anomaly of the gauge theory on $\widetilde{M}$. However, we do not attempt to prove this in the present paper.}. Just as for the bulk Yangian, when $\hbar\neq0$ we can rescaled the level-1 generators to absorb $\hbar$, so all the relations with $\hbar\neq0$ are isomorphic. Again, for comparison to the gauge theory we prefer to keep $\hbar$ explicit.

\medskip

We now show that the coproduct of the twisted Yangian naturally arises from the OPE of line operators in CWY theory on an orbifold. Note that since 
\[
	\Delta_\hbar:\cB(\fg,\fh)\to \cY(\fg) \otimes \cB(\fg,\fh)\,,
\] 
rather than decomposing tensor products of twisted Yangian representations, instead this coproduct allows us to decompose the tensor product of a Yangian representation and a twisted Yangian representation into representations of the twisted Yangian. This suggests we should consider the OPE between a `boundary' Wilson line supported on the orbifold singularity $L=\{x=z=0\}$ and a parallel bulk Wilson line, also supported at $z=0$. 

For definiteness, suppose the bulk Wilson line is labelled by a representation $V$ of $\fg$ and the boundary line operator transforms in a representation $W$ of $\fh$. Classically, as the bulk line operator approaches $L$ we generate a new boundary Wilson line in the tensor product $V\otimes W$, where the representation $V$ is restricted to $\fh\subset\fg$ since $A|_{\fm}=0$ on $L$. As in the bulk, this receives a quantum correction. At $\cO(\hbar)$ the correction comes from the two Feynman diagrams
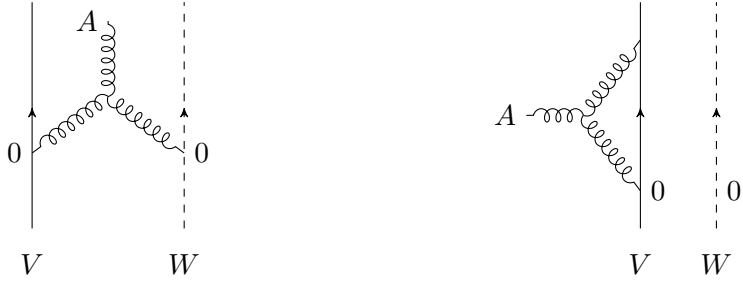
\begin{figure}[th]
	\centering
	\begin{tikzpicture}[baseline]
	\begin{feynman}
	\vertex (nw1);
	\vertex[below=3cm of nw1](sw1);
	\vertex[below=2cm of nw1](w1);
	\vertex[right=2cm of w1](e1);
	\vertex[above=2cm of e1](ne1);
	\vertex[below=1cm of e1](se1);
	\vertex[right=1cm of w1](s1);
	\vertex[above=0.75cm of s1](c1);
	\vertex[above=1cm of c1](n1);
	\vertex[below=0.2cm of sw1]{$V$};
	\vertex[below=0.2cm of se1]{$W$};
	\vertex[left=0cm of n1]{$A$};
	\vertex[right=6cm of ne1](nw2);
	\vertex[below=3cm of nw2](sw2);
	\vertex[below=1.5cm of nw2](e2);
	\vertex[below=1cm of e2](wsw2);
	\vertex[above=1cm of e2](wnw2);
	\vertex[right=1cm of nw2](ne2);
	\vertex[below=3cm of ne2](se2);
	\vertex[left=0.75cm of e2](c2);
	\vertex[left=0.75cm of c2](w2);
	\vertex[left=0cm of w1]{$0$};
	\vertex[right=0cm of e1]{$0$};
	\vertex[right=0cm of wsw2]{$0$};
	\vertex[right=1cm of wsw2]{$0$};
	\vertex[below=0.2cm of sw2]{$V$};
	\vertex[below=0.2cm of se2]{$W$};
	\vertex[left=0cm of w2]{$A$};
	\diagram*{{[edges=fermion] (sw1) --(nw1), (sw2) -- (nw2)}, 
	{[edges = charged scalar] (se1) -- (ne1), (se2) -- (ne2)}, 
	{[edges = gluon] (c1) -- (w1), (c1) -- (e1), (c1) -- (n1), (c2) -- (w2), (c2) -- (wsw2), 
	(c2) -- (wnw2)},};
	\end{feynman}
	\end{tikzpicture}
\end{figure}\\
The diagram on the left is completely analogous to that of the bulk OPE, except that we must compute it with the propagator $\widetilde{\Delta}(w,w')$. The second diagram might appear to simply renormalize the coupling of the bulk Wilson line, an effect already present and accounted for by a classical redefinition. However, as this line approaches $L$, the diagram receives further contributions from self-interaction with its `mirror', an effect not present when the bulk Wilson line is well separated from the boundary.

\medskip

Let's start by calculating the contribution from the first diagram, which is given by
\begin{equation*}
\begin{aligned}
	\frac{\im\hbar}{2\pi}\,f_{abc}\,t_{d,V}\otimes t_{\varepsilon,W}
	\int_{-\infty}^{\infty}\diff s\int_{-\infty}^{\infty}\d t\int_{M_{\rm L}}\d^4w\,
	A^a_y(w)\Big(\widetilde{\Delta}_{\bar zy}^{bd}\big(w;\gamma(s)\big)
	\widetilde{\Delta}_{xy}^{c\varepsilon}\big(w;\ell(t)\big)
	\\
	-\widetilde{\Delta} _{xy}^{bd}\big(w;\gamma(s)\big)
	\widetilde{\Delta}_{\bar zy}^{c\varepsilon}(w;\ell(t)\big)\Big)\,.
	\end{aligned}
\end{equation*}
Here $\gamma(s)$ and $\ell(t)$ are the paths of the bulk and boundary Wilson lines respectively, given explicitly by
\[
	 \gamma(s) = (-\epsilon,s,0,0)~,\qquad \ell(t) = (0,t,0,0)
\]
where $\epsilon$ is the separation between the two lines. Eventually we'll take the $\epsilon\to0$ limit. From our $K$-matrix calculation, we know that the propagator joining a point in the bulk to a point on $L$ simplifies to 
\[
	\widetilde{\Delta}_{ij}^{c\varepsilon}(w,\ell) = 2c^{c\varepsilon}\Delta_{ij}(w,\ell)\,,
\]
in terms of a single propagator on $M$, and furthermore that this vanishes unless the index $c$ lies in $\fh$. With this understanding, we set $c=\gamma$, whereupon the bracketed term in the integrand becomes
\[
	 c^{\gamma\epsilon}\,\frac{1}{\pi}\frac{1}{(x^2 + (y-t)^2 + |z|^2)^2}
	 \Big[2\bar z\widetilde{\Delta}_{\bar zy}^{bd}\big(w;\gamma_1(s)\big) 
	 + x\widetilde{\Delta}_{xy}^{bd}\big(w;\gamma_1(s)\big)\Big]\,.
\]
Substituting the explicit form~\eqref{PropagatoronM} of the propagators into the above and performing the $s$ and $t$ integrals gives
\[
	 c^{\gamma\epsilon}\,\frac{\epsilon}{4}\,\frac{\bar z}{(x^2 + |z|^2)^{3/2}}
	 \bigg[\frac{(c^\sigma)^{bd}}{((x-\epsilon)^2+|z|^2)^{3/2}} 
	 - \frac{c^{bd}}{((x+\epsilon)^2+|z|^2)^{3/2}}\bigg]\,.
\]
Now, noting that $c+c^\sigma\in\mathfrak{h}^{\otimes 2}$, and $c-c^\sigma \in\mathfrak{m}^{\otimes 2}$, we can write the contribution of the first diagram as
\[
 -\frac{\im\hbar}{8\pi}\int_{M_{\rm L}}\d^4w\,A^a_y(w)
 \Big(f_a^{\ \beta\gamma}\,t_{\beta,V}\otimes t_{\gamma,W}\,\mathcal{I}_-(x,z,\bar z;\epsilon) 
 + f_a^{\ \nu\gamma}\,t_{\nu,V}\otimes t_{\gamma,W}\,\mathcal{I}_+(x,z,\bar z;\epsilon)\Big]\,,
\]
where we've introduced the functions
\[
	\mathcal{I}_\pm(x,z,\bar z;\epsilon) = 
	\frac{\epsilon\bar z}{(x^2 + |z|^2)^{3/2}}\bigg[\frac{1}{((x+\epsilon)^2+|z|^2)^{3/2}} 
	\pm \frac{1}{((x-\epsilon)^2+|z|^2)^{3/2}}\bigg]\,.
\]

We now wish to take the limit $\epsilon\to0$. Certainly $\mathcal{I}_\pm$ converge uniformly to 0 on the complement of any neighbourhood of $L$, but are unbounded for sufficiently small $\epsilon$ inside any such neighbourhood. This suggests that in this limit these functions tend to distributions supported on $L$, and constant along $L$ since they are independent of $y$. By dimensional analysis and rotation symmetry in the $z$ direction, this distribution must be proportional to $\del_z\delta^3(x,z,\bar z)$ where $\delta^3(x,z,\bar z)$ is a $\delta$-function satisfying $\int_{\R_{\leq 0}\times\C}\d x\,\d^2 z\,\delta^3(x,z,\bar z)=1$. This allows us to write
$\lim_{\epsilon\to0}\,\mathcal{I}_\pm(x,z,\bar z;\epsilon)  = -\lambda_{\pm}\del_z\big(\delta^3(x,z,\bar z)\big)$, where
\begin{equation}
\label{eq:bOPE1} 
	\lambda_{\pm} = \int_{-\infty}^0\d x\int_{\C}\d^2 z\ z\ \mathcal{I}_{\pm}(x,z,\bar z;1)\,. \end{equation}
We determine the constants $\lambda_{\pm}$ in appendix~\ref{app:inteval}, showing that 
$\lambda_+ = -8\pi \im$ whereas $\lambda_- = -4\pi \im$.

Using these, the contribution of the first Feynman diagram can be written as
\[
-\frac{\hbar}{2}\int_{-\infty}^\infty\d y\,\Big[f_\alpha^{\ \beta\gamma}\,t_{\beta,V}\otimes t_{\gamma,W}\,\del_zA^\alpha_y(\ell(y)) + 2f_\mu^{\ \nu\gamma}\,t_{\nu,V}\otimes t_{\gamma,W}\,\del_zA^\mu_y(\ell(y))\Big]\,
\]
where we've used the properties $[\fh,\fh]\subseteq \fh$  and $[\mathfrak{m},\mathfrak{m}]\subseteq\fh$ of symmetric spaces to simplify the structure constants. However, differentiating the orbifold condition $A=\sigma\cP^*A$ with respect to $z$ $n$ times  and then restricting to $L$, one finds
\[
	\del_z^n A_y^\alpha\big|_L = (-1)^n \,\del_z^n A_y^\alpha\big|_L\,,
\]
for the $\fh$ part of $A_y\big|_L$, whereas
\[
	 \del_z^n A_y^\mu\big|_L = (-1)^{n+1}\,\del_z^n A_y^\mu\big|_L
\]
for the $\fm$ part. In particular, $\del_zA^\alpha_y\big|_L=0$, so the actual contribution of the first Feynman diagram is just
\[
	-\hbar\,f_\mu^{\ \nu\gamma}\,t_{\nu,V}\otimes t_{\gamma,W}\int_{-\infty}^\infty\d y\ 
	\del_zA^\mu_y(\ell(y))
\]
coupling to the single $z$-derivative of the $\fm$ part of the gauge field.

\medskip

Let's now consider the second Feynman diagram, which is given by
\[ 
	\frac{\im\hbar}{2\pi}f_{abc}\,t_{d,V}t_{e,V}\otimes{\bf 1}_W \hspace{-0.3cm}
	\int\limits_{-\infty<s<t<\infty}\hspace{-0.3cm}\d s\,\d t
	\int\limits_{M_{\rm L}}\d^4w\,A^a_y(w)
	\Big(\widetilde{\Delta}_{\bar zy}^{bd}\big(w;\gamma(s)\big)
	\widetilde{\Delta}_{xy}^{ce}\big(w;\gamma(t)\big) 
	- \widetilde{\Delta}_{xy}^{bd}\big(w;\gamma(s)\big)
	\widetilde{\Delta}_{\bar zy}^{ce}(w;\gamma(t)\big)\Big)
\]
with $\gamma(s) = (-\epsilon,s,0,0)$ as before. The combination of  propagators appearing in the integrand is
\begin{multline*}
	\widetilde{\Delta}_{\bar zy}^{bd}\big(w;\gamma_1(s)\big)
	\widetilde{\Delta}_{xy}^{ce}\big(w;\gamma_1(t)\big) - 
	\widetilde{\Delta}_{\bar xy}^{bd}\big(w;\gamma_1(s)\big)
	\widetilde{\Delta}_{\bar zy}^{ce}\big(w;\gamma_1(t)\big) \\
	=  \frac{\epsilon\bar z}{\pi^2}\left[
	\frac{(c^\sigma)^{bd}\,c^{ce}}{((x-\epsilon)^2 + (y-s)^2 + |z|^2)^2((x+\epsilon)^2 + (y-t)^2 + |z|^2)^2}\right.  \\ 
	\qquad\qquad- \left.\frac{c^{bd}\,(c^\sigma)^{ce}}{((x+\varepsilon)^2 + (y-s)^2 + |z|^2)^2((x-\varepsilon)^2 + (y-t)^2 + |z|^2)^2}\right]\,.
\end{multline*}
As before, we decompose this according to the splitting $\fg=\fh\oplus\fm$, with $b\to(\beta,\mu)$, $c\to(\gamma,\nu)$, $d\to(\delta,\rho)$ and $e\to(\varepsilon,\sigma)$. This allows us to write the above equation as 
\[
	\left(c^{\beta\delta}\,c^{\gamma\varepsilon} - c^{\mu\rho}\,c^{\nu\sigma}\right)\,
	\mathcal{J}_-(x,z,\bar z;s-y,t-y;\epsilon) 
	+ \left(c^{\beta\delta}\,c^{\nu\sigma} - c^{\mu\rho}\,c^{\gamma\varepsilon}\right)\,
	\mathcal{J}_+(x,z,\bar z;s-y,t-y;\epsilon)\,,
\]
where we have defined
\[
	\mathcal{J}_\pm(x,z,\bar z;s,t;\epsilon)=\frac{\epsilon\bar z}{\pi^2}
	\left[\frac{1}{((x-\epsilon)^2 + s^2 + |z|^2)^2((x+\epsilon)^2 + t^2 + |z|^2)^2} \pm 
	(s\leftrightarrow t)\right]\,.
\]
Since the gauge field is independent of $s, t$, we can integrate over these variables directly. One finds 
\[
	\int\limits_{-\infty<s<t<\infty} \d s\,\d t \ \mathcal{J}_-(x,z,\bar{z};s,t,\epsilon) =0
\]
using the fact that $\mathcal{J}_-$ is antisymmetric under exchange of $s$ and $t$, and invariant under $s\mapsto-s$, $t\mapsto-t$. On the other hand, the integral of $\mathcal{J}_+$ is non-vanishing, and we find
\[
	\mathcal{K}_+(x,z,\bar{z};\epsilon)\equiv 
	\int\limits_{-\infty<s<t<\infty} \d s\,\d t \ \mathcal{J}_+(x,z,\bar{z};s,t,\epsilon) =
	\frac{1}{4}\frac{\varepsilon \bar z}{((x-\epsilon)^2+|z|^2)^{3/2}((x+\epsilon)^2+|z|^2)^{3/2}}\,.
\]
We now consider taking the $\epsilon\to0$ limit. By an identical argument to that of the first diagram, we deduce that
\[ 
	\lim_{\epsilon\to0}\,\mathcal{K}_+(x,z,\bar z;\epsilon) 
	= -\mu_+\partial_z\big(\delta(x)\delta^2(z,\bar z)\big) \,,
\]
where, as shown in appendix~\ref{app:inteval}, the constant
\begin{equation} 
\label{eq:bOPE2} 
	\mu_+ = \int_{-\infty}^0\d x 
	\int_{\C}\d^2 z\ z\ \mathcal{K}_{+}(x,z,\bar z;1) = -\frac{\pi}{2} \im \,.
\end{equation}
Thus the contribution of the second diagram to the OPE is
\[ 
	\frac{\hbar}{4}\left[f_{a\beta\nu}\,c^{\beta\delta}\,c^{\nu\sigma}\,t_{\delta,V}\,t_{\sigma,V}
	\otimes{\bf 1}_W - f_{a\mu\gamma}\,c^{\mu\rho}\,c^{\gamma\varepsilon}\,
	t_{\rho,V}\,t_{\varepsilon,V}\otimes{\bf 1}_W \right]
	\int_{-\infty}^{\infty}\d y\,\big(\del_zA^a_y\big)(\ell(y)) \,.
\]
Since $[\fh,\fm]\subseteq\fm$, the structure constants appearing here are non-zero only when the index $a$ takes values in $\fm$. The  two terms can then be combined, so that contribution of the second Feynman diagram is
\[ 
	\frac{\hbar}{4}f_{\mu}^{\ \beta\nu}\,\{t_{\beta},t_{\nu}\}_V\otimes{\bf 1}_W
	\int_{-\infty}^{\infty}\d y\,\big(\del_zA^\mu_y\big)(\ell(y))~,
\]
where $\{t_\beta,t_\nu\}$ is the anticommutator, here in the $V$ representation.

Combining the two diagrams, the total contribution to the OPE between a bulk and boundary Wilson line is
\begin{equation} 
\label{eq:bcoprod2}   
	\bigg(-\hbar\,f_\mu^{\ \nu\alpha}\,t_{\nu,V}\otimes t_{\alpha,W} 
	+ \frac{\hbar}{4}\,f_{\mu}^{\ \beta\nu}\,\{t_{\beta},t_{\nu}\}_V\otimes{\bf 1}_W\bigg)
	\int_{-\infty}^{\infty}\d y\ \partial_zA^\mu_y(y)
\end{equation}
at order $\hbar$. Let's now interpret this result.

\medskip

As in section~\ref{sec:Yangian}, the presence of $z$-derivatives of $A$ in~\eqref{eq:bcoprod2} means that, classically, we should consider line operators on $L$ which are labelled by representations of the loop algebra $\fh[[u^2]]\oplus u\mathfrak{m}[[u^2]]$. This is natural from the perspective of the orbifold: the conditions~\eqref{OrbifoldConditions} force
\[ 
	\partial_z^{2m} \!A^\mu_y\big|_L = 0\qquad\hbox{and} \qquad
	\partial_z^{2m+1} \!A^\alpha_y\big|_L = 0\,;
\]
that is, the $\fh$ part of odd $z$-derivatives of $A_y$ and the $\fm$ part of even $z$-derivatives of $A_y$ vanish on $L$. Therefore, expanding the $y$-component of the $\fg[[u]]$-valued gauge field $A^{\fg[[u]]}$ gives 
\[
	A^{\fg[[u]]}_y\big|_L
	= \sum_{m=0}^\infty \frac{t^{(2m)}_\alpha}{(2m)!}\,\del^{2m}_z\!A^\alpha_y\bigg|_L
	+ \sum_{m=0}^\infty \frac{t^{(2m+1)}_\mu}{(2m+1)!}\,\del^{2m+1}_z\!A^\mu_y \bigg|_L
\]
upon restriction to $L$. Thus, $A^{\fg[[u]]}_y|_L$ indeed takes values in $\fh[[u^2]]\oplus u\,\fm[[u^2]]$ with the identifications $b^{(2m)}_\alpha \mapsto t^{(2m)}_\alpha$ and $b^{(2m+1)}_\mu \mapsto t^{(2m+1)}_\mu$.  The single $z$-derivative that appears in the line operator on $L$ in~\eqref{eq:bcoprod2}  shows that, starting from a representation in which all but the level-0 generators vanish, at order $\hbar$ the OPE has generated a non-trivial level-1 coupling. Furthermore, comparing the explicit coupling in~\eqref{eq:bcoprod2} to the deformation in~\eqref{twistedcoprod}, this level-1 coupling is exactly what we expect from the coproduct on the twisted Yangian.

\medskip 

As in the bulk, this strongly suggests that the quantum OPE between a bulk line operator and a line operator on $L$ should be interpreted as realising the coproduct on $\cB(\fg,\fh)$ in the gauge theory. Correspondingly, we should expect line operators on $L$ to be labelled by representations of the twisted Yangian, rather than representations of \smash{$U\big(\mathfrak{h}[[u^2]]\oplus u\,\mathfrak{m}[[u^2]]\big)$}. We will prove that this really is the case in the next section, where it will be convenient to work in the $RTT$ presentation of the twisted Yangian.

\section{The $RTT$ Presentation}
\label{sec:RTT}

In this section we consider the gauge theory realisation of the $RTT$ presentation  integrable systems. In~\cite{Costello:2018gyb} this was used to provide a further proof that bulk line operators are labelled by representations of the Yangian. Similarly, we shall use the $RTT$ presentation of the bYBE to show that line operators supported on the fixed line $L$ transform in representations of the twisted Yangian. An important ingredient in the construction of the Yangian of $\fg = \mathfrak{sl}_n$ in the $RTT$ presentation is the quantum determinant condition, and the corresponding object for its associated twisted Yangians is the Sklyanin determinant. To realise these objects in the gauge theory, it is first necessary to understand the behaviour of line operators that curve in $\Sigma$.

\subsection{The Framing Anomaly}
\label{app:Framing}

In the bulk, line operators which curve in $\Sigma$ were shown in~\cite{Costello:2013zra,Costello:2017dso} to suffer from an anomaly, coming from the Feynman diagram
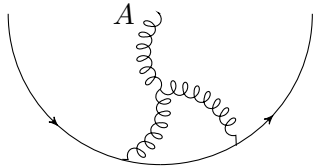
\begin{figure}[h!]
	\centering
	\begin{tikzpicture}[baseline]
	\begin{feynman}
	\vertex (n);
	\vertex [below=1cm of n] (c);
	\vertex [right=2cm of n] (e);
	\vertex [below=2cm of n] (s);
	\vertex [left=2cm of n] (w);
	\vertex [above=0.07cm of s] (uy);
	\vertex [left=0.5cm of uy] (u);
	\vertex [above=0.25cm of s] (vy);
	\vertex [right=1cm of vy] (v);
	\diagram*{{[edges=fermion] (w) -- [quarter right] (s), (s) -- [quarter right] (e)},{[edges=gluon] (c) -- [quarter left] (u), (c) -- [quarter left] (v), (c) -- [quarter left] (n)},} ;
	\vertex[left = 0.2cm of n]{$A$};
	\end{feynman}
	\end{tikzpicture}
\end{figure}\\

\noindent This diagram diverges when all three vertices coincide. As shown in~\cite{Costello:2017dso}, this divergence cannot be (entirely) removed by a renormalization of the Wilson line, and signals a loss of gauge invariance. Instead, the anomaly is cancelled by shifting the spectral parameter so that
\[
z - \frac{\hbar{\bf h}^\vee}{2\pi} \varphi = {\rm const}.\,,
\]
where ${\bf h}^\vee$ is the dual Coxeter number of $\fg$ and $\varphi$ is the angle between the Wilson line and some framing of $\Sigma$, increasing as the Wilson line rotates clockwise. 

Curved line operators on $\widetilde{M}$ also suffer from this anomaly. Above, when computing the $K$-matrices, we considered only straight lines on $\widetilde{M}$. These have no anomaly, despite the fact that when represented in $M_{\rm L}$, they reflect off the boundary with equal angles of incidence and reflection.
It is also possible to consider more general curves on the orbifold provided 
$z-\hbar\,{\bf h}^\vee\,\varphi/2\pi$ remains constant. Note that, when reflected to lie in $M_{\rm L}$, the fact that the $\Z_2$ action takes $z\mapsto-z$ means that the spectral parameter will appear to be shifted in the opposite sense. As ever, there is nothing special about the choice of $\pi$: if a line $\gamma$ has a kink which happens to coincide with $\gamma\cap\pi$, we can always move this kink away from $\pi$ by performing a diffeomorphism of the orbifold. Once the kink lies inside $M_{\rm L}$, the bulk framing anomaly applies.

\subsection{The $RTT$ Presentation in the Bulk}
\label{sec:BulkRTT}

The $RTT$ presentation is an alternative way to construct representations of $\cY(\fg)$, starting from a solution to the Yang-Baxter equation~\cite{Takhtajan:1979iv,drinfeld1986quantum,ChariPressleyBook}. Specifically, if $U$ is the defining representation $\C^n$ of $\fg$, then given an $R$-matrix
\[
	R(z) : U\otimes V \to U\otimes V
\]
we obtain a {\it transfer matrix} $T^i_{\ j,V}(z)\in{\rm End}(V)$ by extracting the $(i,j)^{\rm th}$ entries of $R(z)$ in a basis $\{|1\ra,|2\ra,\ldots,|n\ra\}$ of $U$. As a picture, this is
\begin{figure}[h!]
	\centering
	\begin{tikzpicture}[baseline]
	\begin{feynman}
	\vertex (c);
	\vertex[above=1cm of c] (n);
	\vertex[below=1cm of c] (s);
	\vertex[left=1cm of c] (w);
	\vertex[right=1cm of c] (e);
	\vertex[below=0.5cm of c] (cs);
	\vertex[left=0.5cm of c] (cw);
	\vertex[left=0.75cm of c] (cww);
	\vertex[below=0.75cm of c] (css);
	\diagram*{{[edges=fermion] (s) --[edge label'=$0$] (c), (c) --  (n), (w)
 	-- [edge label=$z$] (c), (c) -- (e)},} ;
	\vertex[below=0.2cm of css] {$V$};
	\vertex[left=0cm of w] {$\langle i|$};
	\vertex[right=0cm of e] {$|j\rangle$};
	\vertex[left=2cm of c] {$T^i_{\ j,V}(z) \ = $};
	\end{feynman}
	\end{tikzpicture}\,.
\end{figure}\\
Expanding the transfer matrix as a power series in $1/z$ one has
\begin{equation}
\label{TransferExpand}
	T^i_{\ j,V}(z) = \delta^i_{\ j}{\bf 1}_V 
	+ \hbar\sum_{m=0}^\infty \frac{(\hat{t}^{(m)})^i_{\ j,V}}{z^{m+1}}\,,
\end{equation}
where we have denoted the coefficient of $1/z^{m+1}$ by \smash{$(\hat{t}^{(m)})^i_{\ j,V}\in\text{End}(V)$}. 

The idea of the $RTT$ presentation is to impose constraints on the $(\hat{t}^{(m)})^i_{\ j,V}$ to ensure that they form a representation of the Yangian $\cY(\fg)$ acting on $V$~\cite{drinfeld1986quantum}. The first condition one needs is that the $R$-matrix used to construct $T$ should solve the Yang-Baxter equation. In terms of the transfer matrix, this is the condition
\begin{figure}[h!]
	\centering
	\begin{tikzpicture}[baseline]
	\begin{feynman}
	\vertex (o){$=$};
	\vertex[left=3cm of o] (l);
	\vertex[above=1.3cm of l] (lbf);
	\vertex[below=1.3cm of l] (lbi);
	\vertex[left=0.75cm of l](ll);
	\vertex[right=0.5cm of l](lc);
	\vertex[above=0.86cm of ll](lcf);
	\vertex[below=0.86cm of ll](lai);
	\vertex[right=1.5cm of l](lr);
	\vertex[above=0.42cm of lr](laf);
	\vertex[below=0.42cm of lr](lci);
	\vertex[right=3cm of o] (r);
	\vertex[above=1.3cm of r] (rbf);
	\vertex[below=1.3cm of r] (rbi);
	\vertex[right=0.75cm of r](rr);
	\vertex[above=0.86cm of rr](raf);
	\vertex[below=0.86cm of rr](rci);
	\vertex[left=1.5cm of r](rl);
	\vertex[left=0.5cm of r](rc);
	\vertex[above=0.42cm of rl](rcf);
	\vertex[below=0.42cm of rl](rai);
	\diagram*{{[edges=fermion] (lai) -- (laf), (lbf) --  (lbi), (lcf) --  (lci), (rai) -- [edge label=$0$] (raf), (rbf) --  (rbi), (rcf) --  (rci)},};
	\vertex[left=0cm of lai]{$V$};
	\vertex[below=0.1cm of lbi]{$|m\ra$};
	\vertex[right=0.1cm of lci]{$|n\ra$};
	\vertex[above=0cm of lbf]{$\langle i|$};
	\vertex[left=0cm of lcf]{$\langle j|$};
	\vertex[left=0cm of rai]{$V$};
	\vertex[below=0.1cm of rbi]{$|m\ra$};
	\vertex[right=0.1cm of rci]{$|n\ra$};
	\vertex[above=0cm of rbf]{$\langle i|$};
	\vertex[left=0cm of rcf]{$\langle j|$};
	\vertex[left=0.1cm of l]{$u$};
	\vertex[right=0.1cm of r]{$u$};
	\vertex[above=0.35cm of lc]{$v$};
	\vertex[below=0.35cm of lc]{$0$};
	\vertex[below=0.35cm of rc]{$v$};
	\end{feynman}
	\end{tikzpicture}
\end{figure}\\

\vspace{-0.3cm}\noindent where the lines supported at $u,v\in\C$ each carry a copy of the same representation $U$. Algebraically, this picture is the equation
\[ 
	R^{ij}_{\ \ k\ell}(u-v)T^k_{\ m,V}(u)T^\ell_{\ n,V}(v) 
	= T^j_{\ \ell,V}(v)T^i_{\ k,V}(u)R^{k\ell}_{\ \ mn}(u-v)\,,
\]
or in short
\begin{equation} 
\label{eq:RTT} 
	R_{12}(u-v)T_{1,V}(u)T_{2,V}(v) = T_{2,V}(v)T_{1,V}(u)R_{12}(u-v)\,,
\end{equation}
where the subscripts $1$ and $2$ indicate on which factor of $U$ the operators act. These are known as the $RTT$ {\it relations}. 

\medskip

From the gauge theory perspective, the transfer matrix is the {\it quantum expectation value} $T^i_{\ j,V}(z) = \la \cW_V[\gamma_{z=0}^{\rm vert}]\cW_U[\gamma_z^{\rm hor}]^i_{\ j}\ra$ of the pair of crossed line operators, again with appropriate states in the $U$ space extracted after computing the correlation function. The coefficients $\hat{t}^{(m)}$ in the expansion~\eqref{TransferExpand} are related to the generators $t^{(m)}$ appearing in the expansion of the classical line operator as a function of the gauge field: at leading order these generators agree, but the $\hat{t}^{(m)}$ receive further contributions from quantum corrections 
to the expectation value of the line operator at each order in $\hbar$. In the $RTT$ presentation, one works directly with the quantum expectation value $T$, rather than the classical line operators. As explained in~\cite{Costello:2018gyb}, the fact that these transfer matrices indeed solve the $RTT$ equations~\eqref{eq:RTT} follows from the construction of the gauge theory.

\medskip 

By themselves, the $RTT$ relations are not sufficient to ensure that bulk line operators transform in representations of $\cY(\fg)$. For the case of $\fg=\mathfrak{sl}_n$, in addition one requires~\cite{Kulish:1981bi,drinfeld1986quantum} 
\[ 
	\epsilon_{i_1i_2\dots i_n}T^{i_1}_{\ j_1,V}(z)T^{i_2}_{\ j_2,V}(z+\hbar)\cdots 
	T^{i_n}_{\ j_n,V}(z+(n-1)\hbar) = \epsilon_{j_1j_2\dots j_n}{\bf 1}_V\,.
\]
as an additional constraint on the transfer matrix, where $\epsilon_{i_1i_2\ldots i_n}$ is the $\mathfrak{sl}_n$-invariant totally antisymmetric tensor. In particular, choosing $j_k=k$, this states that
\begin{equation} 
\label{eq:qdet}
	\epsilon_{i_1i_2\dots i_n}T^{i_1}_{\ 1,V}(z)T^{i_2}_{\ 2,V}(z+\hbar)\cdots 
	T^{i_n}_{\ n,V}(z+(n-1)\hbar) = {\bf 1}_V\,.
\end{equation}
The object on the left is known as the {\it quantum determinant}, $\text{qdet}(T)(z)$ of the transfer matrix. Together with the $RTT$ relations~\eqref{eq:RTT}, the condition $\text{qdet}(T)={\bf 1}_V$ constrains the \smash{$\hat{t}^{(m)}_V$} to furnish a representation of the Yangian $\cY(\fg)$. The role of the quantum determinant condition is essentially to remove the centre of the algebra defined solely by the $RTT$ relations~\cite{molev1996yangians}.

\medskip

In the gauge theory, the quantum determinant arises~\cite{Costello:2018gyb} from a vertex which joins together $n$ Wilson lines in the defining representation $U\cong\C^n$ of $\mathfrak{sl}_n(\C)$. So as to meet at the vertex, each of these Wilson lines must be supported at the same value $z\in\C$ of the spectral parameter. The vertex is
\[ 
	\epsilon_{i_1i_2\dots i_n}\cW_U[\gamma_1]^{i_1}_{\ j_1}\cW_U[\gamma_2]^{i_2}_{\ j_2}\cdots
	\cW_U[\gamma_n]^{i_n}_{\ j_n}
\]
and can be represented by the picture
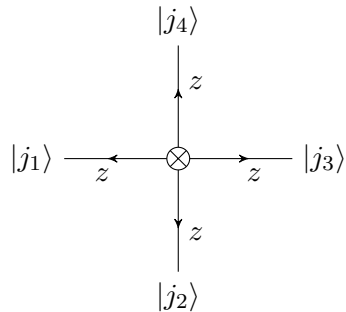
\begin{figure}[th]
	\centering
	\begin{tikzpicture}[baseline]
	\begin{feynman}
	\vertex(c);
	\vertex[above=1.5cm of c] (n);
	\vertex[right=1.5cm of c] (e);
	\vertex[below=1.5cm of c] (s);
	\vertex[left=1.5cm of c] (w);
	\vertex[above=0.67cm of c] (cn);
	\vertex[right=0.67cm of c] (ce);
	\vertex[below=0.67cm of c] (cs);
	\vertex[left=0.67cm of c] (cw);
	\vertex[above=1cm of c] (cnn);
	\vertex[right=1cm of c] (cee);
	\vertex[below=1cm of c] (css);
	\vertex[left=1cm of c] (cww);
	\diagram*{{c [crossed dot] -- [fermion] (n), (c) -- [fermion] (e), (c) -- [fermion] (s), (c) -- [fermion] (w)},};
	\vertex[left=0cm of w]{$|j_1\rangle$};
	\vertex[below=0cm of s]{$|j_2\rangle$};
	\vertex[right=0cm of e]{$|j_3\rangle$};
	\vertex[above=0cm of n]{$|j_4\rangle$};
	\vertex[below=0cm of cww]{$z$};
	\vertex[right=0cm of css]{$z$};
	\vertex[below=0cm of cee]{$z$};
	\vertex[right=0cm of cnn]{$z$};
	\end{feynman}
	\end{tikzpicture}
\end{figure}

\vspace{-0.3cm}\noindent here drawn for the case $n=4$.  More generally, whenever the tensor product of some representations $V_i$ of $\fg$ contains a copy of the trivial representation, Wilson lines in these representations can accumulate at a vertex which extracts the invariant part of $\bigotimes V_i$. In this way, gauge invariant networks can be built up from combinations of vertices and Wilson lines.

In the quantum theory, vertices suffer from anomalies analogous to the framing anomaly for curved Wilson lines. These anomalies can be made to vanish either by fixing the angles between the incoming Wilson lines, or by changing their spectral parameters at order $\hbar$. For totally antisymmetric or symmetric invariants, such as the example of $\epsilon\in\wedge^n \C^n$ for $\mathfrak{sl}_n(\C)$, it's enough to ensure that the angles between the incoming Wilson lines are the same. Our picture above reflects this condition. The conditions for more general invariants can be found in~\cite{Costello:2018gyb}.

Using this vertex, the quantum determinant condition can also be represented pictorially. For example,  in the case $\fg=\mathfrak{sl}_4$, the picture corresponding to~\eqref{eq:qdet} is
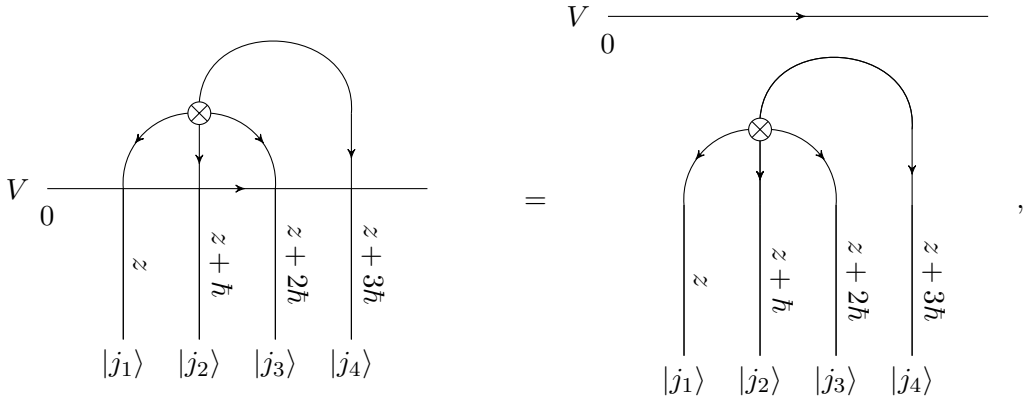
\begin{figure}[h!]
	\centering
	\begin{minipage}{.4\textwidth}
		\begin{tikzpicture}[baseline]
		\begin{feynman}
		\vertex(lc);
		\vertex[right=2cm of lc] (le);
		\vertex[below=1cm of lc] (ls);
		\vertex[left=1cm of ls] (lsw);
		\vertex[right=1cm of ls] (lse);
		\vertex[right=2cm of ls] (lese);
		\vertex[left=1cm of lsw] (lri);
		\vertex[right=1cm of lese] (lrf);
		\vertex[below=2cm of lsw] (lf1);
		\vertex[above=0.25cm of lf1] (lz1);
		\vertex[below=2cm of ls] (lf2);
		\vertex[above=0.25cm of lf2] (lz2);
		\vertex[below=2cm of lse] (lf3);
		\vertex[above=0.25cm of lf3] (lz3);
		\vertex[below=2cm of lese] (lf4);
		\vertex[above=0.25cm of lf4] (lz4);
		\diagram*{lc [crossed dot] -- [fermion, quarter left] (lse), (lc) -- [fermion] (ls), (lc) -- [fermion, quarter right] (lsw), (lc) -- [half left] (le), (le) -- [fermion] (lese), (lri) -- [fermion] (lrf), (lsw) -- (lf1), (ls) -- (lf2), (lse) -- (lf3), (lese) -- (lf4)};
		\vertex[below=0cm of lf1]{$|j_1\rangle$};
		\vertex[below=0cm of lf2]{$|j_2\rangle$};
		\vertex[below=0cm of lf3]{$|j_3\rangle$};
		\vertex[below=0cm of lf4]{$|j_4\rangle$};
		\draw (lsw) -- (lf1) node [midway, above, sloped] (TextNode) {$z$};
		\draw (ls) -- (lf2) node [midway, above, sloped] (TextNode) {$z+\hbar$};
		\draw (lse) -- (lf3) node [midway, above, sloped] (TextNode) {$z+2\hbar$};
		\draw (lese) -- (lf4) node [midway, above, sloped] (TextNode) {$z+3\hbar$};
		\vertex[left=0.1cm of lri]{$V$};
		\vertex[below=0.1cm of lri]{$0$};
		\vertex[right=0.25cm of lri](lt);
		\end{feynman}
		\end{tikzpicture}
	\end{minipage}
	$\qquad=$
	\begin{minipage}{.4\textwidth}
		\begin{tikzpicture}[baseline]
		\begin{feynman}
		\vertex(rc);
		\vertex[right=2cm of rc] (re);
		\vertex[below=1cm of rc] (rs);
		\vertex[left=1cm of rs] (rsw);
		\vertex[right=1cm of rs] (rse);
		\vertex[right=2cm of rs] (rese);
		\vertex[left=1cm of rsw] (rrit);
		\vertex[right=1cm of rese] (rrft);
		\vertex[above=2.5cm of rrit] (rri);
		\vertex[above=2.5cm of rrft] (rrf);
		\vertex[below=2cm of rsw] (rf1);
		\vertex[below=2cm of rs] (rf2);
		\vertex[below=2cm of rse] (rf3);	
		\vertex[below=2cm of rese] (rf4);
		\vertex[above=0.25cm of rf1] (rz1);
		\vertex[above=0.25cm of rf2] (rz2);
		\vertex[above=0.25cm of rf3] (rz3);
		\vertex[above=0.25cm of rf4] (rz4);
		\diagram*{rc [crossed dot] -- [half left] (re), (rc) -- [fermion] (rs), (rc) -- [fermion, quarter left] (rse), (rc) -- [fermion] (rs), (rc) -- [fermion, quarter right] (rsw), (rc) -- [half left] (re), (re) -- [fermion] (rese), (rri) -- [fermion] (rrf), (rsw) -- (rf1), (rs) -- (rf2), (rse) -- (rf3), (rese) -- (rf4)};
		\vertex[below=0cm of rf1]{$|j_1\rangle$};
		\vertex[below=0cm of rf2]{$|j_2\rangle$};
		\vertex[below=0cm of rf3]{$|j_3\rangle$};
		\vertex[below=0cm of rf4]{$|j_4\rangle$};
		\draw (rsw) -- (rf1) node [midway, above, sloped] (TextNode) {$z$};
		\draw (rs) -- (rf2) node [midway, above, sloped] (TextNode) {$z+\hbar$};
		\draw (rse) -- (rf3) node [midway, above, sloped] (TextNode) {$z+2\hbar$};
		\draw (rese) -- (rf4) node [midway, above, sloped] (TextNode) {$z+3\hbar$};
		\vertex[left=0.1cm of rri]{$V$};
		\vertex[below=0.1cm of rri]{$0$};
		\vertex[right=0.25cm of rri](rt);
		\end{feynman}
		\end{tikzpicture}
	\end{minipage},
\end{figure}\\

\vspace{-0.2cm}\noindent where the spectral parameter of the vertex is $z+\hbar$ in each case. 
Note the important role of the framing anomaly in supplying the different arguments of the $T$s in the quantum determinant~\eqref{eq:qdet}: this follows from the fact that all the $U$ lines are parallel as they head out to infinity, and so have to curve so as to make equal angles when they meet at the vertex.

Once again, the virtue of the gauge theory is that the quantum expectation values of the networks of line operators shown on each side of the picture agree by construction. Since the horizontal line operator is at a different location $0\in\C$ than any of the other line operators or the vertex, there are no singularities in any Feynman diagram as one moves between the two configurations. This constitutes a proof that line operators in the quantum theory correspond to representations of the Yangian. The proof is valid to all orders in $\hbar$.

If we choose $V$ to also be a copy of the defining vector representation $U$, the quantum determinant condition imposes a constraint on the simplest rational $R$-matrix. In~\cite{Costello:2018gyb} it was demonstrated that this constraint, together with the YBE, uniquely determines the $R$-matrix to all orders in $\hbar$, given the first-order term. This condition fixes the normalization of the simplest bulk $R$-matrix used in~\cite{Costello:2018gyb} and in the present paper.

\medskip

The $RTT$ presentation is not unique to $\fg=\mathfrak{sl}_n$, or even classical $\fg$~\cite{ChariPressleyBook}. As long as $\fg$ admits a non-trivial irreducible representation lifting to the Yangian, we can perform this construction. The analogue of $\text{qdet}(T)$ is constructed using the invariant tensors of this representation. Together with the $RTT$ relations, the analogous quantum determinant condition defines the Yangian associated to $\fg$. The only choice of $\fg$ for which this construction fails is $\mathfrak{e}_8$, essentially because at the time of writing no suitable representation has been found~\cite{Costello:2018gyb}.

\subsection{The $RTT$ Presentation of Twisted Yangians}
\label{sec:BoundaryRTT}

In section~\ref{sec:TwistedYangians} we showed that, to order $\hbar$, the OPE between a boundary line operator supported on $L$ and a bulk line operator supported at $z=0$ corresponds to the coproduct \smash{$\Delta_\hbar : \cB(\fg,\fh)\to\cY(\fg)\otimes\cB(\fg,\fh)$.} We now use the $RTT$ presentation of $\cB(\fg,\fh)$ to prove that, to all orders in $\hbar$, boundary line operators in the quantum gauge theory are labelled by representations of the twisted Yangian.

The $RTT$ presentation of the twisted Yangian~\cite{molev2002representations,guay2016twisted} starts from a $K$-matrix 
\[
	K(z) : U\otimes W \to U\otimes W
\]
where again $U\cong\C^n$ is the defining representation of $\fg$. The boundary analogue of the transfer matrix is $B^{i}_{\ j,W}(z)\in\text{End}(W)$, obtained by extracting the $(i,j)^{\rm th}$ components of $K$:
\begin{figure}[th]
	\centering
	\begin{minipage}{0.5\textwidth}
	\begin{tikzpicture}[baseline]
	\begin{feynman}
	\vertex (nw);
	\vertex[below=2cm of nw](sw);
	\vertex[below=1cm of nw](w);
	\vertex[right=1.7cm of w](e);
	\vertex[above=1cm of e](ne);
	\vertex[above=1cm of e](u);
	\vertex[below=1cm of e](se);
	\vertex[above = 0.5cm of se](ese);
	\vertex[above = 0.25cm of se](wz);
	\vertex[above=0.25cm of sw](zit);
	\vertex[below=0.25cm of nw](zft);
	\vertex[right=0.43 of zit](zi);
	\vertex[right=0.43 of zft](zf);
	\vertex[above=0.5cm of sw](ut);
	\vertex[right =0.85cm of ut](u);
	\vertex[left=0cm of sw]{$\langle i|$};
	\vertex[left=0cm of nw]{$|j\rangle$};
	\vertex[right=0.2cm of wz]{$0$};
	\vertex[below=0.2cm of wz]{$W$};
	\vertex[below=0.1cm of zf]{$-z$};
	\vertex[below=0.2cm of u]{$z$};
	\vertex[right=0cm of e]{$\tau$};
	\vertex[left=1.5cm of w] {$B^i_{\ j,W}(z) \qquad = $};
	\diagram*{{[edges=fermion] (sw) -- (e), (e) -- (nw), (sw) -- (e)}, {[edges = charged scalar] (se) -- (e), (e) -- (ne)}};
	\end{feynman}
	\end{tikzpicture}
	\end{minipage}\,.
\end{figure}

\noindent 
As with the transfer matrix, in the rational case $B_W(z)$ has a power series expansion in $1/z$ 
\[ 
	B^i_{\ j,W}(z) = \tau^i_{\ j} {\bf 1}_W 
	+ \hbar \sum_{n=0}^{\infty} \frac{(\hat{b}^{(m)})^{i}_{\ j,W}}{z^{n+1}}\,,
\]
with coefficients $\hat{b}^{(m)}$. As in the bulk, in gauge theory, the boundary transfer matrix is the quantum expectation value of a Wilson line in the $U$ representation on $\widetilde{M}$, in the presence of a line operator on $L$ and where we extract the $(i,j)^{\rm th}$ component of the $U$-Wilson line after taking expectation value. Again, at lowest order, the generators $\hat{b}^{(m)}$ agree with the generators $b^{(2m)}_{\alpha,W}$ and $b^{(2m+1)}_{\mu,W}$ appearing in the classical boundary line operator, but the $\hat{b}^{(m)}$ receive further quantum corrections from the expectation value.

\medskip

As in the bulk, constraints on the $B$ ensure that the generators $\hat{b}^{(m)}$ furnish a representation of the twisted Yangian $\cB(\fg,\fh)$ acting on $W$. The twisted Yangian of $\fg=\mathfrak{sl}_n(\C)$ is defined by three relations~\cite{molev2002representations}: a boundary $RTT$ relation, a boundary analogue of the quantum determinant condition known as the Sklyanin determinant and an additional relation known as boundary unitarity.

The analogue of the $RTT$ relations for the boundary transfer matrix $B$ is
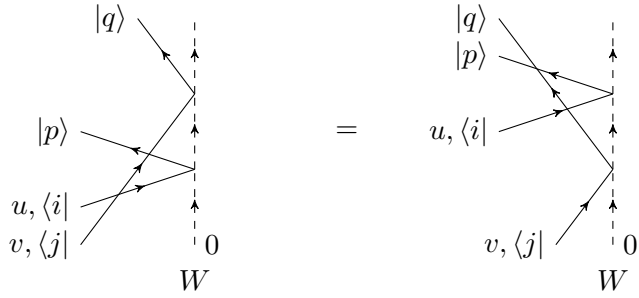
\begin{figure}[th]
	\centering
	\begin{tikzpicture}[baseline]
	\begin{feynman}
	\vertex (o){$=$};
	\vertex[left=2cm of o](l);
	\vertex[above=0.5cm of l](lb);
	\vertex[below=0.5cm of l](la);
	\vertex[above=1.5cm of l](ln);
	\vertex[above=0.5cm of l](luc);
	\vertex[below=0.5cm of l](llc);
	\vertex[below=1.5cm of l](ls);
	\vertex[left=1.5cm of ls](lsw);
	\vertex[left=1.5cm of l](lw);
	\vertex[below=1cm of lw](lai);
	\vertex[left=0.75cm of ln](lbf);
	\vertex[right=3.5cm of o](r);
	\vertex[above=0.5cm of r](ra);
	\vertex[below=0.5cm of r](rb);
	\vertex[above=1.5cm of r](rn);
	\vertex[above=0.5cm of r](ruc);
	\vertex[below=0.5cm of r](rlc);	
	\vertex[below=1.5cm of r](rs);
	\vertex[left=1.5cm of rn](rnw);
	\vertex[left=1.5cm of r](rw);
	\vertex[above=1cm of rw](raf);
	\vertex[left=0.75cm of rs](rbi);
	\diagram*{{[edges=fermion] (lai) --  (la), (la) --  
			(lw), (lsw) -- (lb), (lb) --  (lbf), (rw) -- (ra), (ra) -- (raf), (rbi) -- (rb), (rb) -- (rnw)},{[edges=charged scalar] (ls) -- (llc), (llc) -- (luc), (luc) -- (ln), (rs)--(rlc), (rlc) -- (ruc), (ruc) -- (rn)}};
	\vertex[left=0cm of lsw]{$v,\la j|$};
	\vertex[left=0cm of lai]{$u,\la i|$};
	\vertex[left=0cm of lw]{$|p\ra$};
	\vertex[left=0cm of lbf]{$|q\ra$};
	\vertex[left=0cm of rbi]{$v,\la j|$};
	\vertex[left=0cm of rw]{$u,\la i|$};
	\vertex[left=0cm of raf]{$|p\ra$};
	\vertex[left=0cm of rnw]{$|q\ra$};
	\vertex[right=0cm of ls]{$0$};
	\vertex[below=0.2cm of ls]{$W$};
	\vertex[right=0cm of rs]{$0$};
	\vertex[below=0.2cm of rs]{$W$};
	\end{feynman}
	\end{tikzpicture}
\end{figure}\\

\vspace{-0.3cm}\noindent which just states that the boundary transfer matrix $B$ must be obtained from a $K$-matrix that solves the boundary Yang-Baxter equation. Algebraically, this condition is
\begin{equation} 
	R^{ij}_{\ \ k\ell}(u-v)B^k_{\ m,W}(u)R^{\ell m}_{\ \ np}(u+v)B^n_{\ q,W}(v) 
	= B^j_{\ \ell,W}(v)R^{i\ell}_{\ \ kn}(u+v)B^k_{m,W}(u)R^{nm}_{\ \ qp}(u-v)
\label{eq:RBRB} 
\end{equation}
and is sometimes referred to as the {\it quaternary relation}. As in the bulk, it follows from the construction that any $B$ matrix obtained from the quantum expectation value of line operators in the gauge theory will obey this equation.

The second condition we require of the boundary transfer matrix is
\begin{equation}
\label{eq:BoundaryUnitarity} 
B^i_{\ j,W}(z)B^j_{\ k,W}(-z) = \pm\delta^i_{\ j}{\bf 1}_W\,,
\end{equation}
known as {\it boundary unitarity}\footnote{An analogous relation called {\it bulk unitarity} can be imposed on the transfer matrix $T(z)$, but places no constraints and serves only to give a diagrammatic representation to $T^{-1}(z)$. By contrast, boundary unitarity is essential in defining the twisted Yangian in the $RTT$ presentation.}. Much like the quantum determinant condition in the bulk this condition simply removes central elements from the algebra defined by~\eqref{eq:RBRB}~\cite{molev2002representations,guay2016twisted}. To understand the origin of this relation from gauge theory, consider the following sequence of diagrams
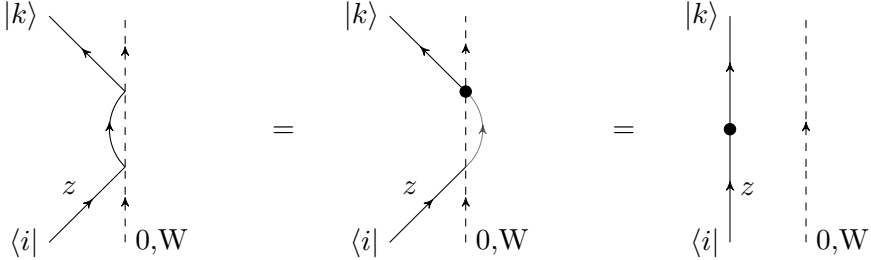
\begin{figure}[ht]
	\centering
	\begin{minipage}{.3\textwidth}
		\begin{tikzpicture}[baseline]
		\begin{feynman}
		\vertex(lb);
		\vertex[above=1.5cm of lb](lbf);
		\vertex[below=1.5cm of lb](lbi);
		\vertex[left=1cm of lbi](lsw);
		\vertex[left=1cm of lbf](lnw);
		\vertex[above=0.5cm of lb](lb2);
		\vertex[below=0.5cm of lb](lb1);
		\vertex[above=0.5cm of lbi](lW);
		\vertex[above=0.25cm of lbi](l0);
		\vertex[left=0.67cm of lW](lU);
		\vertex[left=1cm of l0](lz);
		\diagram*{(lbi) -- [charged scalar,
		] (lb1), (lb1) -- [scalar] (lb2), (lb2) -- [charged scalar] (lbf), (lsw) -- [fermion, edge label=$z$] (lb1), (lb2) -- [fermion] (lnw), (lb1) -- [fermion, quarter left] (lb2)};
		\vertex[left=0cm of lsw]{$\la i|$};
		\vertex[left=0cm of lnw]{$|k \ra$};
		\vertex[right=0cm of lbi]{0,W};
		\end{feynman}
		\end{tikzpicture}
	\end{minipage}
$\hspace{-1cm}=\quad$
	\begin{minipage}{.3\textwidth}
		\begin{tikzpicture}[baseline]
		\begin{feynman}
		\vertex(lb2);
		\vertex[below=0.5cm of lb2](lb);
		\vertex[above=1.5cm of lb](lbf);
		\vertex[below=1.5cm of lb](lbi);
		\vertex[left=1cm of lbi](lsw);
		\vertex[left=1cm of lbf](lnw);
		\vertex[below=0.5cm of lb](lb1);
		\vertex[above=0.5cm of lbi](lW);
		\vertex[above=0.25cm of lbi](l0);
		\vertex[left=0.67cm of lW](lU);
		\vertex[left=1cm of l0](lz);
		\diagram*{(lbi) -- [charged scalar] (lb1), (lb1) -- [scalar] lb2 [dot], (lb2) -- [charged scalar] (lbf), (lsw) -- [fermion, edge label=$z$] (lb1), (lb2) -- [fermion] (lnw), (lb1) -- [fermion, quarter right, opacity=0.5] (lb2)};
		\vertex[left=0cm of lsw]{$\la i|$};
		\vertex[left=0cm of lnw]{$|k \ra$};
		\vertex[right=0cm of lbi]{0,W};
		\end{feynman}
		\end{tikzpicture}
	\end{minipage}
$\hspace{-1cm}=\quad$
	\begin{minipage}{.3\textwidth}
		\begin{tikzpicture}[baseline]
		\begin{feynman}
		\vertex(rw);
		\vertex[below=1.5cm of rw](rsw);
		\vertex[above=1.5cm of rw](rnw);
		\vertex[right=1cm of rnw](rbf);
		\vertex[right=1cm of rsw](rbi);
		\vertex[above=1cm of rbi](rbi1);
		\vertex[above=1cm of rbi1](rbi2);
		\vertex[above=0.5cm of rbi](rW);
		\vertex[above=0.25cm of rbi](r0);
		\vertex[above=0.755cm of rsw](rU);
		\vertex[above=0.25cm of rsw](rz);
		\diagram*{(rsw) -- [fermion, edge label'=$z$] rw [dot] -- [fermion] (rnw), (rbi) -- [scalar] (rbi1), (rbi1) -- [charged scalar] (rbi2), (rbi2) -- [scalar] (rbf)};
		\vertex[left=0cm of rsw]{$\langle i|$};
		\vertex[left=0cm of rnw]{$|k \rangle$};
		\vertex[right=0cm of rbi]{0,W};
		\end{feynman}
		\end{tikzpicture}
	\end{minipage}
\end{figure}\\
\noindent where the dots in the right hand diagrams indicate an insertion of $\tau^2_U$ on the Wilson line. This corresponds to the following sequence of manipulations in the gauge theory
\[ 
	\la\cW[\gamma_i]\tau_U\cW[\gamma_m]\tau_U\cW[\gamma_r]\otimes\cW_b\ra 
	= \la\cW[\gamma_i]\cW[\cP(\gamma_m)]\tau^2_U\cW[\gamma_r]\otimes\cW_b\ra 
	= \la\cW[\gamma'_i]\tau^2_U\cW[\gamma'_r]\otimes\cW_b\ra\,.
\]
Here $\gamma_i$, $\gamma_m$, and $\gamma_r$ are the initial, middle, and final segments of the solid curve in the left hand picture, whereas $\gamma_i'$ and $\gamma_r'$ are the initial and final segments of the solid curve in the right hand picture. All of these curves are contained in $\overline{M}_{\rm L}$. In the first equality we've used equation~\eqref{InnerWilsonLineReflection} to express the central Wilson line in terms of its image under $\cP$, and in the second equality we've performed a diffeomorphism on $\widetilde M$ to pull the whole configuration into $\overline{M}_{\rm L}$. Since $\tau^2\in Z(G)$, for $\fg=\mathfrak{sl}_n$ we have $\tau^2_U=\pm{\bf 1}_U$, giving the relation~\eqref{eq:BoundaryUnitarity}.

The final condition we require (when $\fg=\mathfrak{sl}_n$) is the {\it Sklyanin determinant condition}. This is the boundary analogue of the quantum determinant condition on $T$, and mixes the boundary transfer matrix with the bulk $R$ matrix. It can be expressed as
\begin{equation}
\begin{aligned}
\label{eq:SklyaninDet}
	&\epsilon\, B_{1,W}(z)R_{21}(2z+\hbar)R_{31}(2z+2\hbar)\cdots R_{n1}(2z+(n-1)\hbar)\, 
	B_{2,W}(z+\hbar)R_{32}(2z+3\hbar)\cdots \\ 
	&\qquad \cdots R_{n2}(z+n\hbar)B_{3,W}(z+2\hbar)\cdots \,B_{n,W}(z+(n-1)\hbar)(z) 
	= \epsilon\,{\bf 1}_W\,,
\end{aligned}
\end{equation}
where the numerical suffixes on $B$ and $R$ indicate the particular copy of $U$ in the tensor product $U^{\otimes n}$ on which they act. The totally anti-symmetric $\epsilon$-tensor on the left contracts with the free index on $B_1$, together with the free indices on the $R_{j1}$ appearing immediately to the right of it. The Sklyanin determinant itself, sdet$(B)(z)$, is the contraction of the left hand side of~\eqref{eq:SklyaninDet} with the inverse of $\epsilon$. Thus this equation can equally be written as sdet$(B)={\bf 1}_W$. This condition removes the remaining central elements from the algebra defined by equations~\eqref{eq:RBRB} and~\eqref{eq:BoundaryUnitarity}~\cite{molev2002representations}\footnote{In that paper a different normalization was used for the Sklyanin determinant. This was chosen to be consistent with a particular choice of embedding of the twisted Yangian into the Yangian. From the point of gauge theory this embedding is unnatural. This discrepancy in normalization does not affect the applicability of their results in this context.}.

Much like qdet$(T)$, the Sklyanin determinant condition can be realised in the gauge theory using the $\epsilon$-vertex to join $n$ Wilson lines in the $U\cong\C^n$ representation of $\fg$, each supported at the same value of the spectral parameter. The required sequence of pictures is\\ 
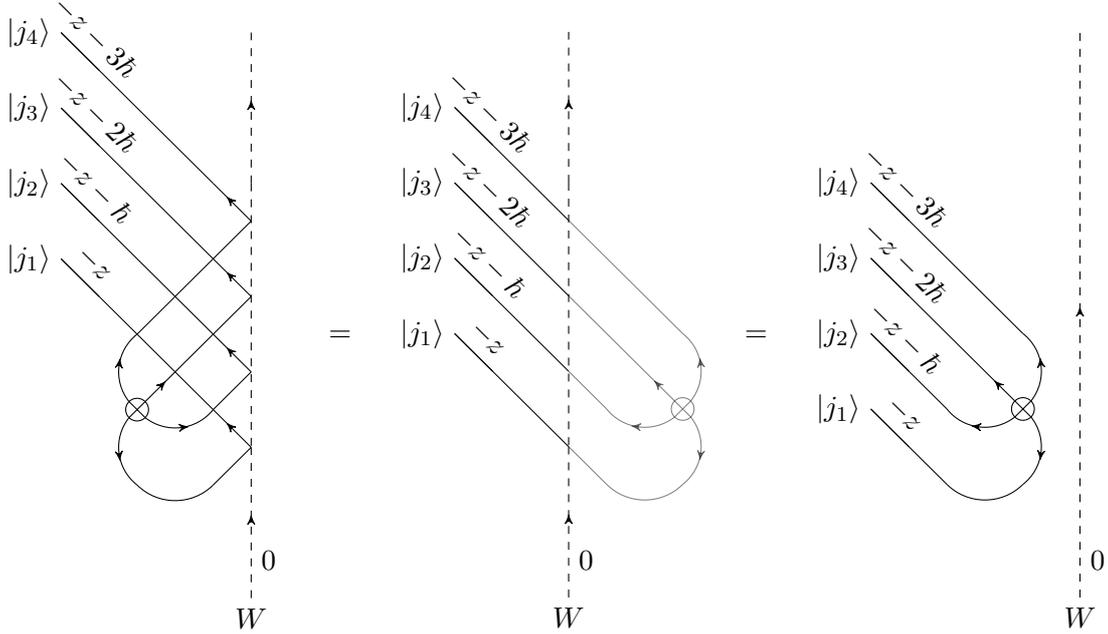
\begin{figure}[ht]
	\centering
	\begin{minipage}{.3\textwidth}
	\vspace{-0.7cm}
	\begin{tikzpicture}[baseline]
	\begin{feynman}
	\vertex(w);
	\vertex[right=1.5cm of w](e);
	\vertex[left=0.5cm of e](c);
	\vertex[below=1cm of c](n);
	\vertex[below=1cm of w](nw);
	\vertex[below=0.5cm of e](ne);
	\vertex[above=0.5cm of e](se);
	\vertex[left=1cm of se](csw);
	\vertex[above=1cm of c](s);
	\vertex[above=1cm of se](sse);
	\vertex[above=1cm of w](sw);
	\vertex[above=1cm of s](ss);
	\vertex[above=1cm of sse](ssse);
	\vertex[above=1cm of ss](sss);
	\vertex[left=1cm of sw](a);
	\vertex[above=1cm of a](t1);
	\vertex[above=2cm of a](t2);
	\vertex[above=3cm of a](t3);
	\vertex[above=4cm of a](t4);
	\vertex[left=0.5cm of sw](b);
	\vertex[above=0.5cm of b](m1);
	\vertex[above=1.5cm of b](m2);
	\vertex[above=2.5cm of b](m3);
	\vertex[above=3.5cm of b](m4);
	\vertex[below=2.5 cm of e](bf);
	\vertex[above=5 cm of e](bi);
	\vertex[above=0.5cm of ssse](bm);
	\vertex[above=0.5cm of bf](bz);
	\vertex[above=1cm of bf](bw);
	\diagram*{w [crossed dot] -- [quarter right, fermion] (nw), (nw) -- [quarter right] (n), (n) -- (ne), (w) -- [quarter right, fermion] (c), (c) -- (se), (w) -- [fermion] (csw) -- (sse), (w) -- [quarter left, fermion] (sw), (sw) -- (ss), (ss) -- (ssse), (ne) -- [fermion] (c), (se) -- [fermion] (s), (sse) -- [fermion] (ss), (ssse) -- [fermion] (sss), (c) -- (t1), (s) -- (t2), (ss) -- (t3), (sss) -- (t4), (bm) -- [charged scalar] (bi), (bm) -- [scalar] (ne), (bf) -- [charged scalar] (ne)};
	\vertex[left=0cm of t1]{$|j_1\rangle$};
	\vertex[left=0cm of t2]{$|j_2\rangle$};
	\vertex[left=0cm of t3]{$|j_3\rangle$};
	\vertex[left=0cm of t4]{$|j_4\rangle$};
		\draw (m1) -- (t1) node [midway, above, sloped] (TextNode) {$-z$};
		\draw (m2) -- (t2) node [midway, above, sloped] (TextNode) {$-z-\hbar$};
		\draw (m3) -- (t3) node [midway, above, sloped] (TextNode) {$-z-2\hbar$};
		\draw (m4) -- (t4) node [midway, above, sloped] (TextNode) {$-z-3\hbar$};
		\vertex[right=0cm of bz]{$0$};
	\vertex[below=0.5cm of bz]{$W$};
	\end{feynman}
	\end{tikzpicture}
\end{minipage}
$\hspace{-0.3cm}=\quad$
	\begin{minipage}{.3\textwidth}
	\begin{tikzpicture}[baseline]
	\begin{feynman}
	\vertex(w);
	\vertex[left=1.5cm of w](e);
	\vertex[right=0.5cm of e](c);
	\vertex[below=1cm of c](n);
	\vertex[below=1cm of w](nw);
	\vertex[below=0.5cm of e](ne);
	\vertex[above=0.5cm of e](se);
	\vertex[right=1cm of se](csw);
	\vertex[above=1cm of c](s);
	\vertex[above=1cm of se](sse);
	\vertex[above=1cm of w](sw);
	\vertex[above=1cm of s](ss);
	\vertex[above=1cm of sse](ssse);
		\vertex[above=1cm of ss](sss);
	\vertex[left=3cm of sw](a);
	\vertex[above=1cm of a](t1);
	\vertex[above=2cm of a](t2);
	\vertex[above=3cm of a](t3);
	\vertex[left=2.5cm of sw](b);
	\vertex[below=0.5cm of b](m0);
	\vertex[above=0.5cm of b](m1);
	\vertex[above=1.5cm of b](m2);
	\vertex[above=2.5cm of b](m3);
 	\vertex[below=2.5 cm of e](bf);
	\vertex[above=5 cm of e](bi);
	\vertex[above=0.5cm of ssse](bm);
	\vertex[above=0.5cm of bf](bz);
	\vertex[above=1cm of bf](bw);
	\diagram*{w [opacity=0.5, crossed dot] -- [quarter left, opacity=0.5, fermion] (nw), (nw) -- [opacity=0.5, quarter left] (n), (n) -- [opacity=0.5] (ne), (w) -- [quarter left, opacity=0.5, fermion] (c), (c) -- [opacity=0.5] (se),(w) -- [opacity=0.5, fermion] (csw) -- [opacity=0.5] (sse), (w) -- [quarter right, opacity=0.5, fermion] (sw), (sw) --[opacity=0.5] (ssse), (ne) -- (a), (se) -- (t1), (sse) -- (t2), (ssse) -- (t3), (bm) -- [charged scalar] (bi), (bm) -- [scalar] (ne), (bf) -- [charged scalar] (ne)};
	\vertex[left=0cm of a]{$|j_1\rangle$};
	\vertex[left=0cm of t1]{$|j_2\rangle$};
	\vertex[left=0cm of t2]{$|j_3\rangle$};
	\vertex[left=0cm of t3]{$|j_4\rangle$};
		\draw (m0) -- (a) node [midway, above, sloped] (TextNode) {$-z$};
		\draw (m1) -- (t1) node [midway, above, sloped] (TextNode) {$-z-\hbar$};
		\draw (m2) -- (t2) node [midway, above, sloped] (TextNode) {$-z-2\hbar$};
		\draw (m3) -- (t3) node [midway, above, sloped] (TextNode) {$-z-3\hbar$};
		\vertex[right=0cm of bz]{$0$};
	\vertex[below=0.5cm of bz]{$W$};
	\end{feynman}
	\end{tikzpicture}
\end{minipage}
$=\quad$
\begin{minipage}{.25\textwidth}
	\begin{tikzpicture}[baseline]
	\begin{feynman}
	\vertex(c);
	\vertex[below=1cm of c](n);
	\vertex[left=1cm of c](w);
	\vertex[above=1cm of c](s);
	\vertex[left=0.5cm of c](cw);
	\vertex[above=0.5cm of cw](sw);
	\vertex[left=1cm of n](nw);
	\vertex[left=1cm of w](t1);
	\vertex[above=1cm of t1](t2);
	\vertex[above=2cm of t1](t3);
	\vertex[above=3cm of t1](t4);
	\vertex[right=0.75cm of c](e);
	\vertex[below=2.5cm of e](bf);
	\vertex[above=5cm of e](bi);
	\vertex[above=0.5cm of bf](bz);
	\vertex[above=1cm of bf](bw);
	\vertex[left=1cm of sw](m2);
	\vertex[below=1cm of m2](m1);
	\vertex[above=1cm of m2](m3);
	\vertex[above=2cm of m2](m4);
	\diagram*{c [crossed dot] -- [quarter left, fermion] (n), (c) -- [quarter right, fermion] (s), (c) -- [quarter left, fermion] (w), (c) -- [fermion] (sw), (n) -- [quarter left] (nw), (nw) -- (t1), (w) -- (t2), (sw) -- (t3), (s) -- (t4), (bf) -- [charged scalar] (bi)};
	\vertex[right=0cm of bz]{$0$};
	\vertex[below=0.5cm of bz]{$W$};
	\vertex[left=0cm of t1]{$|j_1\rangle$};
	\vertex[left=0cm of t2]{$|j_2\rangle$};
	\vertex[left=0cm of t3]{$|j_3\rangle$};
	\vertex[left=0cm of t4]{$|j_4\rangle$};
		\draw (m1) -- (t1) node [midway, above, sloped] (TextNode) {$-z$};
		\draw (m2) -- (t2) node [midway, above, sloped] (TextNode) {$-z-\hbar$};
		\draw (m3) -- (t3) node [midway, above, sloped] (TextNode) {$-z-2\hbar$};
		\draw (m4) -- (t4) node [midway, above, sloped] (TextNode) {$-z-3\hbar$};	
	\end{feynman}
	\end{tikzpicture}
	\end{minipage}
\end{figure}\\

\vspace{-0.3cm}\noindent illustrated here for the case $\fg=\mathfrak{sl}_4$. Note that the spectral parameter at the vertex takes the value $z+2\hbar$ in the left hand diagram, and $-z-2\hbar$ in the two subsequent diagrams. The framing anomaly then shifts the spectral parameters of the Wilson lines as they curve to head out to infinity in parallel (without further crossings). This is exactly what is required for the arguments of the $B$ and $R$ matrices appearing in the Sklyanin determinant~\eqref{eq:SklyaninDet}. Just like in the case of boundary unitarity it is essential to work on the orbifold to deduce the Sklyanin determinant condition, since in going from the second picture to the third we've acted with a diffeomorphism of $\widetilde M$ not fixing $\pi$. It's also important to note that in going from the first picture to the second we've used the fact that $\det\tau_U = 1$ to get rid of the $\tau_U^{-1}$s generated when applying~\eqref{InnerWilsonLineReflection}. Again, the virtue of the gauge theory is that the condition sdet$(B)={\bf 1}_W$ is obeyed by construction on the orbifold $\widetilde{M}$. 

Note that by choosing $W$ to be a representation of the twisted Yangian on $\C$, $B$ reduces to one of the quasi-classical $K$-matrices we found in section~\ref{sec:Kmx}. In this way the Sklyanin determinant condition imposes a constraint on those $K$-matrices. In section~\ref{subsec:unique} we will show that, given a semi-classical $k$-matrix, to all orders in $\hbar$ there is a unique $K$ matrix that obeys both this constraint and the bYBE.

\subsubsection{Symmetry Relations for Lie Algebras of type $B_n$, $C_n$, and $D_n$}
\label{sec:OtherLieAlgebras}

In \cite{molev2002representations} the relations~\eqref{eq:RBRB},~\eqref{eq:SklyaninDet} and~\eqref{eq:BoundaryUnitarity} were used to define the twisted Yangian $\cB(\fg,\fh)$ in the case that $\fg =\mathfrak{sl}_n(\C)$ and $\fh=\mathfrak{sl}_{n-k}(\C)\oplus\mathfrak{sl}_k(\C)\oplus\mathbb{C}$. (In fact the Drinfeld $J$-presentation of the twisted Yangian appearing in \cite{belliard2017drinfeld} is a relatively recent development.) Now let's see if we can apply this construction for other finite dimensional Lie algebras. Boundary unitarity and the boundary $RTT$ equation apply unchanged, so we just need to find the analogues of the Sklyanin determinant for these algebras.

We concentrate on the classical Lie algebras $B_n$, $C_n$ and $D_n$. Of these, $B_n$ and $D_n$ are isomorphic to $\mathfrak{so}_{2n+1}$ and $\mathfrak{so}_{2n}$ respectively, so their defining vector representations admit a symmetric invariant tensor, $\delta\in {\rm Sym}^2V^*$. Similarly $C_n$ is isomorphic to $\mathfrak{sp}_{2n}(\C)$,  so its defining representation admits an antisymmetric invariant tensor $\omega\in\wedge^2V^*$. Since these vertices are totally (anti)symmetric, they both quantize without anomalies provided the two incoming Wilson lines approach the vertex directly opposite one another. We will deal with both cases simultaneously, and refer to the invariant tensor as $\eta$. 

Guided by the construction of the Sklyanin determinant in terms of line operators, we consider the configurations 
\newpage
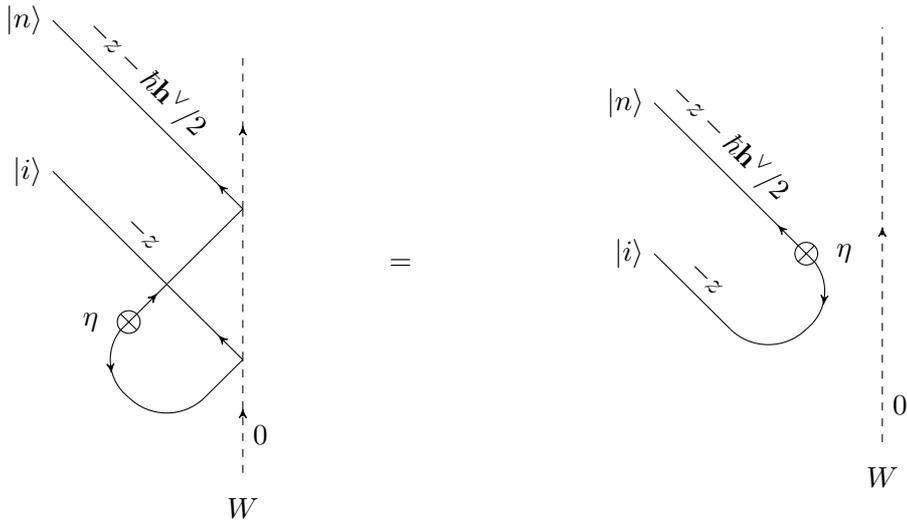
\begin{figure}[ht]
	\centering
	\begin{minipage}{.4\textwidth}
		\begin{tikzpicture}[baseline]
		\begin{feynman}
		\vertex(w);
		\vertex[right=1.5cm of w](e);
		\vertex[left=0.5cm of e](c);
		\vertex[below=1cm of c](n);
		\vertex[below=1cm of w](nw);
		\vertex[below=0.5cm of e](ne);
		\vertex[above=0.5cm of e](se);
		\vertex[left=1cm of se](csw);
		\vertex[above=1cm of c](s);
		\vertex[above=1cm of se](sse);
		\vertex[above=1cm of w](sw);
		\vertex[above=1cm of s](ss);
		\vertex[above=1cm of sse](ssse);
		\vertex[above=1cm of ss](sss);
		\vertex[left=1cm of sw](a);
		\vertex[above=1cm of a](t1);
		\vertex[above=2cm of a](t2);
		\vertex[above=3cm of a](t3);
		\vertex[above=4cm of a](t4);
		\vertex[left=0.5cm of sw](b);
		\vertex[below=2 cm of e](bf);
		\vertex[above=3.5 cm of e](bi);
		\vertex[above=0.5cm of ssse](bm);
		\vertex[above=0.5cm of bf](bz);
		\vertex[above=1cm of bf](bw);
		\diagram*{w [crossed dot] -- [quarter right, fermion] (nw), (nw) -- [quarter right] (n), (n) -- (ne), (w) -- [fermion] (csw) -- (sse), (ne) -- [fermion] (c), (sse) -- [fermion] (ss), 
			(sse) -- [charged scalar] (bi), (ne) -- [scalar] (sse), (bf) -- [charged scalar] (ne)};
		\draw (c) -- (t1) node [midway, above, sloped] (TextNode) {$-z$};
		\draw (ss) -- (t3) node [midway, above, sloped] (TextNode) {$-z-\hbar{\bf h}^\vee/2$};
		\vertex[left=0.5cm of w]{$\eta$};
		\vertex[left=0cm of t1]{$|i\rangle$};
		\vertex[left=0cm of t3]{$|n\rangle$};
		\vertex[right=0cm of bz]{$0$};
		\vertex[below=0.2cm of bf]{$W$};
		\end{feynman}
		\end{tikzpicture}
	\end{minipage}
	$\hspace{-1cm}=\qquad\qquad\qquad$
	\begin{minipage}{.4\textwidth}
		\begin{tikzpicture}[baseline]
		\begin{feynman}
		\vertex(c);
		\vertex[below=1cm of c](n);
		\vertex[left=1cm of c](w);
		\vertex[above=1cm of c](s);
		\vertex[left=0.5cm of c](cw);
		\vertex[above=0.5cm of cw](sw);
		\vertex[left=1cm of n](nw);
		\vertex[left=1cm of w](t1);
		\vertex[above=1cm of t1](t2);
		\vertex[above=2cm of t1](t3);
		\vertex[above=3cm of t1](t4);
		\vertex[right=1cm of c](e);
		\vertex[below=2.5cm of e](bf);
		\vertex[above=3cm of e](bi);
		\vertex[above=0.5cm of bf](bz);
		\vertex[above=1cm of bf](bw);
		\diagram*{c [crossed dot] -- [quarter left, fermion] (n), (c) -- [fermion] (sw), (n) -- [quarter left] (nw),  (bf) -- [charged scalar] (bi)};
		\draw (nw) -- (t1) node [midway, above, sloped] (TextNode) {$-z$};
		\draw (sw) -- (t3) node [midway, above, sloped] (TextNode) {$-z-\hbar{\bf h}^\vee/2$};
		\vertex[right=0.5cm of c]{$\eta$};
		\vertex[right=0cm of bz]{$0$};
		\vertex[below=0.2cm of bf]{$W$};
		\vertex[left=0cm of t1]{$|i\rangle$};
		\vertex[left=0cm of t3]{$|n\rangle$};
		\end{feynman}
		\end{tikzpicture}
	\end{minipage}
\end{figure}

\noindent from which we deduce that
\begin{equation} 
\label{BCDtensorRelation}
\eta_{in} = \eta_{k\ell}\,B^k_{\ j,W}(z)\,R^{\ell j}_{\ \ mi}(2z+\hbar{\bf h}^\vee/2)\,
B^m_{\ n,W}(z+\hbar{\bf h}^\vee/2)\,.
\end{equation}
In this form, this relation closely resembles the Sklyanin determinant condition for $\mathfrak{sl}_n$. We believe it is equivalent to the symmetry relations described in \cite{guay2016twisted} for twisted Yangians of $\fg=B_n$, $C_n$, and $D_n$ in the $RTT$ presentation. The condition \eqref{BCDtensorRelation} is natural from the point of view of the gauge theory.


\medskip

We now briefly turn to the exceptional Lie algebras. In~\cite{Costello:2018gyb} it was demonstrated that Yangians for all of the exceptional Lie algebras with the exception of $\mathfrak{e}_8$ can be constructed using gauge theory. This is done by finding an irreducible representation lifting to the Yangian, determining its invariant tensors, and then constructing the associated vertices. The Yangian can then be defined by the $RTT$ relations and the constrains coming from these vertices. By assuming analogues of the Sklyanin determinant condition for every vertex, together with the quaternary relations and boundary unitarity it is possible to define twisted Yangians for the exceptional Lie algebras.

\subsubsection{Coproducts in the $RTT$ Presentation}
\label{sec:BoundaryRTTCoprod}

A key property of transfer matrices is that concatenation of transfer matrices for representations $V_1$ and $V_2$ should give the transfer matrix for the tensor product:
\begin{equation}
\label{eq:TransferConcatenate}
	T^i_{\ j,V_1\otimes V_2}(z) = T^i_{\ k,V_1}(z)\otimes T^k_{\ j,V_2}(z)\,,
\end{equation}
or equivalently
\newpage
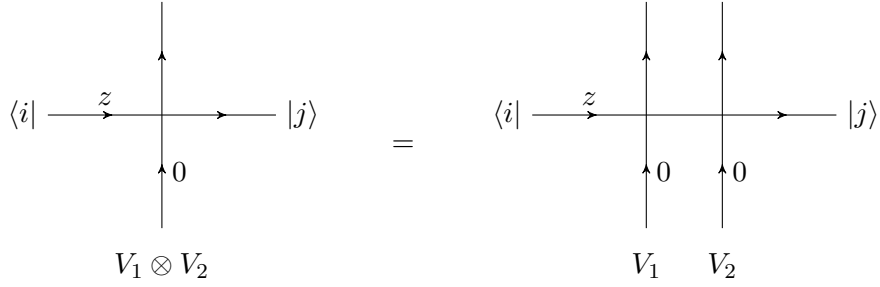
\begin{figure}[ht]
	\centering
	\begin{minipage}{.4\textwidth}
		\begin{tikzpicture}[baseline]
		\begin{feynman}
		\vertex (c);
		\vertex[above=1.5cm of c] (n);
		\vertex[below=1.5cm of c] (s);
		\vertex[left=1.5cm of c] (w);
		\vertex[right=1.5cm of c] (e);
		\vertex[below=0.67cm of c] (cs);
		\vertex[left=0.67cm of c] (cw);
		\vertex[left=1cm of c] (cww);
		\vertex[below=1cm of c] (css);
		\diagram*{{[edges=fermion] (s) --[edge label'=$0$]  (c), (c) --  (n), (w)
				--[edge label=$z$] (c), (c) -- (e)},};
		\vertex[below=0.2cm of s] {$V_1\otimes V_2$};
		\vertex[left=0cm of w] {$\langle i|$};
		\vertex[right=0cm of e] {$|j\rangle$};
		\end{feynman}
		\end{tikzpicture}
	\end{minipage}
	$\hspace{-1cm}=\qquad$
	\begin{minipage}{.4\textwidth}
		\begin{tikzpicture}[baseline]
		\begin{feynman}
		\vertex (c);
		\vertex[left=0.5cm of c] (cl);
		\vertex[right=0.5cm of c] (cr);
		\vertex[above=1.5cm of cl] (nl);
		\vertex[above=1.5cm of cr] (nr);
		\vertex[below=1.5cm of cl] (sl);
		\vertex[below=1.5cm of cr] (sr);
		\vertex[left=1.5cm of cl] (w);
		\vertex[right=1.5cm of cr] (e);
		\vertex[below=0.67cm of cl] (csl);
		\vertex[below=0.67cm of cr] (csr);
		\vertex[left=0.67cm of cl] (cw);
		\vertex[left=1cm of cl] (cww);
		\vertex[below=1cm of cl] (cssl);
		\vertex[below=1cm of cr] (cssr);
		\diagram*{(sl) -- [fermion, edge label'=$0$] (cl), (cl) -- [fermion] (nl), (w)
			-- [fermion, edge label=$z$] (cl), (cr) -- [fermion] (e), (cl) -- (cr), (sr) -- [fermion, edge label'=$0$] (cr), (cr) -- [fermion] (nr)} ;
		\vertex[below=0.2cm of sr] {$V_2$};
		\vertex[below=0.2cm of sl] {$V_1$};
		\vertex[left=0cm of w] {$\langle i|$};
		\vertex[right=0cm of e] {$|j\rangle$};
		\end{feynman}
		\end{tikzpicture}
	\end{minipage}
\end{figure}
\noindent as a diagram. In particular, the Yangian coproduct acts as
\[
	\Delta_\hbar(T^i_{\ j}(z)) = T^i_{\ k}(z)\otimes T^k_{\ j}(z)
\]
on the transfer matrix. As in section~\ref{sec:Yangian}, in the gauge theory this is just the statement of the OPE: topological invariance in $\Sigma$ allows us to treat the two vertical line operators as either arbitrarily far apart, where they have separate identities, or arbitrarily close, where the OPE replaces them by a single line operator. Because $T(z)$ is the full, quantum expectation value of a horizontal and vertical line operator, this form of the tensor product is valid to all orders in $\hbar$.

Much like in the bulk, the tensor product of a Yangian and twisted Yangian representation has a natural interpretation in the $RTT$ presentation. It is given by the equivalence of the following two diagrams
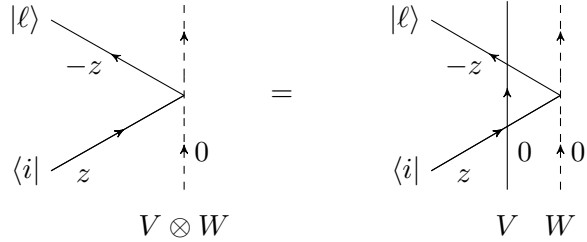
\begin{figure}[th]
	\centering
	\begin{tikzpicture}[baseline]
	\begin{feynman}
	\vertex(o);
	\vertex[right=1cm of o](lnw);
	\vertex[below=2cm of lnw](lsw);
	\vertex[below=1cm of lnw](lw);
	\vertex[right=1.7cm of lw](le);
	\vertex[above=0.25cm of le](ltr);
	\vertex[below=0.25cm of le](ltl);
	\vertex[above=1.25cm of le](lne);
	\vertex[above=1cm of le](lu);
	\vertex[below=1.25cm of le](lse);
	\vertex[left=0.7cm of lne](lt);
	\vertex[left=0.7cm of lse](lb);
	\vertex[above = 0.5cm of lse](lese);
	\vertex[above = 0.25cm of lse](lwz);
	\vertex[above=0.25cm of lsw](lzit);
	\vertex[below=0.25cm of lnw](lzft);
	\vertex[right=0.43 of lzit](lzi);
	\vertex[right=0.43 of lzft](lzf);
	\vertex[above=1cm of lsw](lut);
	\vertex[right=0.65cm of lut](lu);
	\vertex[left=0cm of lsw]{$\langle i|$};
	\vertex[left=0cm of lnw]{$|\ell\rangle$};
	\vertex[below=0.2cm of lse]{$W$};
	\vertex[below=0.1cm of lzi]{$z$};
	\vertex[below=0.1cm of lzf]{$-z$};
	\vertex[above=1cm of lb](lbv);
	\vertex[above=0.5cm of lb](lbo);
	\vertex[below=0.2cm of lb]{$V$};
	\vertex[right=0cm of lbo]{$0$};
	\diagram*{{[edges=fermion] (lsw) -- (le), (le) -- (lnw), (lsw) -- (le), (lb) -- (lt)}, {[edges = charged scalar] (lse) -- [edge label'=$0$](ltl), (ltr) -- (lne)}, (ltl) -- [scalar] (ltr)};
	\vertex[left=4cm of o](rnw);
	\vertex[below=2cm of rnw](rsw);
	\vertex[below=1cm of rnw](rw);
	\vertex[right=1.75cm of rw](re);
	\vertex[above=0.25cm of re](rtr);
	\vertex[below=0.25cm of re](rtl);
	\vertex[above=1.25cm of re](rne);
	\vertex[above=1cm of re](ru);
	\vertex[below=1.25cm of re](rse);
	\vertex[above = 0.5cm of rse](rese);
	\vertex[above = 0.25cm of rse](rwz);
	\vertex[above=0.25cm of rsw](rzit);
	\vertex[below=0.25cm of rnw](rzft);
	\vertex[right=0.43 of rzit](rzi);
	\vertex[right=0.43 of rzft](rzf);
	\vertex[above=0.5cm of rsw](rut);
	\vertex[right=0.65cm of rut](ru);
	\vertex[left=0cm of rsw]{$\langle i|$};
	\vertex[left=0cm of rnw]{$|\ell\rangle$};
	\vertex[below=0.2cm of rse]{$V\otimes W$};
	\vertex[below=0.1cm of rzi]{$z$};
	\vertex[below=0.1cm of rzf]{$-z$};
	\diagram*{{[edges=fermion] (rsw) -- (re), (re) -- (rnw), (rsw) -- (re)}, {[edges = charged scalar] (rse) -- [edge label'=$0$](rtl), (rtr) -- (rne)}, (rtl) -- [scalar] (rtr)};
	\vertex[right=1cm of re]{$=$};
	\end{feynman}
	\end{tikzpicture}
\end{figure}\\
which is again true by construction in the quantum gauge theory. From this we deduce that
\begin{equation}
\label{eq:BoundaryTransferConcatenate}
	B^i_{\ \ell,V\otimes W}(z) = T^i_{\ j,V}(z)\big(T^{-1}\big)^k_{\ \ell,V}(-z)\otimes B^j_{\ k,W}(z)\,,
\end{equation}
where $T^{-1}_V(z)$ is represented by the bulk unitarity relation\vspace{1cm}\\

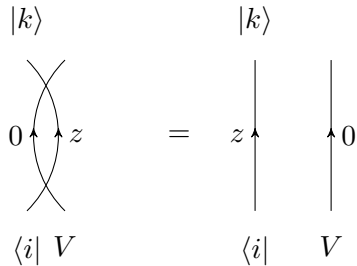
\begin{figure}[h!]
\vspace{-1cm}
\centering
	\begin{tikzpicture}
	\begin{feynman}
		\vertex(o){$=$};
		\vertex[right=2cm of o](re);
		\vertex[left=1cm of re](rw);
		\vertex[above=1cm of re](rne);
		\vertex[below=1cm of re](rse);
		\vertex[above=1cm of rw](rnw);
		\vertex[below=1cm of rw](rsw);
		\diagram*{(rsw) -- [fermion, edge label=$z$] (rnw), (rse) -- [fermion, edge label'=$0$] (rne)};
		\vertex[below=0.2cm of rse]{$V$};
		\vertex[left=2cm of o](lw);
		\vertex[right=0.5cm of lw](le);
		\vertex[above=1cm of le](lne);
		\vertex[below=1cm of le](lse);
		\vertex[above=1cm of lw](lnw);
		\vertex[below=1cm of lw](lsw);
		\vertex[below=2cm of o](s);
		\vertex[below=2.5cm of o](ss);
		\diagram*{(lsw) -- [quarter right, fermion, edge label'={$z$}] (lnw), (lse) -- [quarter left, fermion, edge label={$0$}] (lne)};
		\vertex[below=0.2cm of lse]{$V$};
		\vertex[above=0.2cm of lnw]{$| k\rangle$};
		\vertex[below=0.2cm of lsw]{$\langle i|$};
		\vertex[above=0.2cm of rnw]{$| k\rangle$};
		\vertex[below=0.2cm of rsw]{$\langle i|$};
	\end{feynman}
	\end{tikzpicture}
\end{figure}

\vspace{-0.3cm}\noindent and hence is just the $T$-matrix with the lines crossing in the other orientation with respect to our framing of $\Sigma$.

When $\fg=\mathfrak{sl}_n(\mathbb{C})$. the relation~\eqref{eq:BoundaryTransferConcatenate} is exactly the formula for the tensor product appearing in~\cite{molev2002representations}. It slightly differs from the formula appearing in \cite{guay2016twisted} for the remaining classical Lie algebras, but this discrepancy can be accounted for by writing $T^{-1}(z)$ in terms of $T^\text{t}(z)$ using the analogue of the quantum determinant for these algebras.

\subsection{Uniqueness of the Rational $K$-matrix.}\label{subsec:unique}
At this point we demonstrate that the order $\hbar$ term of the $K$-matrix we derived in section \ref{sec:Kmx} determines it uniquely to all orders in $\hbar$. To do this we make use of the bYBE together with the Sklyanin determinant condition we derived above. The proof follows the same argument employed in \cite{Costello:2018gyb} to prove the uniqueness of the bulk $R$-matrix. Since we will be using the Sklyanin determinant condition this argument will apply only for the defining vector representation $U$ of $\fg=\mathfrak{sl}_N(\C)$. Fortunately, the generalization to other representations and Lie algebras is straightforward, and will be discussed afterwards.\footnote{Note that so far we have only defined $K$-matrices for irreducible representations of $\cY(\fg)$ which lift from representations of $\fg$. We hope to generalize this in the future.}

The proof proceeds in two steps. We start by assuming that we have two rational $K$-matrices  satisfying the bYBE and with Sklyanin determinant 1, which agree up to order $\hbar^{r-1}$ for $r\geq 2$. The two $K$-matrices are hence related by
\begin{equation} 
	K'(z) = K(z) + \frac{\hbar^r}{z^r}\delta k_r + \cO(\hbar^{r+1}) \in\text{End}(U)\,,
\label{eq:kdiff}
\end{equation}
for some $\delta k_r$. In writing the above we have made use of the invariance of the theory under simultaneous rescaling of $z$ and $\hbar$. Substituting this into the Sklyanin determinant condition, one finds that
\begin{equation*}
\begin{aligned}
\text{sdet}(K')(z) &= \\ \text{sdet}(K)(&z) + \varepsilon_{i_1\dots i_N}\frac{\hbar^r}{z^r} \sum_{j=1}^N (\tau_U)^{i_1}_{\ 1}(\tau_U)^{i_2}_{\ 2}\dots(\tau_U)^{i_{j-1}}_{\ j-1} (\delta k_r)^{i_j}_{\ j}(\tau_U)^{i_{j+1}}_{\ j+1}\dots(\tau_U)^{i_N}_{\ N} + \mathcal{O}(\hbar^{r+1})\,,
\end{aligned}
\end{equation*}
where we've used the fact that at lowest order in $\hbar$, the $R$-matrix is equal to ${\bf 1}_{U\otimes U}$ and the $K$-matrix is equal to $\tau_U$. From this we can deduce that
\begin{equation*}
\begin{aligned}
	&\varepsilon_{i_1\dots i_N}\sum_{j=1}^N(\tau_U)^{i_1}_{\ 1}(\tau_U)^{i_2}_{\ 2}
	\dots (\tau_U)^{i_{j-1}}_{\ j-1} (\delta k_r)^{i_j}_{\ j}(\tau_U)^{i_{j+1}}_{\ j+1}
	\dots(\tau_U)^{i_N}_{\ N} \\
	&= \sum_{j=1}^N (\tau^{-1}_U)^j_{\ i_j}(\delta k_r)^{i_j}_{\ j} = \text{Tr}(\tau_U^{-1}\delta k_r) 
	= 0\,.
\end{aligned}
\end{equation*}
where the first equality uses that $\tau_U$ has determinant 1. This shows that  $\tau_U^{-1}\delta k_r$ lies in $\mathfrak{sl}_N(\C)$.

The second step is to substitute equation \eqref{eq:kdiff} into the bYBE, giving
\begin{equation*} 
\begin{aligned}
	& K_2(v)R_{12}(u+v)K_1(u)R_{21}(u-v) + \hbar^r\bigg(\frac{\delta k_{r,2}\tau_{U,1}}{v^r} 
	+ \frac{\tau_{U,2}\delta k_{r,1}}{u^r}\bigg) + \cO(\hbar^{r+1})\\
	&=  R_{12}(u-v)K_1(u)R_{21}(u+v)K_2(v) + \hbar^r\bigg(\frac{\tau_{U,1}\delta k_{r,2}}{v^r} 
	+ \frac{\delta k_{r,1}\tau_{U,2}}{u^r}\bigg) + &\cO(\hbar^{r+1})
\end{aligned}
\end{equation*}
at lowest non-trivial order in $\hbar$. Unfortunately the above holds identically, as $K$ satisfies the bYBE and the $\tau_U$ and $\delta k$ can all commute past one another since they live in different factors of the tensor product. To get a non-trivial constraint we need to expand to order $\hbar^{r+1}$. Note that any terms involving $\delta k_{r+1}$ will not contribute for precisely the same reason that terms involving $\delta k_r$ didn't give a constraint at order $\hbar^r$. Hence the only non-trivial terms come from expanding the $R$-matrices to first order in $\hbar$. Recall from section \ref{subsec:compkmx} that
\[ 
	R(z) = {\bf 1}_{U\otimes U} + \frac{\hbar}{z} C + \cO(\hbar^2)\,,
\]
where $C=c_{U\otimes U} = t_{a,U}\otimes t^a_{\ ,U}$. Note that $C$ is symmetric, so in particular we have $C_{12}=C_{21}=C$. Expanding to order $\hbar^{r+1}$ gives
\begin{align*}
	&\hbar^{r+1}\bigg(\frac{\delta k_{r,2} C \tau_{U,1}}{v^r(u+v)} 
	+ \frac{\delta k_{r,2}\tau_{U,1}C}{v^r(u-v)} + \frac{\tau_{U,2}C\delta k_{r,1}}{(u+v)u^r} 
	+ \frac{\tau_{U,2}\delta k_{r,1}C}{(u-v)u^r}\bigg) + \cO(h^{r+2})\\
	&= \hbar^{r+1}\bigg(\frac{\tau_{U,1}C\delta k_{r,2}}{v^r(u+v)} 
	+ \frac{C\tau_{U,1}\delta k_{r,2}}{v^r(u-v)} + \frac{\delta k_{r,1}C\tau_{U,2}}{(u+v)u^r} 
	+ \frac{C\delta k_{r,1}\tau_{U,2}}{(u-v)u^r}\bigg) + \cO(h^{r+2})\,.
\end{align*}
For this to hold for all $\hbar$, it must hold at order $\hbar^{r+1}$. Furthermore for $r\geq2$ the functions
\begin{equation} 
	\frac{1}{u^r(u+v)}~,\qquad\frac{1}{u^r(u-v)}~,\qquad
	\frac{1}{v^r(u+v)}\qquad\text{and}\qquad\frac{1}{v^r(u-v)}
\label{eq:LIfns} 
\end{equation}
are linearly independent. This means we must have
\begin{align*}
	\delta k_{r,2} C \tau_{U,1} = \tau_{U,1}C\delta k_{r,2}~,\qquad
	\delta k_{r,2}\tau_{U,1}C=C\tau_{U,1}\delta k_{r,2}~, \\
	\tau_{U,2}C\delta k_{r,1} = \delta k_{r,1}C\tau_{U,2}~,\qquad
	\tau_{U,2}\delta k_{r,1}C = C\delta k_{r,1}\tau_{U,2}~.
\end{align*}
It turns out that all four of these relations are equivalent, so we'll concentrate on the first. Let's now write $\delta k_r = \tau_U X_r$, where $X_r\in\mathfrak{sl}_N(\C)$. From this we can deduce that
\[ 
	\delta k_{r,2} C \tau_{U,1} = \tau_{U,2} X_{r,2} C \tau_{U,1} = \tau_{U,1}C \tau_{U,2} X_{r,2} 
	= \tau_{U,1}C\delta k_{r,2}\,.
\]
Now we can exploit the fact that $\tau_{U,1}C\tau_{U,2} = \tau_{U,2}C\tau_{U,1}$, since $C$ is $G$-invariant and $\tau^2\in Z(G)$. This allows us to infer that
\[ 
	\tau_{U,2} X_{r,2} C \tau_{U,1} = \tau_{U,2}C \tau_{U,1} X_{r,2} 
	= \tau_{U,2}C X_{r,2}\tau_{U,1}\,,
\]
which implies that $X_{r,2} C = C X_{r,2}$. On the other hand, using the definition of $C$ we have
\[ 
	X_{r,2} C - C X_{r,2} = t^a_{\ ,U}\otimes[X_r,t_{a,U}]
\]
so we conclude that 
\[
	[X_r,t_{a,U}] = 0\qquad\forall\  t_a\in\fg.
\]
It then follows from Schur's lemma that, since $U$ is an irreducible representation of $\mathfrak{sl}_N(\mathbb{C})$, $X_r$ must vanish. This concludes our proof.

\medskip

This argument generalizes straightforwardly to arbitrary irreducible representations of any simple complex Lie algebra $\fg$. The first part of the proof can be naturally extended to relations derived from arbitrary vertices. In general we find that constraints from vertices are enough to ensure that $\delta k_r \in \tau_V \fg_V$, where by $\fg_V$ we mean the image of $\fg$ in the representation $V$. We use this fact when invoking Schur's lemma at the end of the proof to set $X_r=\tau^{-1}_V\delta k_r$ to 0. Otherwise, the manipulations performed in the second part of the proof go through without incident.

It's also worth noting why this proof fails for $r=1$. The first part of the proof goes through as before, and we learn that $\delta k_1\in \tau_V\fg_V$. However, when $k=1$, the rational functions in~\eqref{eq:LIfns} are no longer linearly independent, since
\[ 
	\frac{1}{u(u+v)} + \frac{1}{u(u-v)} = \frac{1}{v(u-v)} - \frac{1}{v(u+v)}\,.
\]
This means that the constraints we can deduce for $r=1$ are weaker than those for $r\geq2$. Writing $\delta k_1 = \tau_V X$, they are given by
\begin{align*}
\tau_{V,2} X_{2} C \tau_{V,1} + \tau_{V,2} X_{2}\tau_{V,1} C &= \tau_{V,1}C\tau_{V,2} X_{2} + C\tau_{V,1}\tau_{V,2} X_{2}~, \\
\tau_{V,2}C\tau_{V,1} X_{1} + \tau_{V,2} X_{2}\tau_{V,1}C &= \tau_{V,1} X_{1}C\tau_{V,2} + C\tau_{V,1}\tau_{V,2} X_{2}~, \\
\qquad\tau_{V,2}\tau_{V,1} X_{1}C + \tau_{V,2} X_{2}\tau_{V,1}C &= C\tau_{V,1} X_{1}\tau_{V,2}  + C\tau_{V,1}\tau_{V,2} X_{2}~.
\end{align*}
We can manipulate these into a simpler form using the fact that $\tau_{V,1}C\tau^{-1}_{V,1} = \tau^{-1}_{V,1}C\tau_{V,1}= \tau_{V,2}C\tau^{-1}_{V,2}$. It turns out that the third equation amounts to saying that $[X_1,C] + [X_2,C] = 0$, which for $X\in\fg_V$ holds identically, since $c$ is $\fg$-invariant. Applying this fact, the second equation follows from the first. Hence the first equation is the only one we need to worry about. Premultiplying by $\tau^{-1}_{V,1}\tau^{-1}_{V,2}$, we find that it's equivalent to
\[
	\Big[X_2,C + \tau_{V,1}C\tau_{V,1}^{-1}\Big] = \Big[X_2,C + \tau_{V,2}C\tau_{V,2}^{-1}\Big] 
	= 0\,.
\]
Using the definition of $C$ this is the same as
\[ 
	t_{a,V}\otimes\Big[X,t_{a,V} + \tau_Vt_{a,V}\tau^{-1}_V\Big] 
	= 0 \implies \Big[X,t_{a,V} + \tau_Vt_{a,V}\tau^{-1}_V\Big]\qquad\forall \ a\,.
\]
Now we notice that the map $\Pi:\fg\to\fg,~Y\mapsto (Y + \sigma(Y))/2$ for $\sigma=\text{conj}_\tau$ is the projection from $\fg$ onto $\fh$. Thus the above amounts to saying that $X$ must be an element of $\fg$ commuting with everything in $\fh$. This is consistent with the $K$-matrices we found in section \ref{sec:Kmx}. In particular, when $\fh$ contains a copy of $\C$, the $K$-matrices depend on a free parameter appearing in $k$.

This proof fails if $\dim W>1$, where further constraints appear to be required.

\section{Conclusions}
\label{sec:Conc}

In this paper we have studied the mixed topological-holomorphic Chern-Simons theory of Costello-Yamazaki-Witten on an orbifold $\R^2\times\C/\Z_2$. We showed how this leads to a description of lattice integrable systems in the presence of a boundary, and used the gauge theory to obtain and explicit form for the leading correction $k(z)$ to the rational $K$-matrix solving the boundary Yang-Baxter equation, even when the boundary is labelled by a representation $W$ of dimension $>1$. In the case that $\dim W=1$, we showed how our formula generates all known rational solutions to the bYBE (for classical $\fg$) that admit an expansion as a series in $\hbar/z$.

In the quantum theory, line operators were shown to be labelled by representations of the twisted Yangian $\cB(\fg,\fh)$, with its structure as a left co-ideal of $\cY(\fg)$ provided by the OPE of a bulk and boundary line operator. While we checked this statement in the Drinfeld $J$-presentation of~\cite{belliard2017drinfeld} only to order $\hbar$, we then showed that the quantum expectation values of the lines operators obey the boundary $RTT$ relations, boundary unitarity and (for $\fg=\mathfrak{sl}_n$) the Sklyanin determinant condition, or (for $\fg$ a classical Lie algebra) closely related conditions. These conditions imply that quantum line operators are labelled by representations of $\cB(\fg,\fh)$ to all orders. It is interesting to note that the symmetry relations that are natural from the gauge theory perspective are somewhat different to those found in~\cite{guay2016twisted}. In particular, our condition~\eqref{BCDtensorRelation} in the $RTT$-presentation of twisted Yangians for $\fg=B_n$, $C_n$ and $D_n$ appears to be simpler than any we have been able to find in the literature. Finally, we used the $RTT$ relations and the Sklyanin determinant to prove that, for $\dim W=1$, there is a unique quasi-classical $K(z)$-matrix obeying the bYBE for a given classical $k(z)$ matrix.

\medskip

It would be interesting to verify the formulas of~\cite{belliard2017drinfeld} for the twisted Yangian in the Drinfeld $J$-presentation by performing a two-loop calculation, as in \cite{Costello:2017dso}.

\medskip

The results of this paper can be generalized in a number of ways. Most straightforwardly, one can extend the rational $K$-matrices considered here to the case where the involution $\sigma$ is an {\it outer} automorphism of $\fg$. Such outer automorphisms correspond to graph automorphisms of the Dynkin diagram of $\fg$ and generically exchange representations $V\mapsto V^\sigma$. In physics, they play in important role in the case of `soliton non-preserving' scattering off the boundary (see {\it e.g.}~\cite{Arnaudon:2004sd}).

An important direction in which to extend this work is to the construction of trigonometric and elliptic $K$-matrices. These are much less well understood than the rational $K$-matrices we have considered in this paper. The systematic approach the gauge theory provides could play an important role in the discovery and classification of new solutions to the boundary Yang-Baxter equation (perhaps especially in the elliptic case). In the presence of a boundary for $\Sigma$, we expect such $K$-matrices  will again be generated by CWY theory on an orbifold $\widetilde{M} = (\R^2\times C)/\Z_2$, where $C=\C^*$ or $E_\tau$.


\appendix
\newpage

\section{Review of Symmetric Spaces} 
\label{app:symspa}

In this appendix we briefly review symmetric spaces, which are fundamental to the construction of the twisted Yangian.

If $\sigma \in {\rm Aut}(\fg)$ is involutive, we can split $\fg$ as
\[
	\fg=\fh\oplus \fm\,,
\]
where $\fh$ and $\fm$ are the positive and negative eigenspaces of $\sigma$, respectively. One can easily deduce that the Lie bracket on the two summands obeys
\begin{equation} \label{eq:sspace}
	[\fh,\fh] \subseteq \fh\,,\qquad 
	[\fh,\fm] \subseteq \fm
	\qquad\hbox{and}\qquad
	[\fm,\fm] \subseteq \fh\,.
\end{equation}
The first of these relations tells us that $\fh$ is a Lie subalgebra of $\fg$, while the second tells us that the adjoint action of $\fh$ on $\fm$ gives a representation of $\fh$. Conversely, if $\fg$ can be split into a direct sum of $\fh$ and $\fm$ satisfying~\eqref{eq:sspace}, then we say that the pair $(\fg,\fh)$ is a {\it symmetric space}. Such symmetric spaces are in bijection with the involutive automorphisms of $\fg$, since we can simply define an involutive automorphism by $\sigma(\fh) = \fh$ and $\sigma(\fm) = - \fm$.

One can show that $\fh$ and $\fm$ satisfying \eqref{eq:sspace} must be orthogonal, and hence that the Killing form is block diagonalised by this decomposition. This demonstrates that $\fh$ uniquely determines $\fm$. Note that the Killing form on $\fg$ restricted to $\fh$ is not necessarily equal to the Killing form on $\fh$. In general $\fh$ is a reductive Lie algebra, which means it is of the form $\fh = \mathfrak{s}\oplus\mathfrak{t}$ for $\mathfrak{s}$ semisimple and $\mathfrak{t}$ abelian. $\mathfrak{s}$ can then be further decomposed into its simple summands. In practice it turns out that either $\mathfrak{t}=\C$ or it doesn't appear, and that $\mathfrak{s}$ has at most two simple summands.

From now on we use Roman indices to index basis vectors of $\fg$, Greek letters $\alpha, \beta, \gamma, \ldots$ from the beginning of the alphabet to index the basis vectors in $\fh$, and indices $\mu, \nu, \xi, \ldots$ from the middle of the Greek alphabet to label basis vectors in $\fm $. Thus, $\{t_a\}_{a=1}^{\dim\fg}$ forms a basis of $\fg$, $\{t_\alpha\}_{\alpha=1}^{\dim \fh}$ a basis of $\fh$ and $\{t_\mu\}_{\mu = \dim\fh +1}^{\dim\fg}$ a basis of $\fm$.  We further refine our basis of $\fh$ so that $\{t_\alpha\}$ is the union of the generator of $\mathfrak{t}$, if it is present, and bases of the simple summands of $\mathfrak{s}$. Given that $\fh$ and $\fm$ are orthogonal, if we raise or lower a Greek index from the start of the alphabet with the Killing form it will still belong to the start, and similarly for indices from the middle of the Greek alphabet. Using this notation, the conditions in \eqref{eq:sspace} can be expressed 
as
\[ 
	f_{\alpha\beta}^{\ \ \xi} = 0\,,
	\qquad f_{\alpha\nu}^{\ \ \gamma} = 0\,,
	\qquad f_{\mu\nu}^{\ \ \xi} = 0~\,. 
\]
(In fact since we can raise and lower indices with the Killing form the first two conditions are equivalent.) Since $\fg$ is simple, we have
\[ 
	f_{ab}^{\ \ c}f_c^{\ bd} = c_{\fg}\delta_a^{\ d}
\]
where our choice of normalization for the Killing form means that $c_\fg = 2\bf{h}^\vee$. From this we can deduce that
\[ 
	c_{\fg}\delta_\alpha^{\ \delta} = f_{\alpha\beta}^{\ \ \gamma}f_{\gamma}^{\ \beta\delta} 
	+ f_{\alpha\nu}^{\ \ \xi}f_{\xi}^{\ \nu\delta}\,. 
\]
Now, $f_{\alpha\beta}^{\ \ \gamma}f_{\gamma}^{\ \beta\delta}$ is built out of the structure constants of the algebra $\fh$ and the Killing form on $\fg$ restricted to $\fh$. Certainly this restriction defines an $\fh$-invariant bilinear form on $\fh$. Since $\fh$ is reductive, and we've chosen out basis so that it respects the decomposition of $\fh$ into simple and abelian summands, this means that 
\[ 
	f_{\alpha\beta}^{\ \ \gamma}f_{\gamma}^{\ \beta\delta} 
	= c_{(\alpha)} \delta_\alpha^{\ \delta}\,.
\]
Here the $c_{(\alpha)}$ are complex numbers which depend only on which summand of $\fh$ the basis vector $t_{\alpha}$ belongs to. This then implies that
\[ 
	f_{\alpha\nu}^{\ \ \xi}f_{\xi}^{\ \nu\delta} = \bar{c}_{(\alpha)}\delta_\alpha^{\ \delta}\,,
\]
where $\bar{c}_{(\alpha)} = c_{\fg} - c_{(\alpha)}$. These constants appear in the definition of the twisted Yangian in the $J$-presentation~\cite{belliard2017drinfeld}.

\medskip

In this paper, we'll be particularly interested in involutive {\it inner} automorphisms of the classical simple Lie algebras. Inner automorphisms of $\fg$ are given by conjugation by an element of $G/Z(G)$, and for $\sigma$ to be an involution we require $\sigma=\text{conj}_\tau$ where $\tau\in G$ satisfies $\tau^2\in Z(G)$. All such inner automorphisms of classical simple $\fg$, and their associated positive eigenspaces, $\fh$, are listed in the table below, up to conjugation of $\tau$ by an element of $G$. (We have omitted the trivial automorphism, which is allowed for all $\fg$.)
\begin{center}
	\begin{tabular}{|*3{c|}}
		\hline
		$\mathfrak{g}$ & $\tau$ & $\mathfrak{h}$ \\ \hline &&\\ [-5pt]
		$\mathfrak{sl}_n(\mathbb{C})\cong A_{n-1}$&
		$\e^{\im\pi k/n}\begin{pmatrix} {\bf 1}_{n-k} & 0 \\ 0 & -{\bf 1}_k \end{pmatrix}$
		& $\mathfrak{sl}_{n-k}(\mathbb{C})\oplus \mathfrak{sl}_k(\mathbb{C})\oplus \mathbb{C}$ \\[10pt] \hline &&\\ [-5pt]
		$\mathfrak{so}_{2n+1}(\mathbb{C})\cong B_n$ &
		$(-)^k\begin{pmatrix} {\bf 1}_{2n+1-k} & 0 \\ 0 & -{\bf 1}_k \end{pmatrix}$
		& $\mathfrak{so}_{2n+1-k}(\mathbb{C})\oplus \mathfrak{so}_k(\mathbb{C})$ \\[10pt] \hline &&\\ [-5pt]
		\multirow{5}{*}{$\mathfrak{sp}_{2n}(\mathbb{C})\cong C_n$} &
		$\begin{pmatrix} \begin{matrix} {\bf 1}_{(n-k)} & 0 \\ 0 & -{\bf 1}_{k} \end{matrix} & \bigzero \\ \bigzero & \begin{matrix} {\bf 1}_{(n-k)} & 0 \\ 0 & -{\bf 1}_{k} \end{matrix} \end{pmatrix}$
		& $\mathfrak{sp}_{2(n-k)}(\mathbb{C})\oplus\mathfrak{sp}_{2k}(\mathbb{C})$ \\[20pt] \cline{2-3} &&\\ [-5pt] &
		$\begin{pmatrix} 0 & {\bf 1}_{n} \\ -{\bf 1}_{n} & 0 \end{pmatrix}$
		& $\mathfrak{sl}_n(\mathbb{C})\oplus\mathbb{C}$
		\\[10pt] \hline &&\\ [-5pt]
		\multirow{4}{*}{$\mathfrak{so}_{2n}(\mathbb{C})\cong D_n$} &
		$\begin{pmatrix} {\bf 1}_{2(n-k)} & 0 \\ 0 & -{\bf 1}_{2k} \end{pmatrix}$
		& $\mathfrak{so}_{2(n-k)}(\mathbb{C})\oplus \mathfrak{so}_{2k}(\mathbb{C})$\\[10pt] \cline{2-3} &&\\ [-5pt] &  
		$\begin{pmatrix} 0 & {\bf 1}_{n} \\ -{\bf 1}_{n} & 0 \end{pmatrix}$
		& $\mathfrak{sl}_n(\mathbb{C})\oplus\mathbb{C}$\\[10pt] \hline
	\end{tabular}
\end{center}

\newpage

\section{Contribution of Self-Interaction Diagrams to the Classical $k$-Matrix} \label{app:selfinteractions}

In this appendix we find the contribution of the following two Feynman diagrams to the classical $k$-matrix.
\begin{figure}[th]
	\centering
	\begin{tikzpicture}[baseline]
	\begin{feynman}
	\vertex(nw2);
	\vertex[below=2cm of nw2](sw2);
	\vertex[below=1cm of nw2](w2);
	\vertex[right=1.25cm of w2](e2);
	\vertex[above=1cm of e2](ne2);
	\vertex[below=1cm of e2](se2);
	\vertex[left=0.25cm of e2](l2);
	\vertex[above=0.22cm of l2](gi2);
	\vertex[left=1cm of e2](r2);
	\vertex[above=0.782cm of r2](gf2);
	\vertex[left=0cm of sw2]{$z$};
	\vertex[left=0cm of nw2]{$-z$};
	\vertex[left=5cm of nw2](nw3);
	\vertex[below=2cm of nw3](sw3);
	\vertex[below=1cm of nw3](w3);
	\vertex[right=1.25cm of w3](e3);
	\vertex[above=1cm of e3](ne3);
	\vertex[below=1cm of e3](se3);
	\vertex[left=0.25cm of e3](l3);
	\vertex[below=0.22cm of l3](gf3);
	\vertex[left=1cm of e3](r3);
	\vertex[below=0.782cm of r3](gi3);
	\vertex[left=0cm of sw3]{$z$};
	\vertex[left=0cm of nw3]{$-z$};
	\diagram*{{[edges=fermion] (sw2) -- (e2), (e2) -- (nw2), (sw3) -- (e3), (e3) -- (nw3)}, {[edges = scalar] (se2) -- (ne2), (se3) -- (ne3)}, {[edges = gluon] (gi2) -- [half left] (gf2), (gi3) -- [half left] (gf3)},};
	\end{feynman}
	\end{tikzpicture}
\end{figure}\\
The first diagram contributes
\[ t_at_b\tau\int_{-\infty<s<t<0}\diff s\diff t\, \frac{\diff\gamma^i_1}{\diff s}\frac{\diff\gamma^i_1}{\diff t}\tilde\Delta^{ab}_{ij}(\gamma_1(s),\gamma_1(t)) \]
for $\gamma_1(s) = (s\,\cos\theta,s\,\sin\theta,z,\bar z)$, where $\theta$ is the angle of incidence to the normal. Inputting the explicit form~\eqref{Propagator} for the propagator shows that this is
\[ t_at_b(c^{\sigma})^{ab}\tau\frac{4\sin\theta\cos\theta}{\pi}\int_{-\infty<s<t<0}\diff s\diff t\,\frac{\bar z}{(\cos^2\theta\,(s+t)^2 + \sin^2\theta\,(s-t)^2 + 4|z|^2)^2}\,.
\]
Noting the symmetry of the integrand under exchange of $s$ and $t$, and making the substitutions $u=-s-t\cos{2\theta}$ and $v=-t\sin{2\theta}$, this is equal to
\[ t_at_b(c^{\sigma})^{ab}\tau\frac{1}{\pi}\int_{B\subset\R^2}\diff u\diff v\,\frac{\bar z}{(u^2 + v^2 + 4|z|^2)^2}\]
where $B$ is a sector of $\R^2$ subtended by an angle $2\theta$. This integral can be performed directly, giving a contribution
\[ t_at_b(c^{\sigma})^{ab}\tau \frac{1}{4z}\frac{\theta}{\pi} = t_a\tau t^a\frac{1}{4z}\frac{\theta}{\pi}\,.\]
Here we used the fact that $t_a\otimes t_b (c^\sigma)^{ab} = c^\sigma = t_a\otimes\sigma(t_a) = t_a\otimes \tau t_a\tau^{-1}$. To get find the contribution from the second diagram we make the simultaneous replacements $\theta\mapsto-\theta$, $z\mapsto-z$, and $t_a\mapsto -t_a$, under which the contribution of the first diagram is invariant. (Here we're viewing the outgoing Wilson line as an incoming by taking it to be in the dual representation $t_a\mapsto -t_a^{\rm t}$. The reason we don't get any transposes is that we've subsequently swapped what we mean incoming and outgoing back again.) $\tau$ also acts before any interactions in the second diagram, but the identity $\sigma(t_a)\otimes t^a = t_a\otimes\sigma(t^a)$ ensures the color structure of the two diagrams is the same. Hence we conclude that the contributions of the two diagrams are equal, and that in total they give
\[ t_a\tau t^a\frac{1}{2z}\frac{\theta}{\pi}\,.\]
\section{Proof that \eqref{eq:classkmx} Solves the Classical Boundary Yang-Baxter Equation} 
\label{app:cbYBEproof}

Our Feynman diagram computation to find the rational classical $k$-matrix gave the result
\[
	 k(z) = \frac{1}{4z}\,t_{b}\,\tau\,t_{a}\,(\kappa^{-1})^{ab} = \frac{1}{4z}t_{a}\,\tau\,t^{a}
\]
in the simplest case that the boundary Wilson line is switched off. In this appendix we demonstrate that this is indeed a solution of the classical boundary Yang-Baxter equation, or equivalently that the corresponding quasi-classical $K$-matrix satisfies the bYBE up to and including order $\hbar^2$. 

To derive the classical boundary Yang-Baxter equation we simply expand the bYBE to second order in $\hbar$, which gives
\begin{equation}
\begin{aligned}
\label{eq:cbYBE}
	&\tau_{2}r_{12}(u+v)\tau_{1}r_{21}(u-v) + k_2(v)r_{12}(u+v)\tau_{1} 
	+ \tau_{2}r_{12}(u+v)k_1(u)\\
	&\qquad + k_2(v)\tau_{1}r_{21}(u-v) + \tau_{2}k_1(u)r_{21}(u-v) + k_2(v)k_1(u)\\
	&\qquad\qquad\qquad = r_{12}(u-v)\tau_{1}r_{21}(u+v)\tau_{2} + \tau_{1}r_{21}(u+v)k_2(v)
	+  k_1(u)r_{21}(u+v)\tau_{2} \\
	&\qquad\qquad\qquad\qquad
	 + r_{12}(u-v)k_1(u)\tau_{2} + r_{12}(u-v)\tau_{1}k_2(v) + k_1(u)k_2(v)\,.
\end{aligned}
\end{equation}
Note that the quadratic terms in $k$ on either side cancel since $k_1$ and $k_2$ commute with one another. We will also need to recall the solution to the classical $r$-matrix derived using the bulk theory. This is given by
$$ r(z) = \frac{1}{z}\,t_{b}\otimes t_{a}\,(\kappa^{-1})^{ab} = \frac{1}{z}\,t_{a}\otimes t^{a}~.$$
Note that this is invariant under exchange of the two factors in the tensor product, \textit{i.e.} $ r_{12}(z) = r_{21}(z)~.$ Substituting these expressions into the classical boundary Yang-Baxter equation gives
\[
\begin{aligned}
	0&=\frac{4}{u^2-v^2}
	\big(t_a\,\tau\,t_b\otimes t^a\,t^b\,\tau - t_a\,\tau \,t_b\otimes \tau\,t^a\,t^b\big) 
	+ \frac{1}{u(u-v)}
	\big(t_a\,t_b\,\tau\,t^b\otimes t^a\,\tau - t_b\,\tau\,t^b\,t_a\otimes \tau\,t^a\big) \\
	& + \frac{1}{u(u+v)}
	\big(t_b\,\tau\,t^b\,t_a\otimes t^a\,\tau - t_a\,t_b\,\tau\,t^b \otimes \tau\,t^a\big) 
	+\frac{1}{v(u-v)}
	\big(t^a\,\tau \otimes t_a\,t_b\,\tau\,t^b -  \tau\,t^a \otimes t_b\,\tau\,t^b\,t_a \big) 
	\\
	&+ \frac{1}{v(u+v)}
	\big(\tau\,t^a\otimes t_a\,t_b\,\tau\,t^b  - t^a\,\tau \otimes t_b\,\tau\,t^b\,t_a \big)\,.
\end{aligned}
\]
The functions appearing in the above are not linearly dependent, and so we expect some cancellation to take place. First we note that
\[
	t_a\,t_b\,\tau\,t^b = f_{ab}^{\ \ c}\,t^b\,\tau\,t_c + t_b\,t_a\,\tau\,t^b 
	= -f_{a}^{\ bc}\,t_b\,\tau\,t_c + t_b\,t_a\,\tau\,t^b
\]
and
\[
	t_b\,\tau\,t^b\,t_a = - f_{a}^{\ bc}\,t_b\,\tau\,t_c + t_b\,\tau\,t_a\,t^b~.
\]
Applying these identities and using the fact that $t-a\tau\otimes t^a \tau = t_a\tau \otimes \tau t^a$, we find {\it e.g.} that the colour structure of the $u^{-1}(u-v)^{-1}$ term simplifies to
\[ 
	t_a\,t_b\,\tau\,t^b\otimes t^a\,\tau - t_b\,\tau\,t^b\,t_a\otimes \tau\,t^a 
	= f_a^{\ bc}\,t_b\,\tau\,t_c\otimes[\tau,t^a]\,.
\]
with identical considerations simplifying the other colour structures, with the exception of the $(u^2-v^2)^{-1}$ term.  After applying similar simplifications to the other terms, the classical bYBE is the requirement that our solution obeys
\begin{equation}
\label{eq:algebra1}
	0=\frac{2}{u^2-v^2}
	\big(2\,t_a\,\tau\,t_b\otimes t^a\,t^b\,\tau - 2\,t_a\,\tau \,t_b\otimes \tau\,t^a\,t^b 
	+ f_a^{\ bc}\,t_b\,\tau\,t_c\otimes[\tau,t^a] + f_a^{\ bc}\,[\tau,t^a]\otimes t_b\,\tau\,t_c\big)\,.
\end{equation}
Now we expand
\[
\begin{aligned}
	t_a\,\tau\,t_b\otimes t^a\,t^b\,\tau 
	&= f^{ab}_{\ \ c}\,t_a\,\tau\,t_b\otimes t^c\,\tau + t_a\,\tau\,t_b\otimes t^b\,t^a\,\tau \\
	&= f^{ab}_{\ \ c}\,t_a\,\tau\,t_b\otimes t^c\,\tau + \tau\,t_a\,t_b\otimes t^b\,\tau\,t^a\\
	&=f^{ab}_{\ \ c}\,t_a\,\tau\,t_b\otimes t^c\,\tau + f_{ab}^{\ \ c}\,\tau\,t_c\otimes t^b\,\tau\,t^a 
	+ t_a\,\tau\,t_b\otimes \tau\,t^a\,t^b\,.
\end{aligned}
\]
From this we deduce that \eqref{eq:algebra1} is equivalent to
\[ 
	0=2\,f_c^{ab}\,t_a\,\tau\,t_b\otimes t^c\,\tau - 2\,f_c^{\ ab}\,\tau\,t^c\otimes t_a\,\tau\,t_b 
	+ f_a^{\ bc}\,t_b\,\tau\,t_c\otimes[\tau,t^a] + f_a^{\ bc}\,[\tau,t^a]\otimes t_b\,\tau\,t_c\,.
\]
By replacing all of the $f_a^{\ bc}\,t^a$ with commutators, one arrives at
\[
\begin{aligned} 
	0&=t_b\,\tau\,t_c\otimes t^b\,t^c\,\tau - t_b\,\tau\,t_c\otimes t^c\,t^b\,\tau 
	- \tau\,t^b\,t^c\otimes t_b\,\tau\,t_c + \tau\,t^c\,t^b\otimes t_b\,\tau\,t_c \\
	&+ t_b\,\tau\,t_c\otimes\tau\,t^b\,t^c - t_b\,\tau\,t_c\otimes\tau\,t^c\,t^b 
	- t^b\,t^c\,\tau\otimes t_b\,\tau\,t_c + t^c\,t^b\,\tau\otimes t_b\,\tau\,t_c 
\end{aligned}
\]
Using the by now familiar fact that $t_a\,\tau\otimes t^a\,\tau = t_a\,\tau\otimes \tau\,t^a$, this holds identically. Hence the classical $k$-matrix we computed using gauge theory does indeed satisfy the classical boundary Yang-Baxter equation. 

So far we have only explicitly shown that the $k$-matrices obtained in the absence of a line operator on $L$ solves the classical bYBE. The classical $k$-matrix in the presence of such a boundary line operator was computed in section~\ref{subsec:compkmx} to be
\[ k'(z) = k(z)\otimes 1 + \delta k(z) = \frac{1}{4z}t_a\tau t^a\otimes 1 + \frac{2}{z}\tau t_\alpha\otimes t^\alpha = \frac{1}{4z}\tau(t_\alpha t^\alpha - t_{\mu}t^{\mu})\otimes 1 + \frac{2}{z}\tau t_\alpha\otimes t^\alpha\,.\]
Here the first factor in the tensor product corresponds to the bulk Wilson line and the second factor corresponds to the line operator on $L$. This extra factor can be viewed as the space in which the entries of the $k$-matrix take values. As such this extra factor also appears in the classical boundary Yang-Baxter equation. Substituting $k'(z)$ into the classical boundary Yang-Baxter equation, and using~\eqref{eq:cbYBE} to cancel the terms involving $k(z)$, we're left with
\begin{equation}
\begin{aligned}
\label{eq:cbYBEW}
&  \delta k_{23}(v)r_{12}(u+v)\tau_{1} 
+ \tau_{2}r_{12}(u+v)\delta k_{13}(u) + \delta k_{23}(v)\tau_{1}r_{21}(u-v)\\
&\qquad + \tau_{2}\delta k_{13}(u)r_{21}(u-v) + \delta k_{23}(v)\delta k_{13}(u) + k_2(v)\delta k_{13}(u) + \delta k_{23}(v)k_{13}(u) \\
&\qquad\qquad\qquad = \tau_{1}r_{21}(u+v)\delta k_{23}(v)
+  \delta k_{13}(u)r_{21}(u+v)\tau_{2} + r_{12}(u-v)\delta k_{13}(u)\tau_{2}\\
&\qquad\qquad\qquad\qquad
+ r_{12}(u-v)\tau_{1}\delta k_{23}(v) + \delta k_{13}(u)\delta k_{23}(v) + k_1(u)\delta k_{23}(v) + \delta k_{13}(u)k_2(v)\,.
\end{aligned}
\end{equation}
This can be simplified by noting that $[k_1(u),\delta k_{23}(v)] = [k_2(u),\delta k_{13}(v)]=0$. Note, however, that the term quadratic in $\delta k$ no longer vanishes. In fact
\[ [\delta k_{13}(u),\delta k_{23}(z)] = \frac{4}{uv}\tau t_\alpha\otimes\tau t_{\beta}\otimes[t^\alpha,t^{\beta}] = \frac{4}{uv}f^{\alpha\beta\gamma}\tau t_{\alpha}\otimes \tau t_{\beta}\otimes t_{\gamma}\,.\]
We can directly compute
\[
\begin{aligned}
&\tau_2r_{12}(u+v)\delta k_{13}(u) - \delta k_{13}(u)r_{21}(u+v)\tau_2 = 
\\&\qquad \frac{2}{u(u+v)}(t_\alpha\tau t_\beta\otimes \tau t^\alpha\otimes t^\beta + t_\mu\tau t_\beta\otimes \tau t^\mu\otimes t^\beta - \tau t_\alpha t_\beta \otimes t^\beta\tau \otimes t^\alpha - \tau t_\alpha t_\mu \otimes t^\mu\tau \otimes t^\alpha) = \\
&\qquad\qquad\frac{2}{u(u+v)}(f^{\alpha\beta\gamma}\tau t_\alpha\otimes\tau t_\beta\otimes t_\gamma - f^{\mu\nu\alpha}\tau t_{\mu}\otimes \tau t_{\nu}\otimes t_\alpha)\,,
\end{aligned}
\]
and similarly find
\[
\begin{aligned}
&\delta k_{23}(v)r_{12}(u+v)\tau_1 - \tau_1r_{21}(u+v)\delta k_{23}(v) = \frac{2}{v(u+v)}(f^{\alpha\beta\gamma}\tau t_\alpha\otimes\tau t_\beta\otimes t_\gamma - f^{\mu\nu\alpha}\tau t_{\mu}\otimes \tau t_{\nu}\otimes t_\alpha)\,,\\
&\tau_2\delta k_{13}(u)r_{21}(u-v) - r_{21}(u-v)\delta k_{13}(u)\tau_2 = -\frac{2}{u(u-v)}(f^{\alpha\beta\gamma}\tau t_\alpha\otimes \tau t_\beta\otimes t_\gamma + f^{\mu\nu\alpha}\tau t_\mu\otimes\tau t_\nu\otimes t_\alpha)\,,\\
&\delta k_{23}(v)\tau_1r_{12}(u-v) - r_{12}(u-v)\tau_1\delta k_{23}(v) = \frac{2}{v(u-v)}(f^{\alpha\beta\gamma}\tau t_\alpha\otimes \tau t_\beta\otimes t_\gamma + f^{\mu\nu\alpha}\tau t_\mu\otimes\tau t_\nu\otimes t_\alpha)\,.
\end{aligned}
\]
Substituting all of this directly into~\eqref{eq:cbYBEW}, it becomes equivalent to
\[
\begin{aligned}
&\bigg(\frac{2}{u(u+v)} + \frac{2}{v(u+v)}\bigg)(f^{\alpha\beta\gamma}\tau t_\alpha\otimes\tau t_\beta\otimes t_\gamma - f^{\mu\nu\alpha}\tau t_{\mu}\otimes \tau t_{\nu}\otimes t_\alpha)~+ \\
&\qquad\bigg(\frac{2}{v(u-v)}-\frac{2}{u(u-v)}\bigg)(f^{\alpha\beta\gamma}\tau t_\alpha\otimes \tau t_\beta\otimes t_\gamma + f^{\mu\nu\alpha}\tau t_\mu\otimes\tau t_\nu\otimes t_\alpha)~= \\
&\qquad\qquad\frac{4}{uv}f^{\alpha\beta\gamma}\tau t_{\alpha}\otimes \tau t_{\beta}\otimes t_{\gamma}\,.
\end{aligned}
\]
This clearly holds identically, and so the formula we derived in section~\ref{subsec:compkmx} for the classical $k$-matrix is indeed a solution of the classical boundary Yang-Baxter equation.
\section{Evaluation of Equations \eqref{eq:bOPE1} and \eqref{eq:bOPE2}} \label{app:inteval}

In this appendix we evaluate the integrals involved in computing the OPE of a bulk and boundary Wilson line in section \ref{sec:TwistedYangians}.

Let's start by evaluating the two integrals in equation \eqref{eq:bOPE1}. They are given by
\[
	 \lambda_{\pm} = \int_{-\infty}^0\d x\int_{\C}\d^2 z\,z\,{\mathcal I}_{\pm}(x,z,\bar z;1)\ ,
\]
where
\[
	{\mathcal I}_\pm(x,z,\bar z;1) = \frac{\bar z}{(x^2 + |z|^2)^{3/2}}
	\bigg(\frac{1}{((x+1)^2+|z|^2)^{3/2}} \pm \frac{1}{((x-1)^2+|z|^2)^{3/2}}\bigg)\,.
\]
We start with the easier of the two integrals
\[
\begin{aligned}
	\lambda_+ &= \int_{\R}\d x\int_{\C}\d^2 z\,
	\frac{|z|^2}{(x^2 + |z|^2)^{3/2}((x-1)^2+|z|^2)^{3/2}} \\
	&= -4\pi\im \int_{\R}\d x\int_{0}^\infty\d r\,\frac{r^3}{(x^2 + r^2)^{3/2}((x-1)^2+r^2)^{3/2}}~.
\end{aligned}
\]
where in the first equality we have mapped $x\mapsto-x$ in the first term of $\mathcal{I}_+$ to get an integral over all of $\R$, and in going to the second line we have performed the integral over the phase of $z=r\e^{\im\theta}$. We can perform the resulting integral using Feynman parametrization 
\[
	\frac{1}{A^\alpha B^\beta} = \frac{\Gamma(\alpha+\beta)}{\Gamma(\alpha)\Gamma(\beta)}			
	\int_0^1\d t\,\frac{t^{\alpha-1}(1-t)^{\beta-1}}{(tA+(1-t)B)^{\alpha+\beta}}~.
\]
Applying this to our integral we arrive at
\[
	\lambda_+ = -32\im\,\int_0^1\d t\,t^{1/2}(1-t)^{1/2}\int_{\R}\d x\int_{0}^\infty\d r\,
	\frac{r^3}{(r^2 + t(x-1)^2 + (1-t)x^2)^3}~.
\]
Now using the fact that substituting $r=a\tan\varphi$ gives
\[
	\int_0^\infty\d r\,\frac{r^3}{(r^2+a^2)^3} 
	=\frac{1}{a^2}\int_0^{\pi/2} \d\varphi\,\cos\varphi\,\sin^3\varphi =  \frac{1}{4a^2}~,
\]
we can perform the integral over $r$ to obtain
\[
	\lambda_+ = -8\im\int_0^1\d t\,t^{1/2}(1-t)^{1/2}\int_{\R}\d x\,\frac{1}{t(x-1)^2 + (1-t)x^2}~.
\]
Noticing that $t(x-1)^2 + (1-t)x^2 = x^2-2xt+t = (x-t)^2+t(1-t)$, we recognise that the integral over $x$ is elementary, giving
\[
	\lambda_+ = -8\pi \im \int_0^1\d t\,\frac{t^{1/2}(1-t)^{1/2}}{t^{1/2}(1-t)^{1/2}} 
	= -8\pi \im~,
\]
which is the value used in the main text. 

The second of the two integrals in equation \eqref{eq:bOPE1} does not contribute to the OPE, but we include it anyway for completeness. It can be written as
\[
\begin{aligned}
	\lambda_- &= \int_{\R}\d x\int_{\C}\d^2 z\,
	\frac{\text{sgn}(x)\,|z|^2}{(x^2 + |z|^2)^{3/2}((x-1)^2+|z|^2)^{3/2}} \\
	&= -4\pi \im\int_{\R}\d x\int_0^\infty\d r\,
	\frac{\text{sgn}(x)\,r^3}{(x^2 + r^2)^{3/2}((x-1)^2+r^2)^{3/2}}\,,
\end{aligned}
\]
where as for $\lambda_+$ we have mapped $x\mapsto-x$ in the first term of $\mathcal{I}_-$ to get an integral over all of $\R$. The major difference here is the appearance of the function $\text{sgn}(x)$ in the integrand. This doesn't prevent us from performing the integral over $r$ in exactly the same way as before, giving
\[
	\lambda_- = -8\im\int_0^1\d t\,t^{1/2}(1-t)^{1/2}\int_{\R}\d x\,
	\frac{\text{sgn}(x)}{(x-t)^2+t(1-t)}~.
\]
The integral over $x$ is marginally more involved than before and gives
\[
\begin{aligned}
	\int_{\R}\d x\,\frac{\text{sgn}(x)}{(x-t)^2+t(1-t)} 
	&= \frac{1}{t^{1/2}(1-t)^{1/2}}
	\Bigg(\bigg[\arctan\bigg(\frac{x-t}{t^{1/2}(1-t)^{1/2}}\bigg)\bigg]^\infty_0
	-\bigg[\arctan\bigg(\frac{x-t}{t^{1/2}(1-t)^{1/2}}\bigg)\bigg]^0_{-\infty}\Bigg)\\
	&= \frac{2}{t^{1/2}(1-t)^{1/2}}\arctan\bigg(\frac{t^{1/2}}{(1-t)^{1/2}}\bigg)~.
\end{aligned}
\]
Thus we are left with
\[
	\lambda_- = -16\im \int_0^1\d t\,\arctan\bigg(\frac{t^{1/2}}{(1-t)^{1/2}}\bigg) 
	= -16\im\int_0^{\pi/2}\d\varphi\,\varphi\sin{2\varphi} = -4\pi \im ~\,.
\]
We conclude that
\[ 
	\lambda_{\pm} = -(6\pm2)\pi \im\,,
\]
which were the values used in the main text.

Finally, we evaluate the integral in equation~\eqref{eq:bOPE2}. We have
\[
\begin{aligned}
	\mu_+ &= \frac{1}{4}\int_{-\infty}^0\d x\int_{\C}\d^2 z\,
	\frac{|z|^2}{((x-1)^2+|z|^2)^{3/2}((x+1)^2+|z|^2)^{3/2}} \\
	&=\frac{1}{8}\int_{\R}\d x\int_{\C}\d^2 z\,
	\frac{|z|^2}{((x-1)^2+|z|^2)^{3/2}((x+1)^2+|z|^2)^{3/2}}~.
\end{aligned}
\]
Making the substitutions $x = 2\tilde x - 1$ and $z=2\tilde z$, this becomes 
\[
	\mu_+ = \frac{1}{16}\int_{\R}\d \tilde x\int_{\C}\d^2 \tilde z\,
	\frac{|\tilde z|^2}{((\tilde x-1)^2+|\tilde z|^2)^{3/2}({\tilde x}^2+|\tilde z|^2)^{3/2}} 
	= \frac{\lambda_+}{16}~.
\]
We conclude that
\[ 
	\mu_+ = - \frac{\pi}{2} \im
\]
which was used in the text.

\vspace{1cm}

\noindent {\large{\bf Acknowledgments}}

It is a pleasure to thank Nick Dorey, Paul Fendley, David Tong and Jack Williams for helpful discussions. This work has been partially supported by STFC consolidated grant ST/P000681/1. The work of RB is supported by EPSRC studentship EP/N509620/1.

\vspace{1cm}

\bibliographystyle{JHEP}
\bibliography{references}

\end{document}